\documentclass[prb,twocolumn,showpacs,superscriptaddress]{revtex4-1}
\usepackage[dvipdfmx]{graphicx}
\usepackage{amsmath,amssymb,bm}

\begin{document}

\title{
Field-induced SU(4) to SU(2) Kondo crossover  in a half-filling nanotube dot:  \\ 
spectral and finite-temperature properties
}

\author{Yoshimichi Teratani}
\affiliation{
Department of Physics, Osaka City University, Sumiyoshi-ku, 
Osaka, 558-8585, Japan
}

\author{Rui Sakano}
\affiliation{
The Institute for Solid State Physics, 
the University of Tokyo, Kashiwa, Chiba 277-8581, Japan
}

\author{Tokuro Hata }
\affiliation{
Department of Physics, Osaka University, Toyonaka, 
Osaka 560-0043, Japan
}

\author{Tomonori Arakawa}
\affiliation{
Department of Physics, Osaka University, Toyonaka, 
Osaka 560-0043, Japan
}
\affiliation{
Center for Spintronics Research Network, Osaka University, Toyonaka, Osaka  560-8531, Japan
}

\author{Meydi Ferrier}
\affiliation{
Department of Physics, Osaka University, Toyonaka, 
Osaka 560-0043, Japan
}
\affiliation{
Laboratoire de Physique des Solides, CNRS,
Universit\'{e} Paris-Sud, 
Universit\'{e} Paris Saclay, 
91405 Orsay Cedex, 
France
}

\author{Kensuke Kobayashi}
\affiliation{
Department of Physics, Osaka University, Toyonaka, 
Osaka 560-0043, Japan
}
\affiliation{
Institute for Physics of Intelligence and Department of Physics, The University of Tokyo, Tokyo 113-0033, Japan
}

\author{Akira Oguri}
\affiliation{
Department of Physics, Osaka City University, Sumiyoshi-ku, 
Osaka, 558-8585, Japan}
\affiliation{ 
Nambu Yoichiro Institute of Theoretical and Experimental Physics, 
Sumiyoshi-ku, 
Osaka, 558-8585, Japan
}

\date{\today}

\begin{abstract} 
We study finite-temperature properties of the Kondo effect in a carbon nanotube (CNT) quantum dot 
using the Wilson numerical renormalization group (NRG).
In the absence of magnetic fields, four degenerate energy levels of the CNT consisting of spin and orbital degrees of freedom give rise to the SU(4) Kondo effect.
We revisit the universal scaling behavior of the SU(4) conductance for quarter- and half-filling in a wide temperature range. 
We find that the filling dependence of the universal scaling behavior at low temperatures $T$ can be explained clearly with an extended Fermi-liquid theory.
This theory clarifies that a $T^{2}$ coefficient of conductance becomes zero at quarter-filling whereas
the coefficient at half-filling is finite.
We also study a field-induced crossover from the SU(4) to SU(2) Kondo state observed at the half-filled CNT dot.
The crossover is caused by the matching of the spin and orbital Zeeman splittings, which lock two levels among the four 
at the Fermi level even in magnetic fields $B$.
We find that the conductance shows the SU($4$) scaling behavior at $\mu_{B}B<k_{B}T_{K}^{\mathrm{SU(4)}}$ 
and it exhibits the SU($2$) universality at $\mu_{B}B\gg k_{B}T_{K}^{\mathrm{SU(4)}}$, where $T_{K}^{\mathrm{SU(4)}}$ is 
the SU($4$) Kondo temperature.
To clarify how the excited states evolve along the SU(4) to SU(2) crossover,  
we also calculate the spectral function.  
The results show that the Kondo resonance width of the two states locked at the Fermi level becomes sharper with increasing fields.
The spectral peaks of the other two levels moving away from the Fermi level merge with atomic limit peaks for $\mu_{B}B \gtrsim k_{B}T_{K}^{\mathrm{SU(4)}}$.

\end{abstract}


\maketitle

\section{Introduction}
\label{sec:introduction}

Quantum dots provide an ideal testbed
to investigate
strong correlations between the electrons in localized levels 
and the conduction electrons in reservoirs.
Kondo effect\cite{Kondo-Resistance-Minimum,HewsonBook}
is a typical many-body phenomenon that occurs also in quantum dots 
having local spin degrees of freedom.
The Kondo effect in quantum dots has been studied 
theoretically\cite{Glazman_Kondo,T_K_Ng_phys_rev_lett} and 
experimentally.\cite{Goldhaber_Gordon_qd_kondo_phys_rev_lett,van2000kondo}
In addition to the spin degrees of freedom,
carbon nanotube (CNT) quantum dots 
have also the orbital (valley) degrees of freedom, 
corresponding to clockwise and counter clockwise orbitals 
around the nanotube axis.\cite{RMP-Kouwenhoven}
Four energy levels consisting of the spin and orbital degrees of freedom
give rise to the SU(4) Kondo effect 
in the case where the four localized states are degenerate.
\cite{Sasaki2000,Choi2005,Izumida1998,Broda2003,Anders-SU4,SU4_Kondo_ferro_Weymann,Nishikawa_emergence_SU4_Kondo}
A number of experiments for non-equilibrium transport have observed the SU(4) Kondo effect.\cite{Makarovski, Goldhaber-Gordon2014, Franceschi2005,Gotz_Schimid_Kondo_in_CNT,
Meydi-Nat-Phys,Hata_san_cnt_sc}
Perturbations such as spin-orbit coupling $\Delta_\mathrm{SO}$, 
valley mixing $\Delta_{K,K^{\prime}}$ and magnetic fields $B$ break the SU($4$) symmetry.  
Effects of such perturbations on the Kondo state are theoretically studied
for instance, using  
Wislon's numerical renormalization group (NRG)
approach\cite{WilsonRMP,Wilson-NRG,Wilson-NRG2}
 which has been extended to explore  transport coefficients 
and spectral functions with very high accuracy.
\cite{AndersSchiller2005,Anders-Complete-Fock,WeichselbaumVonDelft,Anders-SU4}

The main purpose of the present paper is to clarify the finite temperature properties of the Kondo effect in CNT dots. 
In the first half of this paper, we study the scaling behavior of the SU($4$) conductance at 
quarter- and half-filling.
Although the scaling behavior has been studied,
\cite{Anders-SU4, Goldhaber-Gordon2014, GalpinLoganAnders2010, MantelliMocaZarandGrifoni, Bas_SU4_scaling} 
 we revisit it with an extended microscopic Fermi-liquid 
 theory 
which describes transport phenomena at low-temperatures.\cite{Mora_Fermi_liquid_Kondo_2009,Mora_Fermi_liquid_2015,Filippone_Fermi_liquid_in_magnetic_field,Oguri_Higher_Order_Non_Equilibrium,teratani2020fermi} 
It has recently been shown 
that the $T^{2}$ coefficient $C_{T}$ for the conductance is 
determined in terms of five Fermi-liquid parameters: electron filling, two linear susceptibilities, and two non-linear susceptibilities which are defined with respect to the equilibrium ground state.
Calculating these parameters using the NRG, 
 we  successfully explain the filling dependence of the scaling with the description.  
Specifically, we explore the filling dependence of a $T^{2}$ coefficient $C_{T}$ and find that 
$C_{T}$ becomes zero at quarter-filling whereas it is finite at half-filling.
  
In the second half of this paper,
we examine a field induced crossover from SU($4$) to SU($2$) Kondo state at half-filling nanotube dot.\cite{Meydi-Phys-Rev-Lett, Teratani-JPSJ}
At the valley where the crossover occurs, the SU(4) Kondo resonance emerges in the absence of magnetic field, 
because $\Delta_{K,K^{\prime}}$ and $\Delta_\mathrm{SO}$ are smaller than the SU(4) Kondo energy scale, $k_{B}T_{K}^\mathrm{SU(4)}$.
The field induced crossover is different from the other crossover occurring at quarter-filling.\cite{GalpinLoganAnders2010, Grifoni_PRB2015, MantelliMocaZarandGrifoni,Lopes-SU4-SU2-Crossover, Busser2007,Lim,Kleeorin_Meir_reemergence_SU4_SU2,Bas_Aligia_SU4_thermo,
Krychowski2018}
Specifically, this crossover at half-filling occurs in a situation where two localized levels among the four remain the Fermi level even in magnetic fields
while the other two levels move away from the Fermi level.
The situation realizes if the spin Zeeman splitting coincides the orbital splitting. 
Such coincidence can reasonably occur in real CNT dots.

In the previous work, 
we have studied
the crossover occurring in this situation, by calculating quasi-particle 
parameters such as phase shift $\delta$, wave function renormalization factor $Z$ and Wilson ratio $R$.\cite{Teratani-JPSJ}
We have found that 
the applied magnetic fields enhance the electron correlations. 
For instance,  
as magnetic fields increase,
the renormalization factor 
$Z$ decreases 
from the SU(4) value to the SU(2) value and 
thus the Kondo energy scale $T_{K}$ decreases.
Our NRG results 
are in good agreement with the experimental observations.\cite{Meydi-Phys-Rev-Lett}

In this paper, we calculate the temperature dependence of the conductance 
in magnetic fields to clarify the crossover in a wide range of temperature $T$.
We show that a temperature scale $T^{*}$ around which the conductance shows  $\log T$ dependence
decreases with increasing magnetic fields.
This decrease of $T^{*}$ becomes clearer in a strong Coulomb interaction case
and agree with the field dependence of $Z$.
We also examine the scaling behavior of the conductance.
Whereas the conductance follows the SU(4) scaling at $\mu_{B}B< k_{B}T_{K}^{\mathrm{SU(4)}}$, it shows the SU(2) scaling at $\mu_{B}B\gg k_{B}T_{K}^{\mathrm{SU(4)}}$.

In addition to the conductance, we calculate the level resolved spectral functions in magnetic fields.
The component for the doubly degenerate levels shows that 
the Kondo resonance width becomes sharper with increasing magnetic fields.
This field dependence of the width corresponds to that of $T^{*}$.
Spectral weights of the other two states transfer from the Fermi level,
and the peaks merge with atomic limit peaks.

This paper is organized as follows. 
In the next section, 
we describe the microscopic Fermi-liquid theory and the NRG approach to CNT dots.
In Sec.\ \ref{sec:CNT_dot_energy_level}, we examine the scaling behavior of the SU($4$) conductance 
at quarter- and half-filling.
We discuss how the quasi-particle parameters evolve during the field induced crossover in Sec.\ \ref{b_dependence_quasi}.
We present the NRG results of conductance 
and discuss the influence of magnetic fields on 
the temperature dependence of conductance in Sec.\ \ref{Conductance_finT}.
The spectral functions in increasing magnetic fields are shown in Sec.\ \ref{Spectral_properties_along_the_field-induced_crossover}.
Summary is given in Sec.\ \ref{sec:summary}.

\section{Formulation}
 \label{sec:model_for_CNT}

Transport properties of carbon nanotube quantum dots are  
determined by a linear combination of four one-particle levels,  
consisting of the spin ($\uparrow$,$\downarrow$) and valley (K, K') degrees of freedom.
The structure of these four states staying near the Fermi level 
 depend on the inter-valley scattering, 
the spin-orbit  coupling, and the Zeeman splittings 
of the spin and orbital degrees of freedom. 
We introduce the Anderson model for the CNT dot in this section 
 using a diagonal form for these local levels. 
We provide a more microscopic description specific to a real CNT dot,
in which experiments have measured the current and corresponding noise with high accuracy.
The renormalized parameters that characterize the low-energy 
Fermi-liquid properties and details of the NRG calculations 
are also described in this section. 

\subsection{Anderson model for  CNT quantum dots}
 \label{sec:Hamiltonian}

We start with an Anderson impurity model 
for a CNT dot,  which  has  $N=4$ internal degrees of freedom 
and is  connected to two noninteracting leads:  
\begin{align}
& \mathcal{H}\,=\, 
 \mathcal{H}_{d}^{0} 
+ \mathcal{H}_{U}^{} 
+\mathcal{H}_{c}
+\mathcal{H}_{T},  
\label{eq:Hamiltonian} 
\\ 
&\mathcal{H}_{d}^{0} \,= 
\sum_{m=1}^N
 \epsilon_m^{}  
d_{m}^{\dagger} d_{m}^{} , 
\qquad 
 \mathcal{H}_{U}^{} =
 \,U \! 
\sum_{m< m'} n_{dm} n_{dm'} , 
\label{eq:Hd} 
\\
&\mathcal{H}_{c}\,=\, 
 \sum_{\nu=L,R}
\sum_{m=1}^N
\int_{-D}^D \! d\varepsilon \, 
\varepsilon\, 
c_{\nu,\varepsilon m}^{\dagger} 
c_{\nu,\varepsilon m}^{} 
\label{eq:HC}
\\
&\mathcal{H}_{T}^{} =
\sum_{\nu=L,R}\sum_{m=1}^N v_{\nu}^{}
\left( 
\psi_{\nu,m}^{\dagger} d_{m}^{} + 
d_{m}^{\dagger} \psi_{\nu,m}^{} 
\right) , 
\label{eq:HT} 
 \\
&\psi_{\nu,m} 
\,\equiv\,
 \int_{-D}^D \! d\varepsilon \,\sqrt{\rho_c^{}} \, 
c_{\nu,\varepsilon, m}^{} , 
 \qquad n_{dm}^{} \equiv  d_m^\dagger d_m^{} . 
\end{align}
Here,  $d_{m}^{\dagger}$ and $d_{m}^{}$ are  
the creation and annihilation operators, respectively,  
for an electron with energy $\varepsilon_{m}$ 
in the $m$-th discrete one-particle eigenstate of the dot ($m=1,2,\ldots, N$).
We shall also call $m$ the ``{\it flavour}'' in the following.
The Coulomb interactions $U$ between 
the electrons occupying the dot levels  are assumed to be independent of  $m$, 
 assuming that the intra- and inter-valley Coulomb repulsions to be identical.
We  also assume that  Hund's rule coupling can be neglected.  
This is consistent with the CNT dot in which  
 the field-induced SU($4$) to SU(2) Kondo crossover 
has been observed\cite{Meydi-Nat-Phys}, 
and also with the other nanotube dots.\cite{Makarovski} 
$\mathcal{H}_{c}$ describes conduction bands in the leads on the left and right.
$\nu=L$ and $\nu=R$ respectively denote the left and right lead.
The conduction electrons are assumed to carry the flavour index  $m$, 
and the continuous energy states in the bands are normalized such that       
$\{ c_{\nu,\varepsilon m}^{}\,,\, c_{\nu',\varepsilon' 
m'}^{\dagger} \} = \delta(\varepsilon-\varepsilon')\, 
\delta_{\nu\nu'}\delta_{mm'}$.  
The Fermi level  is chosen to be   $\varepsilon_{F} =  0$,  
 at the center of  the flat conduction bands with the width $2D$.  
$\mathcal{H}_{T}^{}$ describes charge transfer between the dot and leads.
We assume that the  tunneling matrix element $v_{\nu}$ 
is independent of flavour $m$, 
which can also be  justified 
for a class of CNT dots.\cite{Meydi-Nat-Phys,Meydi-Phys-Rev-Lett}
In this situation, the resonance width due to these hybridizations 
 becomes  $\Delta =  \Delta_{L}+\Delta_{R}$  
with $\Delta_{\nu} \equiv \pi\rho_{c}v_{\nu}^{2}$,  
and  $\rho_{c}=1/2D$ the density of states of the conduction band. 
Unless other wise stated, we set the Boltzmann constant $k_{B}$ to unity i.e., $k_{B}=1$.

This Hamiltonian $\mathcal{H}$ has the SU($N$) symmetry when 
all the impurity energies are equal  $\epsilon_{m}\equiv\epsilon_{d}$.
This is because $\mathcal{H}_{T}$ that describes charge transfer preserves 
the flavour index $m$, 
and the Coulomb interaction 
is determined essentially by the total number of impurities electrons  
$\widehat{N}_{d}\equiv \sum_{m} n_{dm}$ 
 as $\mathcal{H}_{U}=(U/2)\,\widehat{N}_{d}\,(\widehat{N}_{d}-1)$.
In Appendix \ref{sec:Symmetries_of_Hamiltonian}, we explain the symmetry of this Hamiltonian in more detail.

\subsection{Fermi-liquid parameters}
\label{sec:FL_prameters}

The equilibrium retarded Green $G_{m}^r(\omega)$ is an useful tool 
to study transport coefficients of  quantum dots. 
$G_{m}^r(\omega)$ is given by
\begin{align}
G_{m}^r(\omega, T)\,&\equiv \, -i \int_0^{\infty}\! dt \, 
e^{i (\omega +i0^+ )t} 
\left\langle \{ d_{m}^{}(t)\, , \,d_{m}^{\dagger}(0) \} \right\rangle
\label{definition-Gm}
\\
&=\,\frac{1}{\omega\,-\,\epsilon_{m}\,+\,i\Delta\,-\,\Sigma_{m}(\omega,T)}, \label{Dyson_eq} \\
A_{m}^{}(\omega, T)\,&\equiv 
\, -\, \frac{1}{\pi}\, \mbox{Im}\,G_{m}^r(\omega,T) .
\label{def_A_m}
\end{align}
Here, $\langle \mathcal{O} \rangle \equiv 
\mbox{Tr}\left[\mathcal{O}\, e^{-\mathcal{H}/T}\right]/\,
\Xi $ with $\Xi \equiv \mbox{Tr}\,e^{-\mathcal{H}/T}$ 
is the thermal average of an observable  $\mathcal{O}$. 

The phase shift $\delta_{m}$  is a  primary parameter 
that characterizes the Fermi-liquid ground state. 
 It  is determined by the value of the self-energy $\Sigma_{m}(\omega,T)$ 
at  $\omega=T=0$,
\begin{align}
\delta_{m}\,=\,\frac{\pi}{2}\,-\,\tan^{-1}\biggl[\frac{\epsilon_{m}\,+\,\Sigma_{m}(0,0)}{\Delta}\biggr].
\label{phase_shift_def}
\end{align}
The Friedel sum rule also  relates  the phase shift $\delta_m$ 
to the occupation number $\langle n_{d m} \rangle$, \cite{Langreth}
\begin{align}
\langle n_{d m} \rangle\,=\,\frac{\partial \Omega}{\partial \epsilon_{m}}\xrightarrow{T\to 0}\,\frac{\delta_{m}}{\pi}.
\label{Friedel_sum_rule}
\end{align}
Here,  $\Omega = -T\ln \mathrm{e}^{-\mathcal{H}/T}$ is the free energy. 
The phase shift  also determines the value of  the  spectral function at  $\omega=T=0$, 
\begin{align}
A_{m}(0,0)\,=\,\frac{\sin^{2}\delta_{m}}{\pi\Delta}. 
\label{zero_T_spectral}
\end{align}

The linear-susceptibilities $\chi_{m_{1},m_{2}}$ are also  important 
parameters that  determine the Fermi-liquid properties: 
\begin{align}
 \chi_{m_{1},m_{2}}^{}  
\, \equiv   \,  
\int_0^{1/T} \!\!  d \tau 
\left\langle  \delta n_{d m_{1}}(\tau)\,\delta  n_{d m_{2}}\right\rangle.    
\label{two_body_cor_def}
\end{align}
Here, $\delta  n_{d m}\equiv n_{d m}-\langle n_{d m} \rangle$ is the fluctuation of the occupation number, and $\delta n_{d m}(\tau)\,=\,\mathrm{e}^{\tau \mathcal{H}}\delta n_{d m} \mathrm{e}^{-\tau \mathcal{H}}$.
The derivative of the self-energy with respect to the impurity level also gives the susceptibilities,
\begin{align}
\chi_{m_{1},m_{2}}\,&=\,-\frac{\partial^{2}\Omega}{\partial \epsilon_{m_{1}}\partial \epsilon_{m_{2}}}\,=\,-\,\frac{\partial \langle n_{d m_{1}}\rangle}{\partial \epsilon_{m_{2}}} \notag \\
&\xrightarrow{T\to 0}\,A_{m_{1}}(0,0)\left(\delta_{m_{1},m_{2}}\,+\,\frac{\partial \Sigma_{m_{1}}(0,0)}{\partial \epsilon_{m_{2}}}\right).  
\end{align}

The Ward-Takahashi identities relate the linear-susceptibilities to the wave function renormalization factor $Z_{m}$ and the vertex function $\Gamma_{m m'; m' m}(0,0;0,0)$, \cite{ShibaKorringa,Yoshimori} 
\begin{align}
\chi_{m,m}\,&=\,\frac{A_{m}(0,0)}{Z_{m}},\ \ \frac{1}{Z_{m}}\,\equiv\,1-\left.\frac{\partial \Sigma_{m}(\omega,0)}{\partial \omega}\right|_{\omega=0},    \label{chi_parallel_ward_takahashi}  \\
\chi_{m,m'}\,&=\,-A_{m}(0,0)\,A_{m'}(0,0)\,\Gamma_{m m'; m' m}(0,0;0,0).
\label{chi_anti_parallel_ward_takahashi}
\end{align}
$\Gamma_{m m'; m' m}
(\omega,\omega';\omega',\omega) $ 
is the $m\neq m'$ component of  the vertex correction, 
defined at  $T=0$ for the causal Green's functions.  
Note that the intra-level components for $m=m'$  
vanish  at zero frequencies,  $\Gamma_{m m;m m}(0,0;0,0)=0$, 
because of the fermionic antisymmetrical properties, i.e.,  the Pauli exclusion principle.
The renormalized level position $\widetilde{\epsilon}_{m}$ and corresponding resonance width $\widetilde{\Delta}_{m}$ 
are determined by $Z_{m}$,
\begin{align}
\widetilde{\epsilon}_{m} \equiv Z_{m} 
\, \bigl[ \,\epsilon_{m}+\Sigma_{m}(0)
\, \bigr], 
\qquad \quad 
\widetilde{\Delta}_{m} \equiv Z_{m}\Delta. \label{renormalized_level_and_width}
\end{align}

The residual interaction between quasi-particles $\widetilde{U}_{m,m'}$ 
is also an essential Fermi-liquid parameter:\cite{HewsonRPT2001} 
\begin{align}
\widetilde{U}_{m,m'}\,\equiv\,Z_{m}Z_{m^{\prime}}
\Gamma_{m m';m' m}(0,0;0,0) \;.
\label{def_residual_interaction}
\end{align}
The Wilson ratio $R_{m,m'}$
corresponds to  a dimensionless residual interaction,\cite{hewson2004renormalized} 
and it  generally depends on  $m$ and $m'$:  
\begin{align}
R_{m,m'}^{}\,\equiv \,1\,+
\sqrt{\widetilde{A}_{m}^{}\,\widetilde{A}_{m'}^{}} 
\ \widetilde{U}_{m,m'}^{} \;.
\label{def_Wilson_ratio}
\end{align}
Here, $\widetilde{A}_{m}\equiv A_{m}(0,0)/Z_{m}$ is the density of states of the quasi-particles.
Using Eqs.\eqref{chi_parallel_ward_takahashi}-\eqref{def_residual_interaction}, 
$R_{m,m'}$ can also be expressed in terms of the linear susceptibilities,
\begin{align}
R_{m,m'}\,-\,1\,=\,-\, \frac{\chi_{m,m'}}{\sqrt{\chi_{m,m}\,\chi_{m',m'}}}.
\label{Wilson_ratio_linear_sus}
\end{align}
we calculate the renormalization factor $Z_{m}$ and residual interaction $\widetilde{U}_{m,m'}$ using the NRG,
and deduce the Wilson ratio $R_{m,m'}$ from Eq.\ \eqref{def_Wilson_ratio}.
We note that the linear susceptibility $\chi_{m,m}$ determines 
the $T$-linear specific heat $\mathcal{C}_{\mathrm{dot}}$ due to the impurity, 
\begin{align}
\mathcal{C}_{\mathrm{dot}} = \gamma \,T, \qquad \gamma \,=\, \frac{\pi^2}{3} \sum_m  
\chi_{m,m} \;. \label{specific_heat}
\end{align}
Furthermore, fluctuations of the impurity electron filling are described by the 
charge susceptibility $\chi_c^{} \equiv \sum_m \chi_{c,m}^{}$,  
which can also be expressed in terms of  
 $\widetilde{A}_{m}$ and  $\widetilde{U}_{m,m'}$
using Eqs.\ \eqref{chi_parallel_ward_takahashi}   
and \eqref{chi_anti_parallel_ward_takahashi}:  
\begin{align}
\chi_{c,m}^{} \,&\equiv \  
- \sum_{m'}
\frac{\partial \left\langle n_{dm}\right\rangle}{\partial \varepsilon_{m'}}
\,=\,\sum_{m'}\,\chi_{m,m'} 
\nonumber \\
&= \ 
\widetilde{A}_{m}^{}(0) \left[ 
\, 1 - \sum_{m' (\neq m)}
 \widetilde{U}_{m,m'} \widetilde{A}_{m'}^{}(0) 
\, \right] . 
\label{eq:chi_c}
\end{align}
The last line shows that the  residual interaction $\widetilde{U}_{m,m'}$ 
 reduces the free-quasiparticle contributions given by 
the first term in the right-hand side.

\subsection{Conductance and non-linear susceptibilities}
\label{sec:linear_conductance_and_non-linear_susceptibilities}
The conductance $g_\mathrm{tot}^{}$ through a quantum dot can be 
expressed in the Landauer form, \cite{Landauer,MeirWingreen,HershfieldDaviesWilkins}
\begin{align}
g_\mathrm{tot}^{} \, = & \    \sum_{m=1}^{N} g_{m} ,
\\ 
g_{m}\,= &  \   
\frac{e^{2}}{h}\,
\frac{4\Delta_{L}\Delta_{R}}{\Delta_{L}+\Delta_{R}}
\int_{-\infty}^{\infty} \! 
d\omega\,\left(-\frac{\partial f(\omega)}{\partial \omega} \right)\,
\pi A_{m}^{}(\omega,T)  . 
\label{conductance_finT}
\end{align}
Here, $f(\omega)=1/(e^{\omega/T}+1)$ is the Fermi distribution function.
It has recently been clarified that 
 low-temperature behavior of conductance up to order $T^2$ 
can be determined completely by an extended Fermi-liquid theory.
\cite{Mora_Fermi_liquid_2015,Filippone_Fermi_liquid_in_magnetic_field,Oguri_Higher_Order_Non_Equilibrium} 
This formulation is also applicable to the multilevel Anderson impurity model  
at arbitrary electron fillings, and the low-temperature expansion can be expressed 
in the following form,  for symmetric tunneling couplings $\Delta_L=\Delta_R$,  
\cite{Oguri_Higher_Order_Non_Equilibrium,teratani2020fermi}  
\begin{align}
g_{m}\,&=\,\frac{e^{2}}{h}\,\left[
\,\sin^{2}\delta_{m}\,+\,c_{T,m}\,\bigl(\pi T \bigr)^{2}\,+\,\cdots \right]. 
\label{g_m_low_T} 
\end{align} 
Here, the first term in the right-hand side  
corresponds to the ground-state value 
that is determined by the transmission probability  $\sin^{2}\delta_{m}$.
The coefficient $c_{T,m}$ for the term of order $T^2$ consists of two parts:
\cite{Mora_Fermi_liquid_2015,Filippone_Fermi_liquid_in_magnetic_field,Oguri_Higher_Order_Non_Equilibrium,teratani2020fermi} 
\begin{align}
c_{T,m}^{} \,&=\,  \frac{\pi^2}{3} \Bigl[\,w_{T,m}\,+\,\theta_{T,m}\, \Bigr], 
\label{C_T_general} 
\\
w_{T,m}\,&=\,-\cos 2 \delta_{m}\biggl(\chi_{m,m}^2+ 2 \sum_{m'(\neq m)}\chi_{m,m'}^2 \biggr), 
\label{W_t_general} 
\\
\theta_{T,m}\,&=\,\frac{\sin 2\delta_{m}}{2\pi}\,\biggl(\chi_{m,m,m}^{[3]}+ \sum_{m'(\neq m)}\chi_{m,m',m'}^{[3]}\biggr). 
\label{Theta_t_general} 
\end{align}
Here,  $w_{T,m}$ represents the two-body contributions  
which are determined by the linear susceptibilities $\chi_{m,m'}$.  
The other part  $\theta_{T,m}$ represents 
the three-body contributions, 
determined by  the non-linear susceptibilities,
\cite{Oguri_Higher_Order_Non_Equilibrium,teratani2020fermi} 
\begin{align}
\chi_{m_{1},m_{2},m_{3}}^{[3]} \equiv\, -\!\!\int_{0}^{\beta}\!\!\!d\tau_{1}\!\!\int_{0}^{\beta}\!\!\!d\tau_{2}\bigl\langle T_{\tau}\delta n_{d m_{3}}(\tau_{3})\delta n_{d m_{2}}(\tau_{2})\delta n_{d m_{1}} \bigr\rangle.
\label{three_body_cor}
\end{align}
Here, $T_{\tau}$ is the imaginary time ordering operator.
This correlation function corresponds also to the derivative of linear susceptibilities:  
\begin{align}
\chi_{m_{1},m_{2},m_{3}}^{[3]}\,=\,-\frac{\partial^{3}\Omega}{\partial\epsilon_{m_{1}}\partial\epsilon_{m_{2}}\partial\epsilon_{m_{3}}}\,=\,\frac{\partial \chi_{m_{2},m_{3}}}{\partial \epsilon_{m_{1}}}.
\label{non_linear_susceptibility}
\end{align}
We note that $\chi_{m_{1},m_{2},m_{3}}^{[3]}$ vanishes 
in the case at which the system has both the particle-hole and time-reversal symmetries.

\subsection{NRG approach to the spectral function and transport coefficients}
\label{sec:NRG_section}

The NRG  has successfully been applied to 
multi-orbital quantum dots including CNT dots 
since a seminal work of Izumida {\it et al\/}.
\cite{Izumida1998,izumida2001kondo, Broda2003, Lim, Grifoni_PRB2015, A.Oguri-1/N-1-away-half, NishikawaCrowHewson1,Stadler-Interleaved-NRG}
Using the NRG, we calculate the renormalized parameters 
such as the phase shifts of electrons $\delta$, 
the wavefunction renormalization factor $Z$, 
and the Wilson ratio $R$.
\cite{NozieresFermiLiquid,Yosida-Perturbation,Yamada-Perturbation}
The present work uses the NRG 
to calculate not only  the renormalized parameters, 
but also the linear conductance $g$  and the spectral function $A_{m}^{}(\omega,T)$.  
\cite{WilsonRMP,Wilson-NRG,Wilson-NRG2}

The key approximation of the NRG is the  logarithmic discretization of the conduction band.
A dimensionless parameter $\Lambda$ ($>1$) controls the discretization.  
The noninteracting part of the discretized  Hamiltonian 
$\mathcal{H}_d^0 + \mathcal{H}_T + \mathcal{H}_c$ 
is  transformed into  a one-dimensional tight-biding chain 
with exponentially decaying  hopping matrix elements
 $t_n \sim D\Lambda^{-n/2}$. 
Then, the total Hamiltonian  $\mathcal{H}$ including the interactions  
can be diagonalized iteratively by adding the states on the tight-biding chain, 
 starting from the impurity site. 
Owing to the exponential decay of $t_n$, 
high energy states can be discarded at each successive step 
without affecting  low-lying energy states so much.  
Although this iteration itself is an artificial procedure, 
it can be  interpreted as a process to probe  lower energy scale,  
step by step,  stating from  high-energy scale.\cite{WilsonRMP} 
We briefly explain the process of the NRG iteration in Appendix \ref{Overview_of_NRG_iterations}.
Furthermore, we can deduce the quasi-particles parameters 
$Z_{m}$,  $\widetilde{\epsilon}_{m}$, and $\widetilde{U}_{m,m'}^{}$, 
from asymptotic behaviors of the low-lying eigenvalues near 
the NRG fixed point (see Appendix \ref{Calculation_method_of_Fermi-liquid_parameters}).\cite{Wilson-NRG,Wilson-NRG2,hewson2004renormalized}

In the present work, we explore the CNT dots 
in a case where the four-fold degeneracy is not completely lifted  
by the magnetic field but still a double degeneracy remains 
for the reason which will be discussed  in more detail 
in Sec.\ \ref{subsec:CNT_Zeeman_matching}.

Our NRG code uses the U(1)$\otimes$[SU(2)$\otimes$U(1)]$\otimes$U(1) symmetries 
to explore the SU(4) to SU(2) Kondo crossover.
The U(1)$\otimes$U(1)$\otimes$U(1)$\otimes$U(1) symmetries are used to examine 
how perturbations, i.e., the valley mixing and spin-orbit interaction, affect the crossover because 
the perturbations break the SU(2) symmetry.   
The NRG calculations are carried out  keeping typically the lowest $4100$ states 
in the truncation procedure,  choosing the discretization parameter 
to be $\Lambda=6.0$.
Furthermore,   the spectral function and temperature-dependent  
 conductance \cite{Z-average-Oliveira,Costi-1ch-mag-finT}  
are obtained using the complete Fock-space basis algorithm, 
\cite{AndersSchiller2005,Anders-Complete-Fock,WeichselbaumVonDelft} 
together with 
the Oliveira {\it z averaging\/}.\cite{MYoshidaWhitakerOliveira,Z-average-Oliveira} 
For obtaining the spectral functions, we also calculate 
the higher-order correlation function in order to directly deduce 
the self-energy $\Sigma_{m}(\omega,T)$.
\cite{Bulla-Self-Energy}
These supplemental techniques for the dynamical correlations functions 
are also described in Appendix \ref{sec:NRG_calculations}.

\section{Field-induced SU(4) to SU(2) crossover of  Kondo singlet state
}

\label{sec:CNT_dot_energy_level}

\subsection{Microscopic description for the CNT-dot levels}

The Hamiltonian $\mathcal{H}$  
 defined in Eqs.\ \eqref{eq:Hamiltonian}--\eqref{eq:HT}  
are described using a representation in which 
 the dot part  $\mathcal{H}_d^{0}$ has already been diagonalized.     
However, to see a microscopic background, 
the other basis set using the spin ($\uparrow$, \!$\downarrow$) 
and valley  ($\mathrm{K}$, \!\! $\mathrm{K'}$) degrees of freedom 
is more suitable.

The valley degrees of freedom capture a magnetic moment along the CNT axis 
because of the cylindrical geometry of the CNT.  
This orbital moment couples to an external magnetic field parallel to the CNT 
axis.\cite{RMP-Kouwenhoven,MantelliMocaZarandGrifoni,Grifoni_PRB2015,Izumida2015,Izumida2016} 
The four levels of  the CNT dot are also coupled each other 
 through the spin-orbit interaction $\Delta_\mathrm{SO}$ 
and the valley mixing $\Delta_\mathrm{KK'}^{}$. 
The dot part of the Hamiltonian can be expressed 
in the following form,  using the dot-electron operator 
$ \psi^{\dagger}_{d: \ell s}$ 
for  orbital  $\ell$ ($= \mathrm{K}, \mathrm{K'}$)  
and  spin  $s$ ($= \uparrow, \downarrow $), 
\begin{align}
&\mathcal{H}_d^{0} \equiv  \, 
 \sum_{\ell, \ell'}
\sum_{s,s'}
\psi^{\dagger}_{d: \ell s}  \, 
H_{d:\ell s, \ell' s'}^0 \,   \psi^{}_{d: \ell' s'}
\ =\  \bm{\psi}^{\dagger}_{d} \, \bm{H}_d^0   \bm{\psi}^{}_{d}
\;. 
\label{eq:Hd0_micro}
\end{align}
The matrix  $\bm{H}_d^0 \equiv \{H_{d:\ell s, \ell' s'}^0\}$ is given by 
\cite{RMP-Kouwenhoven,MantelliMocaZarandGrifoni,Grifoni_PRB2015} 
\begin{align}
&
\!\!\!\!\!\!\!\!\!\!\!\!\!\!\!\!\!
\bm{H}_d^0 
=  
 \varepsilon_d^{}  \bm{1}_\mathrm{s} \!\otimes\! \bm{1}_\mathrm{orb}
 + \frac{\Delta_\mathrm{KK'}}{2} 
\bm{1}_\mathrm{s} \!\otimes\! \bm{\tau}^x \! 
+ \frac{\Delta_\mathrm{SO}}{2}  \bm{\sigma}^z \!\otimes\! \bm{\tau}^z \! 
 - \overrightarrow{\bm{\mathcal{M}}} \cdot \vec{b},
\label{eq:H_d0_mat_1}
\\
 \overrightarrow{\bm{\mathcal{M}}} \, \equiv  & \  
-\frac{1}{2} \, g_\mathrm{s} \, \vec{\bm{\sigma}} 
\!\otimes\! \bm{1}_\mathrm{orb}
- g_\mathrm{orb} \,
\bm{1}_\mathrm{s} \!\otimes\! \bm{\tau}^z  
 \,\vec{e}_{z}^{} 
.
\label{eq:dot_magnetization}
\end{align}
Here, $\bm{\sigma}^j=\{ \sigma_{ss'}^j\}$ and 
 $\bm{\tau}^j=\{ \tau_{\ell\ell'}^j\}$ 
for $j=x,y,z$ are the Pauli matrices 
for the spin and the valley pseudo-spin spaces, respectively.  
Correspondingly,  $\bm{1}_\mathrm{s}=\{ \delta_{ss'}\}$ 
and  $\bm{1}_\mathrm{orb}=\{ \delta_{\ell\ell'}\}$ 
are the corresponding unit  matrices.
In a finite magnetic field  $\vec{b}\equiv \mu_B \vec{B}$, 
where $\mu_B$ is the Bohr magneton, 
both the spin and orbital moments contribute to  
the magnetization  $\overrightarrow{\bm{\mathcal{M}}}$.     
The g-factor for the spin  is  $g_\mathrm{s}= 2.0$. 
The orbital moment couples  to the magnetic field along the nanotube axis, 
and the orbital Zeeman splitting  
is given by  $\pm g_\mathrm{orb} |\vec{b}| \cos \Theta$. 
Here,  $g_\mathrm{orb}$ is the orbital Land\'{e} factor and 
 $\Theta$ is the angle of the magnetic field relative to the nanotube axis, 
which is chosen as the $z$-axis with  $\vec{e}_{z}^{}$ the unit vector.
The one-particle Hamiltonian of the dot levels  $\bm{H}_d^0$ can be 
 diagonalized via the unitary transform with the matrix   
$\bm{\mathcal{U}}_d^{}
 =(\bm{u}_{1}^{},\bm{u}_{2}^{},\bm{u}_{3}^{},\bm{u}_{4}^{})$, 
constructed with the eigenvectors,  
\begin{align}
\bm{H}_d^0\, \bm{u}_m =& \  \epsilon_m \,\bm{u}_m \;,
\qquad \quad \ \bm{u}_m^\dagger \cdot \bm{u}_{m'}^{} \,= \, \delta_{mm'} 
\;. 
\label{eq:one_particle_level_CNT_eigen0}
\end{align}
The operator $d^{}_{m}$ that annihilates an electron in the eigenstate 
with energy $\epsilon_m$ can be expressed 
in a linear combination of  $\psi^{}_{d: \ell s}$'s 
via this transform with $\bm{\mathcal{U}}_d^{}$. 

For  $\Delta_\mathrm{SO}=\Delta_\mathrm{KK'}=0$, 
 the spin component of the magnetization becomes parallel to the field  
 $\vec{e}_\Theta^{} = \cos \Theta \,\vec{e}_{z}^{} 
+ \sin \Theta \,\vec{e}_{x}^{}$ while the  orbital component 
is in the direction along the nanotube axis ($\Theta \leq \pi/2$). 
Then,  the eigenvalues of $\bm{H}_d^0$ can be written as  
$\epsilon_1 = 
\varepsilon_d - (g_\mathrm{orb }\cos\Theta +g_\mathrm{s}/2) b$, 
$\,\epsilon_2 =  
\varepsilon_d - (g_\mathrm{orb }\cos\Theta -g_\mathrm{s}/2) b$, 
$\,\epsilon_3 =  
\varepsilon_d + (g_\mathrm{orb }\cos\Theta -g_\mathrm{s}/2) b$, 
and 
$\epsilon_4 =  
\varepsilon_d + (g_\mathrm{orb }\cos\Theta +g_\mathrm{s}/2) b$. 
The eigenstates, for $m=1,2,3$ and $4$, correspond to     
$|\mathrm{K'}\! \downarrow_{\vec{b}}\rangle$,    
$|\mathrm{K'}\! \uparrow_{\vec{b}}\rangle$,  
$|\mathrm{K}\! \downarrow_{\vec{b}}\rangle$ and  
$|\mathrm{K}\! \uparrow_{\vec{b}}\rangle$, respectively,   
with  $\uparrow_{\vec{b}}$  and $\downarrow_{\vec{b}}$    
the spin defined with respect to the direction along the field $\vec{b}$. 
Therefore, the thermal average of $ \overrightarrow{\bm{\mathcal{M}}}$ can be written in the form
\begin{align}
& \langle 
\bm{\psi}^{\dagger}_{d} \, 
\overrightarrow{\bm{\mathcal{M}}}   \bm{\psi}^{}_{d}
\rangle \,
=  \,  
\mathcal{M}_\mathrm{orb}  \, \vec{e}_z 
+  \mathcal{M}_\mathrm{s} \, \vec{e}_\Theta^{}\;,
\label{eq:magnetization_example1}
\\
& \ \mathcal{M}_\mathrm{orb} 
\,= \, 
g_\mathrm{orb}
\Bigl[\,
\left \langle n_{d1} \right \rangle 
-\left \langle n_{d4} \right \rangle 
+\left \langle n_{d2} \right \rangle 
-\left \langle n_{d3} \right \rangle 
\,\Bigr], 
\label{eq:magnetization_example2}
\\
& \quad \,  \mathcal{M}_\mathrm{s} 
\,=  \,  
\frac{g_\mathrm{s}}{2}\,
\Bigl[\,
\left \langle n_{d1} \right \rangle 
-\left \langle n_{d4} \right \rangle 
-\left \langle n_{d2} \right \rangle 
+\left \langle n_{d3} \right \rangle 
\,\Bigr].
\label{eq:magnetization_example3}
\end{align}

In more general cases, the matrix for  $\overrightarrow{\bm{\mathcal{M}}}$ 
with respect to one-particle eigenvector $\bm{u}_m^{}$, 
 can be expressed in the following forms using the Feynman theorem,  
\begin{align}
\vec{K}_{m}
\equiv\,
\bm{u}_m^{\dagger} \,
\overrightarrow{\bm{\mathcal{M}}} 
\,\bm{u}_m^{}  
\, = \,  - \,\bm{u}_m^{\dagger} 
\frac{\partial \bm{H}_d^0}{\partial \vec{b}} 
 \,\bm{u}_m^{}  
\,=\, - \,\frac{\partial \epsilon_{m}}{\partial \vec{b}} .
\end{align}
The thermal average of the magnetization 
 $\overrightarrow{\mathcal{M}}$ 
can be written in terms of  these matrix elements, 
\begin{align}
\overrightarrow{\mathcal{M}} \,\equiv \,
\left \langle 
 \bm{\psi}^{\dagger}_{d} \, \overrightarrow{\bm{\mathcal{M}}} 
\,
 \bm{\psi}^{}_{d} 
\right \rangle
\,=\, 
\sum_m 
\vec{K}_{m} 
\left \langle 
n_{d,m}
\right \rangle . 
\label{eq:magnetization}
\end{align}
Therefore, the magnetic susceptibility
$\chi_{\mathcal{M}}^{\mu\nu} \equiv 
\partial \mathcal{M}^{\mu}/\partial b_\nu$, 
which is not isotropic for nano-tube dots, 
can be expressed in the following form,
\begin{align}
\chi_{\mathcal{M}}^{\mu\nu}
 \, = & \    
\sum_m \frac{\partial K_{m}^{\mu} 
}{\partial b_\nu}
\,\left \langle 
n_{d,m}
\right \rangle 
+ \sum_{m,m'} 
K_{m}^{\mu}\,
\frac{\partial \epsilon_{m'}}{\partial b_\nu}
\frac{\partial  \left \langle n_{d,m}\right \rangle 
}{\partial \epsilon_{m'}}
\nonumber \\
\,=& \   
\sum_m \frac{\partial K_{m}^{\mu} 
}{\partial b_\nu}
\,\left \langle 
n_{d,m}
\right \rangle 
 \ + \sum_{m,m'}
K_{m}^{\mu}\,
K_{m'}^{\nu}\,
\chi_{m,m'} \;. 
\nonumber \\
 \,=& \   
 \sum_m \frac{\partial K_{m}^{\mu}}{\partial b_\nu}
 \,\left \langle n_{d,m}\right \rangle 
 + \sum_{m} K_{m}^{\mu}\, K_{m}^{\nu}\,\widetilde{A}_{m}^{} 
 \nonumber \\
 & \ - \sum_{m\neq m'} K_{m}^{\mu}\, K_{m'}^{\nu}\,
 \widetilde{A}_{m}^{}\,\widetilde{A}_{m'}^{} \,\widetilde{U}_{mm'}.
\end{align}
 Here, the last term in the right-hand side represents 
 the contributions of the residual interaction, or the vertex corrections. 
Since the Wilson ratio $R_{m,m'}$ given in Eq.\ \eqref{def_Wilson_ratio} depends on the residual interaction,
$R_{m,m'}$ relates to the magnetic susceptibilities.

\subsection{CNT level structure  \& field-induced crossover}
\label{subsec:CNT_Zeeman_matching}

In recent experiments reported 
in Refs.\ \onlinecite{Meydi-Nat-Phys, Meydi-Phys-Rev-Lett}, 
nonlinear current and current noise were measured
for a CNT dot with the orbital Land\'{e} factor 
$g_\mathrm{orb} \approx 4$  at finite magnetic fields 
with an angle  $\Theta \approx 75^{\circ}$.
These values of $g_\mathrm{orb}$ and  $\Theta$ 
imply that  the magnitude of the 
orbital Zeeman splitting becomes  almost the same as the spin Zeeman splitting 
in this particular situation,  
\begin{align}
g_\mathrm{orb} \cos \Theta \,\approx\, \frac{1}{2} \,g_\mathrm{orb}  \  =\, 1 \,.
\label{eq:MatchingCondition_ORG}
\end{align}
This situation can take place for CNT dots 
as the orbital Land\'{e} factor  $g_\mathrm{orb}$   
depends significantly on the diameter of nanotube 
and  takes a value   around $g_\mathrm{orb}\sim 10$.\cite{RMP-Kouwenhoven}
In the case where this matching of the orbital and spin Zeeman splittings is satisfied,    
the energy level of the dot has a double degeneracy 
which remains unlifted in magnetic fields:
\begin{align}
\epsilon_{1}=\varepsilon_{d} - 2b,  
\quad 
\epsilon_{2} \equiv \epsilon_{3} \, = \varepsilon_{d}, \quad  
\epsilon_{4}=\varepsilon_{d} +2b. 
\label{eq:MatchingCondition}
\end{align}
In this case  the occupation numbers of the degeneracy 
become the same   
$\left \langle n_{d2} \right \rangle 
=\left \langle n_{d3} \right \rangle$.
Thus, both the orbital and spin magnetizations 
are determined by the occupation numbers of the other two levels  $m=1$ and $m=4$:  
    $\mathcal{M}_\mathrm{orb} = g_\mathrm{orb}\, \mathcal{M}_{14}$ 
and $\mathcal{M}_\mathrm{s} = \frac{g_\mathrm{s}}{2} \mathcal{M}_{14}$, 
with  
\begin{align}
\mathcal{M}_{14}
\,\equiv& \ 
\left \langle n_{d1} \right \rangle 
-\left \langle n_{d4} \right \rangle \;.
\label{eq:Magnetization_14}
\end{align}

We note that $\Delta_\mathrm{SO}$  and $\Delta_\mathrm{KK'}$ 
are less important for 
the examined CNT dot.\cite{Meydi-Nat-Phys,Meydi-Phys-Rev-Lett,Teratani-JPSJ}
In this situation,  the system has an SU(2) rotational symmetry 
defined with respect to the degenerate states in the middle,
and  the U(1) symmetry that conserves the sum of  
the occupation numbers  $n_{2}^{}+n_{3}^{}$,  
in addition to the other two  U(1) symmetries corresponding 
to  $n_1$ and $n_4$ for the levels $m=1$ and $m=4$, respectively. 
 Therefore, the SU(4) symmetry that the  total Hamiltonian has  
 at zero field breaks down to the 
 U(1)$_{m=1}$$\otimes$[SU(2)$\otimes$U(1)$]_{m=2,3}$$\otimes$U(1)$_{m=4}$ 
 symmetry at finite magnetic fields. 
This SU(2) symmetric part plays a central role in 
the field-induced SU(4) to SU(2) Kondo crossover, 
occurring at half-filling point $N_{d} =2$. 
At this point, due to the matching condition given in Eq.\ \eqref{eq:MatchingCondition},
the Hamiltonian $\mathcal{H}$ is invariant under an extended electron-hole transformation 
described by Eq.\ \eqref{extended_particle_hole_symmetry}.

\begin{figure}[h]

\begin{minipage}{0.8\linewidth} 
\begin{center}
\includegraphics[width=\linewidth]{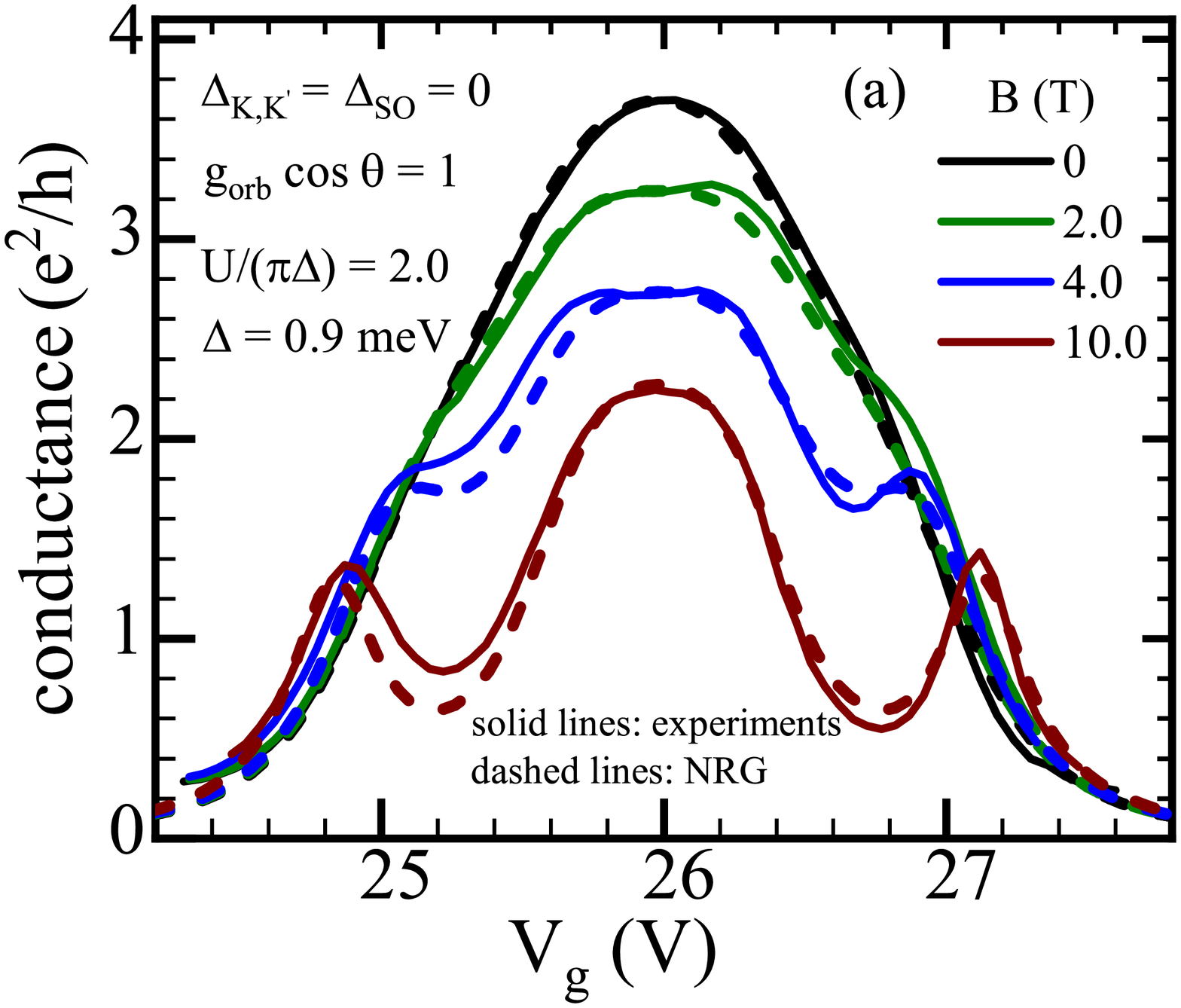}
\end{center}
\end{minipage}

\begin{minipage}{0.8\linewidth} 
\begin{center}
\includegraphics[width=\linewidth]{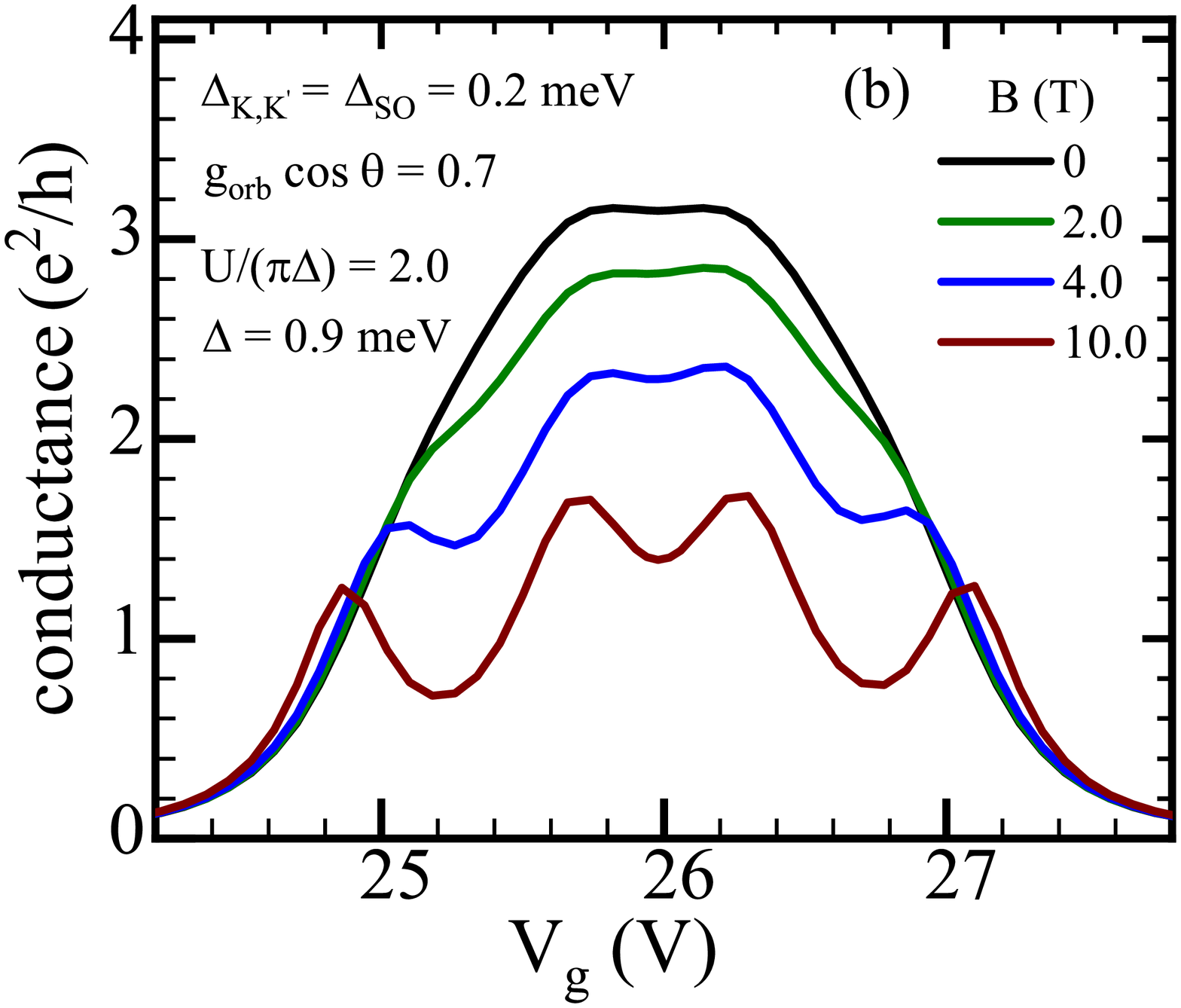}
\end{center}
\end{minipage}

\caption{
(a) and (b) plot the zero temperature conductance as functions of experimental gate voltages $V_\mathrm{g}$ 
for four values of magnetic fields: $B=0,2,4$, and $10$ T.
In the CNT dot where the experiments observed the crossover,
the experimental $V_\mathrm{g}$ relates to theoretical 
the theoretical $\epsilon_d$ via a linear relation: 
$V_\mathrm{g}= (0.8\,\epsilon_{d}/U + 27.18)\,$ in units of volt.
The asymmetric factor in Eq.\ \eqref{conductance_finT} and the Coulomb interaction are respectively $4\Delta_L\Delta_R/\Delta^2=0.92$ and $U/(\pi\Delta)=2.0$ for each two figure.
Here, $\Delta\equiv \Delta_L+\Delta_R =0.9\, \mathrm{meV}$.
In (a), the dashed lines and solid lines respectively 
represent NRG and experimental results.  
The NRG results plotted in (a) are for parameters: $\Delta_\mathrm{SO}=\Delta_{K,K^\prime}=0$, $g_\mathrm{orb}\cos\theta=1$. 
The experimental results have been obtained  
at  $T=16$ mK,\cite{Meydi-Nat-Phys,Meydi-Phys-Rev-Lett}
which is much lower than  $T_{K}^\mathrm{SU(4)} = 4.3$ K, 
the Kondo temperature for the half-filled case (at $V_\mathrm{g} \simeq 26$ mV).  
(b) plots only NRG results by solid lines. 
The parameters for the results are $\Delta_\mathrm{SO}=\Delta_{K,K^\prime}=0.2$ meV, $g_\mathrm{orb}\cos\theta=0.7$.
}
 \label{fig:Conductance_T=0}
\end{figure}

\begin{figure}[h]

\begin{center}
\includegraphics[width=0.75\linewidth]{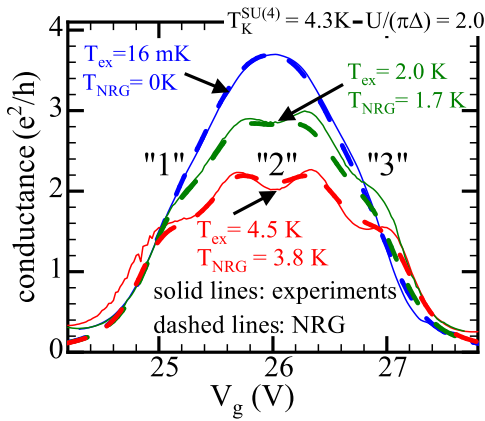}
\end{center}

\caption{ 
Zero-field, $B=0$, conductance at finite temperatures:    
experimental (solid line) and NRG  (dashed line) results are plotted vs $V_\mathrm{g}$.
The experimental data  has been obtained   
at  $T=16$ mK, 2 K, and 4.5 K.\cite{Meydi-Nat-Phys} 
For NRG calculations,  slightly lower temperatures  $T=$ 0 K, 1.7 K, and 3.8 K are chosen. 
The other parameters are the same as those used for  Fig.\  \ref{fig:Conductance_T=0}.
The numbers ``3'', ``2'', and ``1'' shown in these two figures represent the electron filling $N_{d}$ at corresponding $V_{g}$.
} 
 \label{GU2-SU4-ed-dependence}
\end{figure}

\subsection{Comparison of NRG and experiments results: 
gate-voltage dependence at finite $B$ or  $T$}

We have shown in the previous work\cite{Meydi-Phys-Rev-Lett,Teratani-JPSJ} that the level scheme, given in Eq.\ \eqref{eq:MatchingCondition},  
nicely explains the field-induced SU(4) to SU(2) Kondo crossover 
observed at half-filling  $N_{d} =2$ where two electrons occupy the local levels 
of the CNT dot.
The Coulomb interaction for this CNT dot is estimated to be  $U \approx 6$ meV,
and  the hybridization energy is  $\Delta \equiv \Delta_L + \Delta_R \approx 0.9$ meV. 
Its asymmetric factor is  $4\Delta_L\Delta_R/(\Delta_L+\Delta_R)^2 =0.92$. 
The energy scales the interaction as $U/(\pi\Delta)=2.0$. 
The SU(4) Kondo temperature for the scaled $U$ is $T_{K}^{\mathrm{SU(4)}}/\Delta=0.41$ which corresponds to $T_{K}=4.3$ K in a real scale.
These values of $U$ and $\Delta$ are 
much larger than the value of the vally mixing $\Delta_{KK'}$ and than that of the spin-orbit interaction $\Delta_\mathrm{SO}$.
The experimental values of $\Delta_{KK'}$ and $\Delta_\mathrm{SO}$ are $\Delta_\mathrm{KK'} \simeq \Delta_\mathrm{SO} \approx 0.2$ meV.
We  note that these values depend 
on the detailed structures of carbon nanotube samples.
Furthermore, it has also estimated in the experiment that the orbital Zeemann coupling as 
$g_\mathrm{orb}\cos\theta  \simeq  0.7$,\cite{Meydi-Phys-Rev-Lett} 
which is not so far from the matching condition $g_\mathrm{orb}\cos\theta  =1$,
described in Eq.\ \eqref{eq:MatchingCondition_ORG}.

We fist of all discuss the magnetoconductance for the case  $g_\mathrm{orb}\cos\theta = 1$ and $\Delta_{KK'}=\Delta_\mathrm{SO}=0$,  
at which an ideal SU(4) to SU(2) Kondo crossover occurs.
Figure \ref{fig:Conductance_T=0}(a) compares the NRG results of the conductance at $T=0$ with the experimental results obtained at $T=16$ mK. 
The values of $U$ and $\Delta$ for the NRG calculations are the experimental values. 
The comparisons which have been done also in Ref.\  \onlinecite{Meydi-Phys-Rev-Lett} 
show that the NRG results nicely agree with the experimental results.      
We can clearly see in this figure that 
the Kondo ridge emerges near half-filling  $V_\mathrm{g} \simeq 26 \mathrm{V}$ 
in the absence and presence of magnetic fields.
Its height reduces from the SU(4) value  $3.68\ \mathrm{e}^2/h$ to the SU(2) value $1.84\ \mathrm{e}^2/h$ 
as magnetic field increases.    
This reduction of the height implies that the observed  SU(2) behavior is caused by the doubly degenerate states,  
which are labeled as $m=2$ and $m=3$ in Eq.\ \eqref{eq:MatchingCondition},
are shifted towards the Fermi level in the half-filled case 
where $\epsilon_d=-3U/2$. 
Furthermore, the additional sub-peaks, which emerge outside the Kondo ridge  
for large magnetic fields $B \gtrsim 4$ T, can also be regarded as 
the resonances  corresponding to the other 
two non-degenerate levels, labeled as  $m=1$ and $m=4$.  
Note that the Kondo ridges corresponding to the $1/4$ and $3/4$ fillings 
are not so pronounced at $B=0$ because the Coulomb interaction for this CNT dot 
  $U/(\pi\Delta)=2.0$ is not very large.

We also examine the case where  the matching condition 
described in Eq.\ \eqref{eq:MatchingCondition_ORG} cannot be satisfied, 
taking the parameters such that  $\Delta_{KK'} = \Delta_\mathrm{SO} = 0.2$ meV 
and  $g_\mathrm{orb}\cos\theta=0.7$\footnote{Supplenental Material of Ref.\ \onlinecite{Meydi-Phys-Rev-Lett} 
at http://link.aps.org/supplemental/10.1103/PhysRevLett.118.196803}. 
At zero field $b=0$, the level structure is given by 
\begin{align}
&\epsilon_{1}=\epsilon_{2}=\epsilon_{d}-\frac{1}{2}\sqrt{\Delta_{KK'}^{2}+\Delta_\mathrm{SO}^{2}}, \label{epsilon_1_epsilon_2_DeltaKKprime_SO} \\
&\epsilon_{3}=\epsilon_{4}=\epsilon_{d}+\frac{1}{2}\sqrt{\Delta_{KK'}^{2}+\Delta_\mathrm{SO}^{2}}. \label{epsilon_3_epsilon_4_DeltaKKprime_SO}
\end{align} 
The  value of the gap due to  the spin-orbit interaction and valley mixing
is $\sqrt{\Delta_{KK'}^{2}+\Delta_\mathrm{SO}^{2}} \simeq 0.28\, \mathrm{meV}$, which is smaller than the resonance width. 
Figure \ref{fig:Conductance_T=0}(b) shows the NRG results, 
obtained taking the other parameters the same as 
 those for Fig.\ \ref{fig:Conductance_T=0}(a), 
i.e.\ $U/(\pi\Delta)=2.0$, $\Delta=0.9$ meV.
The broad peak emerges also in Fig.\ \ref{fig:Conductance_T=0}(b) although 
its peak value $3.14\,\mathrm{e}^{2}/h$ is smaller than that the value $3.68\,\mathrm{e}^{2}/h$ which 
is for the case of $\Delta_{KK'}=\Delta_\mathrm{SO}=0$ and $g_\mathrm{orb}\cos\theta=1$. 
The flat structure is still preserved in small magnetic fields up to $B\approx \ 5\, T$.
This preservation shows that 
the SU(4) to SU(2) Kondo crossover is robust against the perturbations. 
The larger magnetic fields $B\gtrsim5\ T$ split the peak into two peaks.
Furthermore, other two sub peaks grow with increasing $B$, 
and thus the four distinct peaks emerge. 
Figure \ref{fig:Conductance_T=0}(b) shows that 
the magnetoconductance for $B=10$ T has the four peaks.

Our NRG results for the CNT dots,  reported so far, 
were restricted to ground-state properties.
The present work sheds light also on the finite-temperature and dynamic properties of the Kondo crossover. 
Figure \ref{GU2-SU4-ed-dependence} 
compares  the experimental results\cite{Meydi-Phys-Rev-Lett} 
 of the zero-field conductance  measured at  $T=$ 16 mK, 2 K, and 4.5 K 
with the corresponding NRG results,   
 calculated for slightly lower  temperatures $T=$ 0 K, 1.7 K, and 3.8 K   
 to demonstrate how these comparisons work at the best.
We see a reasonable agreement  between the theoretical results and experimental results.
The height of the  Kondo ridge emerging near half-filling $V_{g}\simeq 26\, V$
decreases as temperature increases.  
At  temperatures of order  $T \sim T_{K}^\mathrm{SU(4)}=4.3$ K, 
Four peaks corresponding to the Coulomb oscillation emerge.
This agreement also indicates that the theory of the SU(4) Kondo effect can explain 
the experimental results of the conductance.

\subsection{Scaling behavior of SU($4$) conductance at quarter and half-filling}
\label{Scaling_behavior_of_SU4_conductance}
In this section, we examine scaling behavior of the $\mathrm{SU(4)}$ conductance as functions of temperature
especially at quarter-filling $N_{d} =1$ and half-filling $N_{d} =2$.
In Fig.\ \ref{GU2-SU4-ed-dependence}, the valley is quarter and half-filled at gate voltages,
$V_{g}\simeq 25\, V$ and $V_{g}\simeq 26\, V$, respectively.
The first value of $V_{g}$ corresponds to the theoretical value $\epsilon_{d}/U=-1/2$, 
and the second one corresponds to $\epsilon_{d}/U=-3/2$.

Since the $T^{2}$ coefficient $c_{T,m}$ of conductance given in Eq.\ \eqref{C_T_general} depends 
on linear and non-linear susceptibilities,
we discuss properties of these susceptibilities to examine the scaling.
In the $\mathrm{SU(\it{N})}$ symmetric case at which the $N$ impurity levels are degenerate $\epsilon_{m}\equiv\epsilon_{d}$, the linear susceptibilities have only two independent elements:
a diagonal element $\chi_{m,m}$ and the off-diagonal element $\chi_{m,m'}$ 
for  $m\neq m'$.
The diagonal element determines a characteristic energy scale $T^{*}$,
\begin{align}
T^{*}\,\equiv\,\frac{1}{4\,\chi_{m,m}}. \label{T_star}
\end{align}
In the particle-hole symmetric case $\epsilon_{d}/U \to -(N-1)/2$,  
it corresponds the renormalized width of the Kondo resonance 
as  $T^{*}  \rightarrow (\pi/4) \widetilde{\Delta}$.
At any arbitrary electron fillings,
$T^{*}$ scales the impurity specific heat defined in Eq.\ \eqref{specific_heat} 
as $\mathcal{C}_{\mathrm{dot}}=\frac{N \pi^{2}}{12}(T/T^{*})$.  

In the SU($N$) symmetric case, 
the Wilson ratio $R$ defined in Eq.\ \eqref{def_Wilson_ratio} 
also takes a simplified form $R-1=-\chi_{m,m'}/\chi_{m,m}$,  
where the flavour index has been dropped in the left-hand side.  
The Wilson ratio takes the maximum value 
in the strong coupling limit $U \to \infty$ at integer filling points:  
\begin{align}
R_{\mathrm{SU}(N)}^\mathrm{max}\,-\,1\, \xrightarrow{\,U\to \infty,\, \,
 N_d \to \, \mathrm{integer} \,}\, \frac{1}{N-1}.
\label{R_MAX_SUN}
\end{align}
This is because the charge fluctuations are suppressed in this limit, 
and the charge susceptibility defined in  Eq.\ \eqref{eq:chi_c} 
vanishes:  $\chi_{c,m} =  \chi_{m,m}+(N-1)\chi_{m,m'} \to 0$. 

The non-linear susceptibilities have three independent elements 
in the SU($N$)  symmetric case for $N>2$: 
$\chi_{m,m,m}^{[3]}$, $\chi_{m,m',m'}^{[3]}$, and $\chi_{m,m',m''}^{[3]}$, 
where $m\neq m'$, $m'\neq m''$, and $m\neq m''$.
We note that the derivative of the diagonal linear susceptibility $\chi_{m,m}$ 
with respect to $\epsilon_{d}$ can be related to the first two elements, as  
\begin{align}
\frac{\partial \chi_{m,m}^{}}{\partial  \epsilon_{d}^{}}
\,=\,\chi_{m,m,m}^{[3]}\,+\,(N-1)\,\chi_{m,m',m'}^{[3]}.
\label{eq:3body_multi_relation_2_1}
\end{align}

We next revisit  
the temperature dependence of conductance for the SU($N$) Anderson model, 
which has been previously studied in detail by Anders {\it et al\/},\cite{Anders-SU4} 
taking in a recent Fermi-liquid viewpoint.\cite{Mora_Fermi_liquid_2015,Filippone_Fermi_liquid_in_magnetic_field,Oguri_Higher_Order_Non_Equilibrium,teratani2020fermi} 
The  low-energy expansion, 
given in  Eqs.\eqref{g_m_low_T} and \eqref{C_T_general}, takes 
the following form in the SU($N$) symmetric case, 
\begin{align}
g_\mathrm{tot}\,&=\,\frac{N\,e^{2}}{h}\Biggl[ \sin^{2}\delta\,-\,C_{T}\,\left(\frac{\pi T}{T^{*}} \right)^{2}\,+\,\cdots\,\Biggr], \label{g_tot_SUN} \\
C_{T}\,&\equiv\,\frac{\pi^{2}}{48}\,\bigl(W_{T}\,+\,\Theta_{T} \bigr). \label{C_T_SUN}
\end{align}
The  two-body contributions  $W_{T}$ and three-body contributions  $\Theta_{T}$,  
defined respectively in Eqs.\ \eqref{W_t_general} and \eqref{Theta_t_general},  can be 
rewritten in the following form using also Eq.\ \eqref{eq:3body_multi_relation_2_1},    
\begin{align}
W_{T}\,&\equiv\,-\left[1\,+\,2\,\left(N\,-\,1\right)\left(R\,-\,1 \right) ^{2} \right]\cos{2\delta}, \label{W_T_SUN}\\
\Theta_{T}\, & =\,\frac{\sin{2\delta}}{2\pi}\frac{1}{\chi_{m,m}^{2}}\,\frac{\partial \chi_{m,m}^{}}{\partial  \epsilon_{d}^{}}. \label{Theta_T_SUN_Derivative}
\end{align}

\begin{figure}[h]
\begin{minipage}{0.85\linewidth} 
\begin{center}
\includegraphics[width=\linewidth]{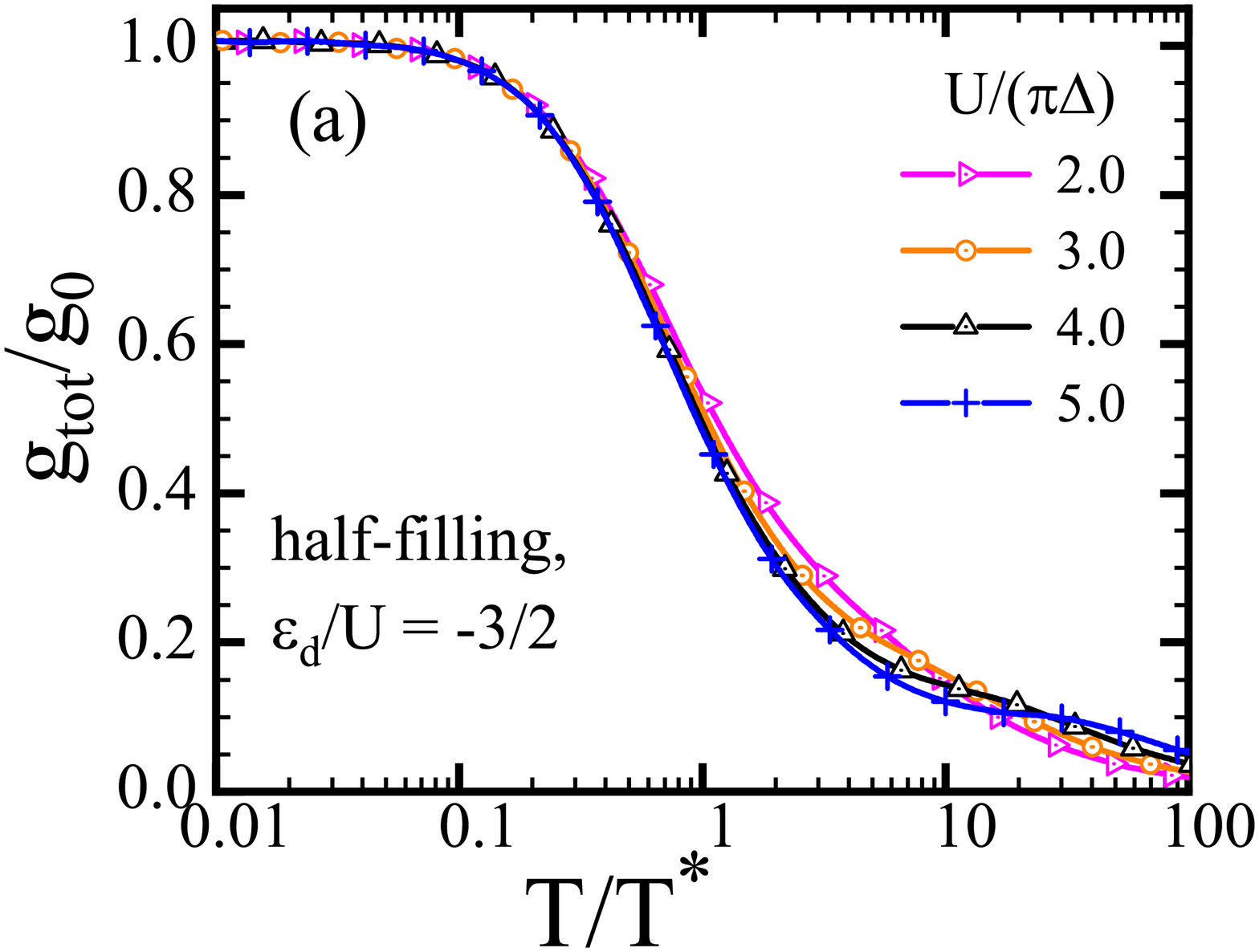}
\end{center}
\end{minipage}

\vspace{0.01\linewidth}
\begin{minipage}{0.85\linewidth} 
\begin{center}
\includegraphics[width=\linewidth]{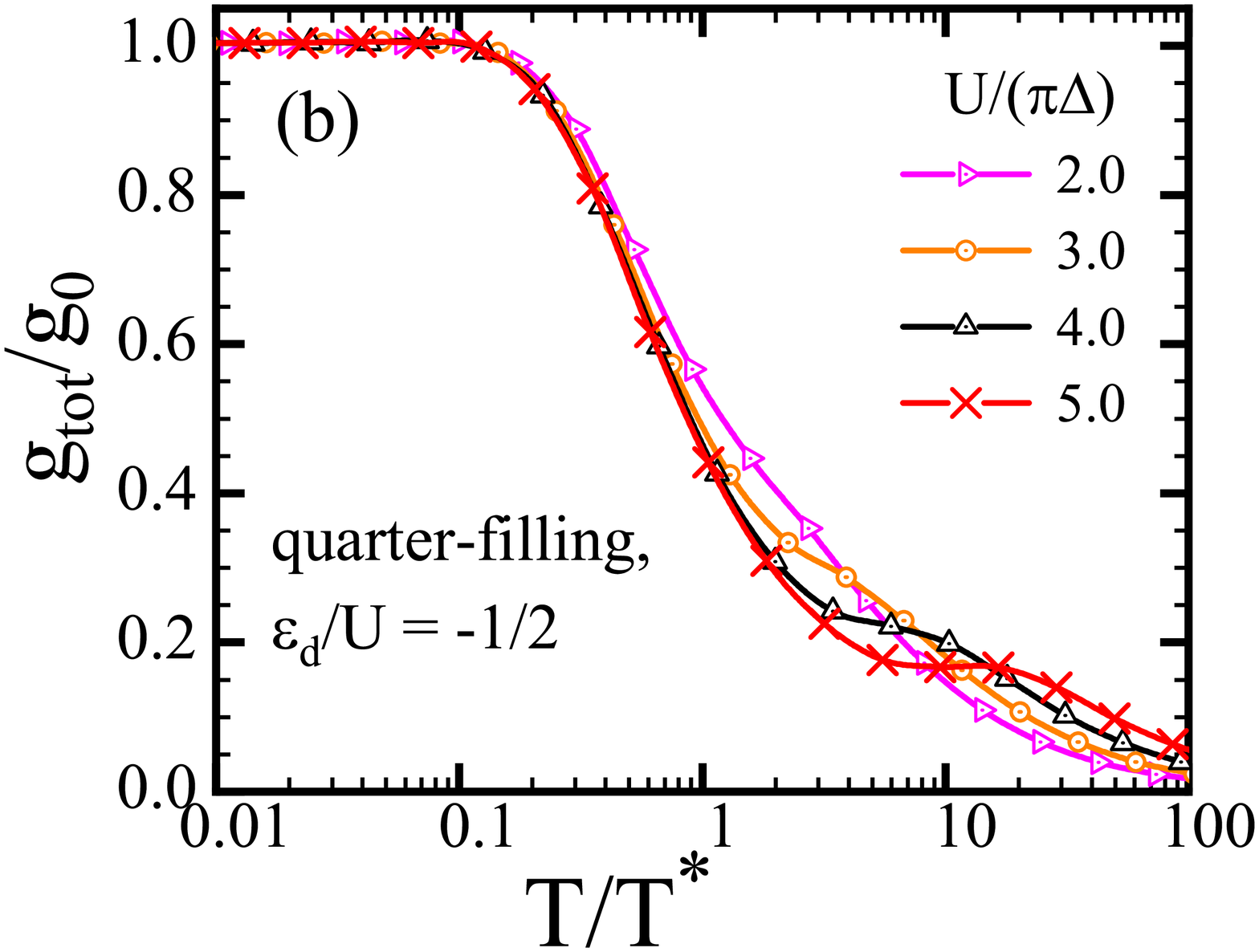}
\end{center}
\end{minipage}

\vspace{0.01\linewidth}
\begin{minipage}{0.85\linewidth} 
\begin{center}
\includegraphics[width=\linewidth]{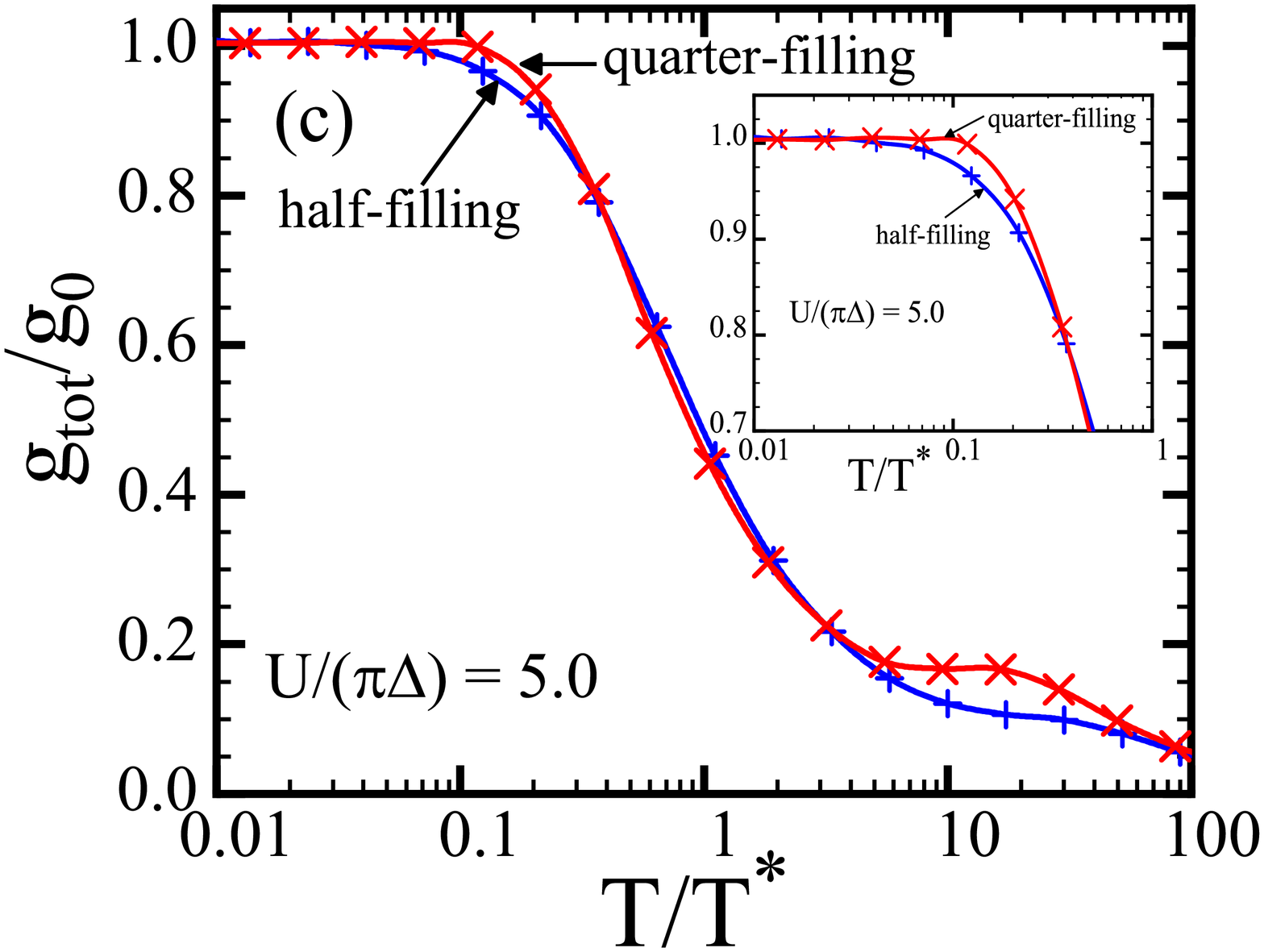}
\end{center}
\end{minipage}

\caption{SU(4) conductance curves are plotted as functions of temperature $T$.
(a) and (b) show the results at half-filling and quarter-filling, respectively.
In these figures, the conductance curves are plotted for four values of the 
interaction, $U/(\pi\Delta)=2.0,3.0, 4.0,$ and $5.0$. 
(c) shows the curves of half-filling and quarter-filling 
for the largest value $U/(\pi\Delta)=5.0$.
The inset of (c) is an enlarged view for $0.01\le T/T^{*} \le 1$. 
This inset clearly shows that the $T^{2}$ coefficient $C_{T}$ given in Eq.\ \eqref{C_T_SUN} 
vanishes at quarter-filling whereas that at half-filling does not.
In (a)-(c), the x-axis is scaled by the Kondo temperature $T^{*}\equiv 1/(4\chi_{m,m})$.
The values of $T^{*}/\Delta$ at half-filling are $0.41, 0.29, 0.20,$ and $0.13$ for 
$U/(\pi\Delta)=2.0,3.0, 4.0,$ and $5.0$, respectively.
Similarly, the values at quarter-filling are $0.82, 0.57, 0.37,$ and $0.23$.
The y-axis is normalized by the zero temperature values of conductance, $g_{0}=(4e^{2}/h)\sin^{2}\delta$.
At half-filling, $\sin^{2}\delta\equiv 1$ for any value of $U$, 
and at quarter-filling,  $\sin^{2}\delta=0.56, 0.55, 0.54,$ and $0.54$.
} 
 \label{SU4_Conductance_Scaling}
\end{figure}

Figures \ref{SU4_Conductance_Scaling}(a) and \ref{SU4_Conductance_Scaling}(b) show the temperature dependence of $g_\mathrm{tot}$ at half-filling and quarter-filling, respectively.
We choose four values of the interaction, $U/(\pi\Delta)=2.0, 3.0, 4.0,$ and $5.0$. 
The first value is the experimental value for the valley where
the SU(4) Kondo effect occurs.
In other valleys, the experimental values of $U/(\pi\Delta)$ can be larger than $2.0$,
and we also consider the larger interaction cases.
The temperatures are scaled by the Kondo energy scales $T^{*}$ defined in Eq.\ \eqref{T_star} for each $N_{d}$ and $U$. 
In each of the two figures, we find that the scaled conductance curves collapse into a
single curve over a wide range of temperatures $T\lesssim T^{*}$ for $U/(\pi\Delta)\gtrsim 3.0$,
and thus the conductance shows the universality for each filling.

To clarify the filling dependence of the universality, we replot the 
curves of quarter and half filling in Fig.\ \ref{SU4_Conductance_Scaling}(c). For the two curves, we choose the largest $U$ among the four, $U/(\pi\Delta)=5.0$.
We find that whereas these two curves almost overlap each other around $T\simeq T^{*}$, the conductance of quarter-filling is slightly larger than that of half-filling at low-temperatures $T<T^{*}$, especially around $T\simeq 0.1 T^{*}$.
The inset of Fig.\ \ref{SU4_Conductance_Scaling}(c) clearly shows the different behavior depending on the filling.

This filling dependence of the scaling can be explained by the Fermi-liquid theory\cite{Mora_Fermi_liquid_2015,Filippone_Fermi_liquid_in_magnetic_field,Oguri_Higher_Order_Non_Equilibrium,teratani2020fermi}.
We have recently clarified using also the NRG 
that  the derivative of the diagonal susceptibility 
$|\partial \chi_{m,m}/\partial \epsilon_{d}|$
becomes much smaller than $\chi_{m,m}^{2}$ 
in a wide region of electron fillings 
 $1\lesssim N_{d} \lesssim 3$ for large interactions $U \gg \Delta$.
\cite{teratani2020fermi}
From this result it follows that the three-body contributions 
 $\Theta_{T}$ 
given by Eq.\ \eqref{Theta_T_SUN_Derivative} 
 almost vanish in the same region, 
and the $T^2$ conductance in this case is determined 
by the two-body contributions, as  
\begin{align}
C_{T} \simeq -\left(\pi^{2}/48\right) \left[ 1\,+\,6\,(R-1)^{2} \right]\, 
\cos \left(\pi N_{d}/2 \right).
\end{align}
Thus, $C_{T}$ becomes $0$ at quarter filling $N_{d} =1$,
whereas it is finite $C_{T}=\left(\pi^{2}/48\right) [1+6(R-1)^{2} ]$  
at half-filling $N_{d} =2$. 
Therefore, the conductance at $N_{d} =1$ persists 
the zero temperature value at   $ 0\leq T/T^{*}\simeq 0.1$ 
as shown in the inset of Fig.\ \ref{SU4_Conductance_Scaling}(c).
At $N_{d} =2$, the coefficient approaches 
$C_{T}^{\infty}=5\pi^{2}/144 \simeq 0.3427$  
at $U \to \infty$. This is because the Wilson ratio for $N=4$ saturates 
to the strong coupling limit value $R_\mathrm{SU(4)}^\mathrm{max}-1\to1/3$.
Already at $U/(\pi\Delta)=3.0$,  the coefficient is given by $C_{T}\simeq 0.34$,  
which is very close to the strong-coupling value $C_{T}^{\infty}$. 

\clearpage

\begin{widetext}

\begin{figure}[h]
\centering
\begin{minipage}{0.32\linewidth} 
\begin{center}
\includegraphics[width=\linewidth]{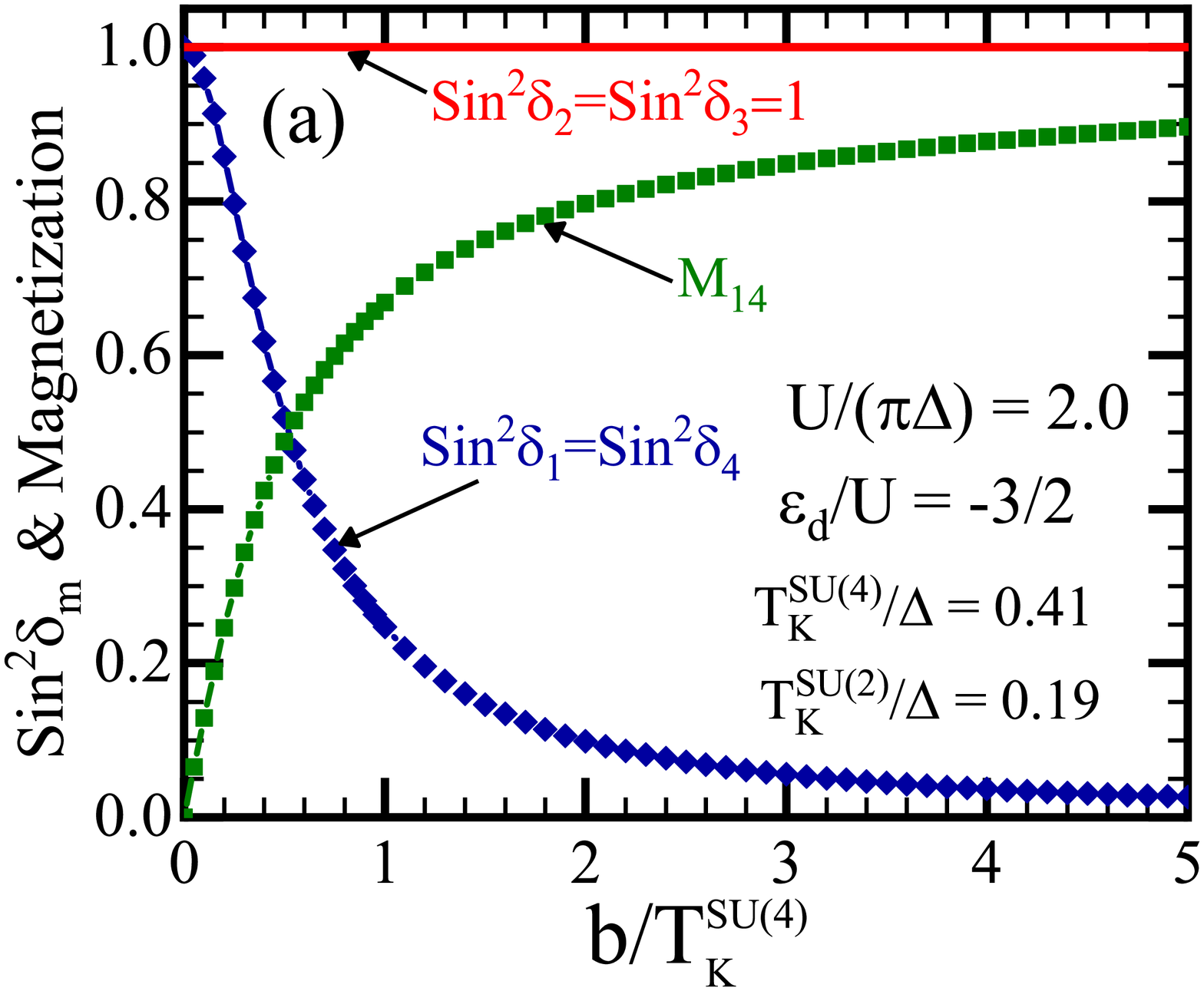}
\end{center}
\end{minipage}
\vspace{0.04\linewidth}
\begin{minipage}{0.32\linewidth} 
\begin{center}
\includegraphics[width=\linewidth]{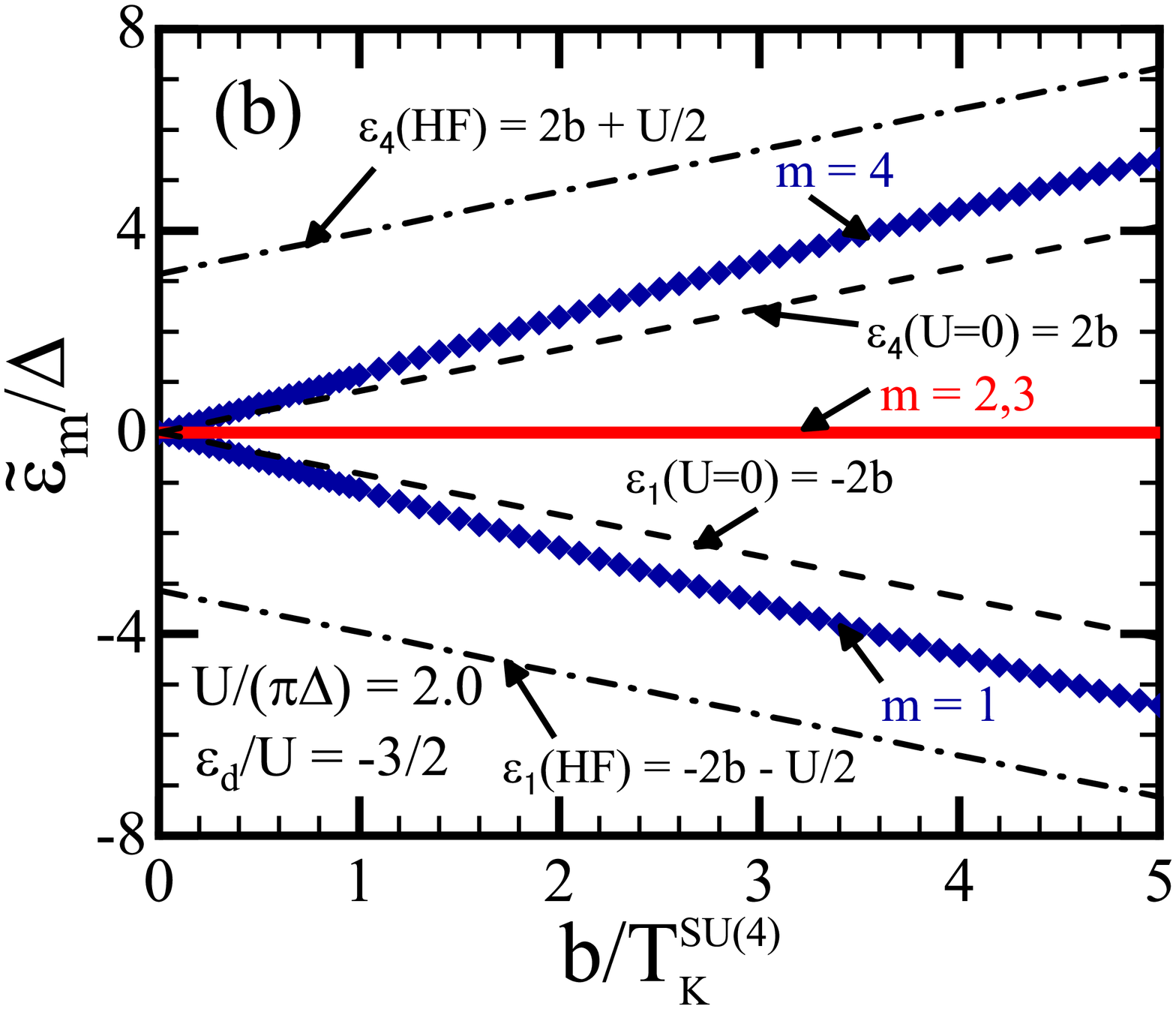}
\end{center}
\end{minipage}
\begin{minipage}{0.32\linewidth} 
\begin{center}
\includegraphics[width=\linewidth]{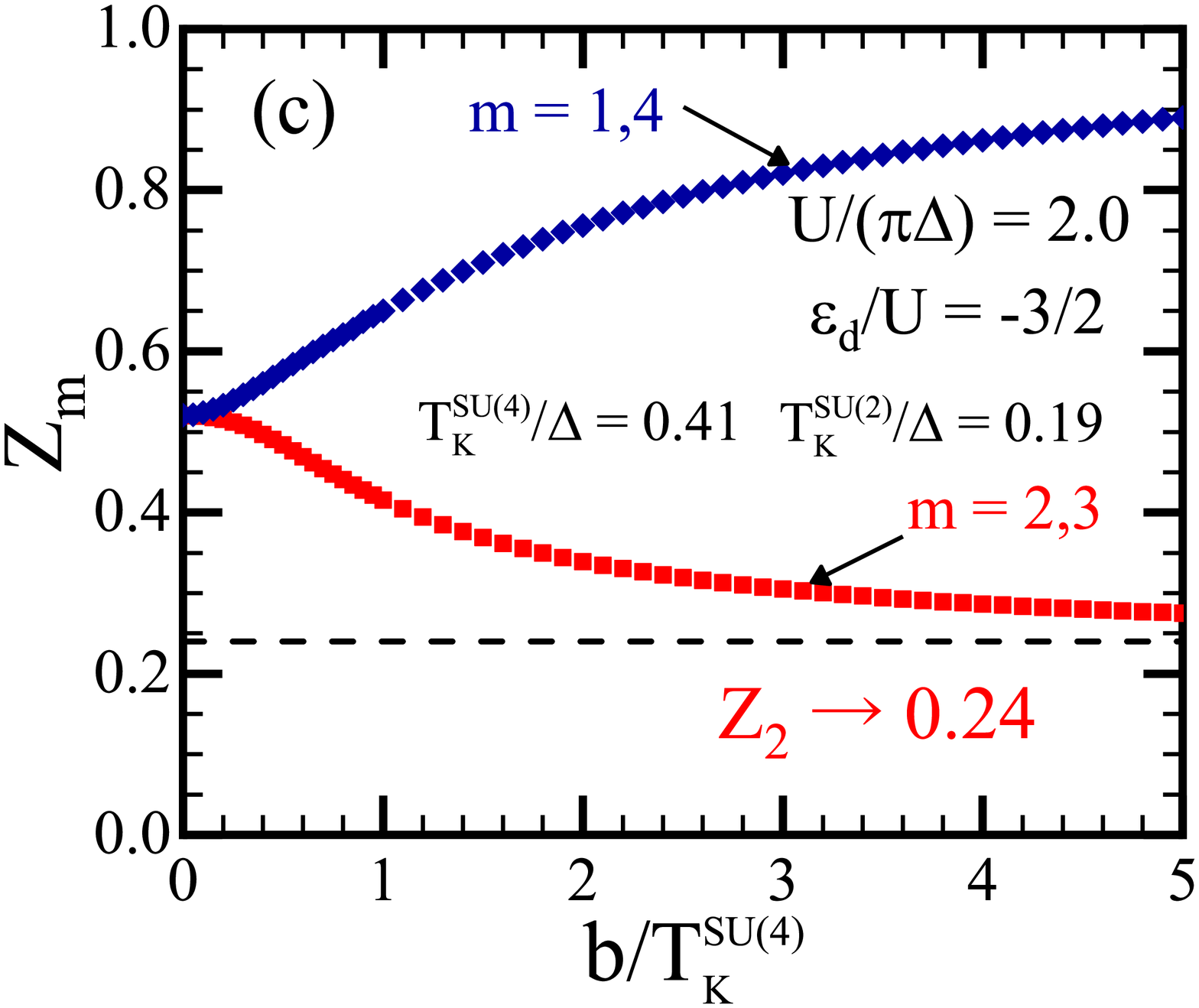}
\end{center}
\end{minipage}

\begin{minipage}{0.32\linewidth} 
\begin{center}
\includegraphics[width=\linewidth]{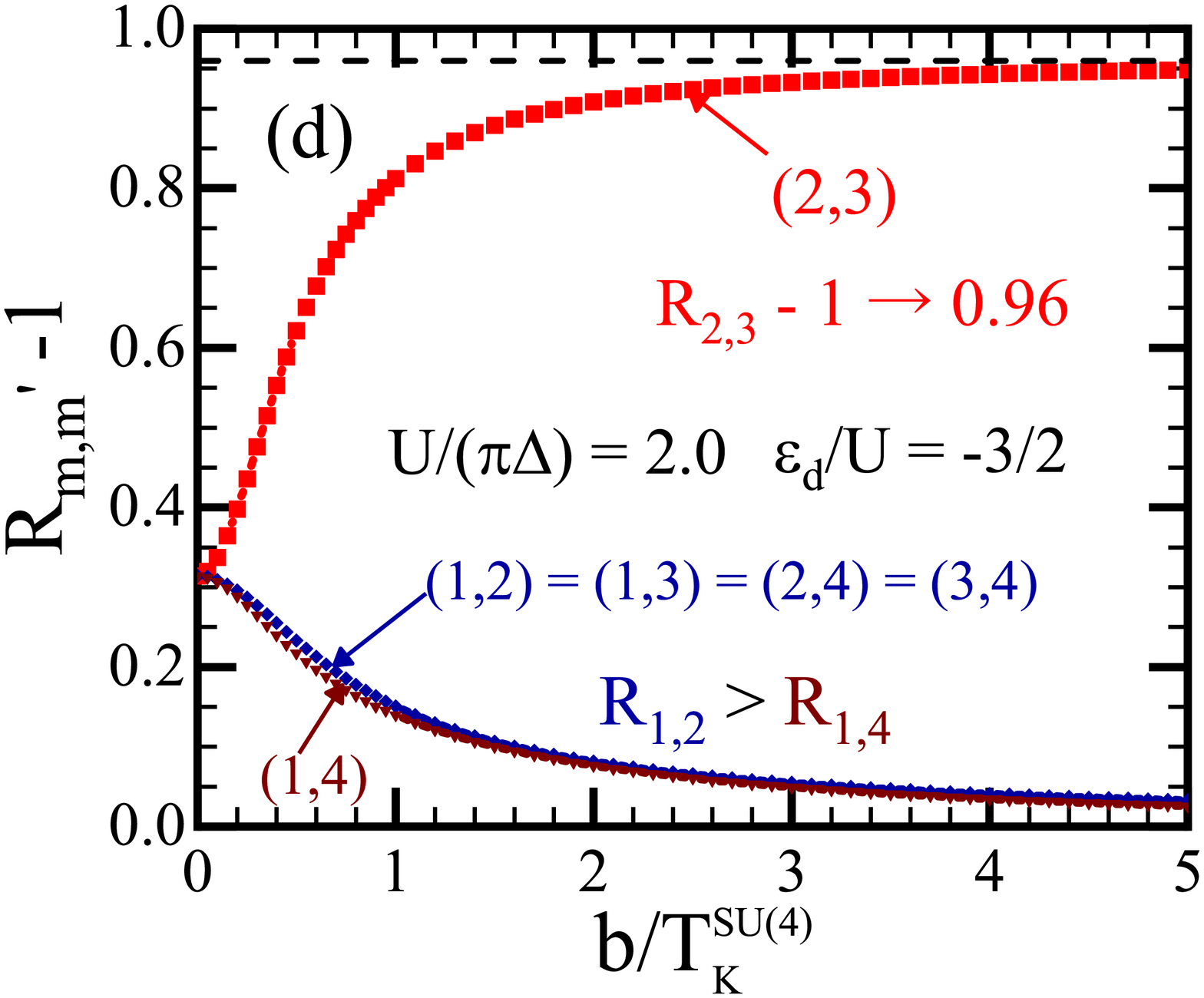}
\end{center}
\end{minipage}
\vspace{0.04\linewidth}
\begin{minipage}{0.32\linewidth} 
\begin{center}
\includegraphics[width=\linewidth]{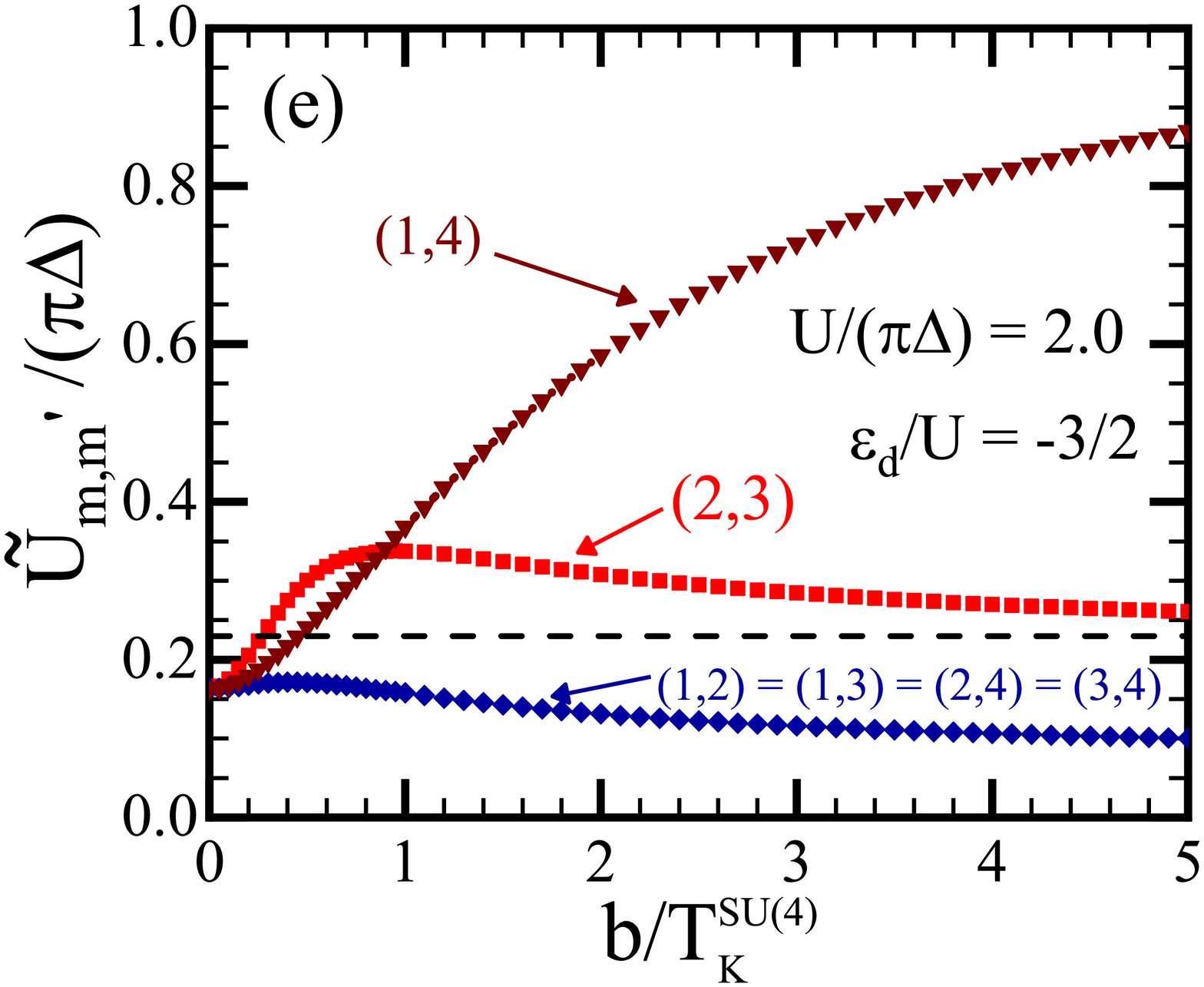}
\end{center}
\end{minipage}
\begin{minipage}{0.32\linewidth} 
\begin{center}
\includegraphics[width=\linewidth]{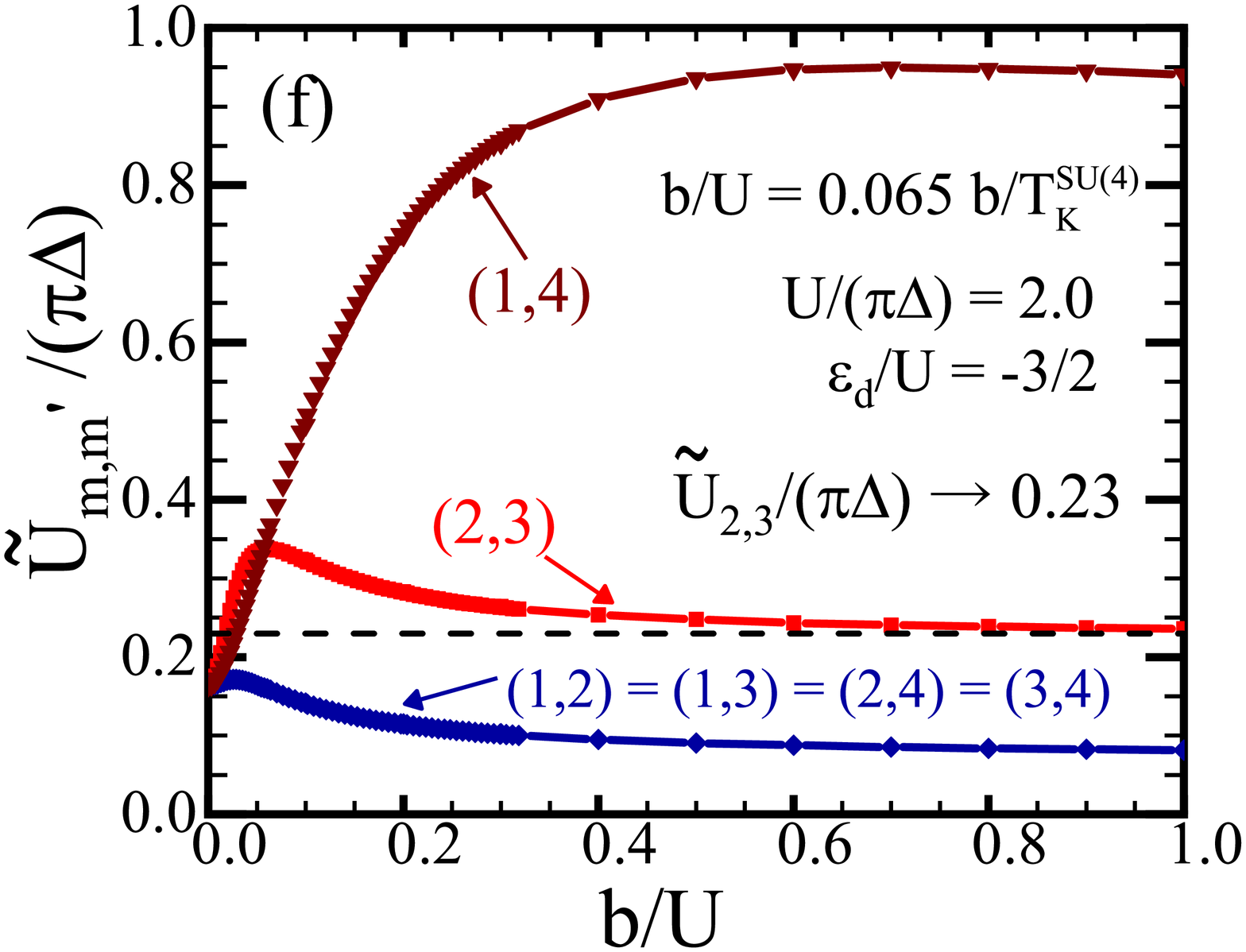}
\end{center}
\end{minipage}

\caption{(a) $\sin^{2}\delta_{m}$ and magnetization $\mathcal{M}_{14} 
=\langle n_{d1}\rangle  - \langle n_{d4} \rangle$, 
(b) renormalized level position $\widetilde{\varepsilon}_{m}$,
(c) renormalization factor $Z_{m}$,
(d) Wilson ratio $R_{m,m'}-1$, and (e) residual interaction $\widetilde{U}_{m,m'}$ 
are plotted as functions of magnetic field $b$ at half-filling $\varepsilon_{d}/U=-3/2$ for $U/(\pi\Delta)=2.0$. the x axis in (a)-(e) is scaled by the SU(4) Kondo temperature $T_{K}^{\mathrm{SU(4)}}=0.41\Delta=(0.065 U)$ determined at $b=0$.
In (f) the axis is scaled by $U$ for examining the behavior of $\widetilde{U}_{m,m'}$ at larger magnetic fields $b>T_{K}^{\mathrm{SU(4)}}$. 
The dot levels $\epsilon_{m}$ are chosen in a such way that is described in Eq.\ \eqref{eq:MatchingCondition}.  
In (b), the dashed lines indicate the bare Zeemann splitting, 
and  the dash-dotted lines indicate the mean-field splitting $\varepsilon_{1}^\mathrm{HF}=-(2b+U/2)$ and 
$\varepsilon_{4}^\mathrm{HF}=(2b+U/2)$. 
In a similar way, the dashed lines in (c)-(f) indicate the SU(2) symmetric values of $Z_{2} \to 0.23$, $R_{2,3} \to 1.96$, and $\widetilde{U}_{2,3}/\pi\Delta \to 0.026$, respectively.
} 

\label{b-dependent-FL-U2}
\end{figure}

\end{widetext}

\section{Evolution of quasi-particles along the field-induced crossover}
\label{b_dependence_quasi}

The Fermi-liquid parameters for renormalized quasi-particles
describe low energy properties of quantum dots.
The phase shift $\delta_{m}$, which is the primary parameter, corresponds to zero temperature transmission probability.
The renormalization factor $Z_{m}$ and the Wilson ratio $R_{m,m'}$
are also important parameters to examine the higher-order properties of the Fermi-liquid states. 
We describe how these and related parameters
evolve along a field induced crossover from the 
 SU(4) to SU(2) Kondo singlet state.
The crossover occurs for the dot levels defined in Eq.\ \eqref{eq:MatchingCondition}.  
Specifically, we consider the half-filled case 
 corresponding to the point  $V_\mathrm{g}\simeq 26$ V 
in the  middle of the Kondo ridge seen in Fig.\  \ref{fig:Conductance_T=0}.
The center of the dot levels is chosen to be  $\varepsilon_{d}=-(3/2)U$, 
and thus the average number of electrons in the dot levels conserves 
in a way such that  $\langle n_{d2}\rangle = \langle n_{d3} \rangle =1/2$ and
 $\langle n_{d1}\rangle  + \langle n_{d4} \rangle =1$ at finite magnetic fields. 

We examine two different values for the Coulomb interaction in the following: 
 (i) $U/(\pi\Delta) =2.0$ and (ii) $U/(\pi\Delta)=4.0$.
The first  one,  (i), simulates the situation of the CNT dot, 
in which the field-induced crossover has been observed 
and the parameters have been estimated as 
$U\approx 6$ meV and $\Delta \approx 0.9$ meV.
\cite{Meydi-Nat-Phys,Meydi-Phys-Rev-Lett}
We can see more clearly the renormalization effects due to strong correlations    
in the second case (ii).

\subsection{
Fermi-liquid parameters for the real CNT dot 
}
\label{quasi_for_weak_U}
First of all, we consider the case $U/(\pi \Delta)=2.0$ 
that is estimated by the recent experiments.
The NRG results for this case are shown in Fig.\ref{b-dependent-FL-U2}.
Figure \ref{b-dependent-FL-U2}(a) shows 
the  transmission probability $\mathcal{T}_{m}(0)=\sin^{2}\delta_{m}$ 
and magnetization
  $\mathcal{M}_{14} \equiv \langle n_{d1}\rangle  - \langle n_{d4} \rangle $, 
as a function of magnetic field $b$ at half-filling. 
The degenerate levels, $m=2$ and $m=3$, keep their positions just on the Fermi level 
for finite magnetic fields, and  show  
the unitary limit transport $\sin^{2}\delta_{m} =1$ 
as $\delta_{2}=\delta_{3}=\pi/2$.
The magnetic field partly lifts the degeneracy and the other two states, $m=1$ and $m=4$.
For these orbitals,  $\sin^{2}\delta_{m}$ decreases as magnetic field increases.
The magnetization $\mathcal{M}_{14}$, which  in the present case is determined by 
the occupation number or these two levels,  increases as the magnetic field increases.
It saturates to  $\mathcal{M}_{14} \to 1$ in the limit of $b\rightarrow\infty$, 
and the charge fluctuations are suprressed as 
  $\langle n_{d,1}\rangle  \rightarrow 1$ and
  $\langle n_{d,4}\rangle  \rightarrow 0$.

Figure \ref{b-dependent-FL-U2}(b) 
shows the renormalized resonance level position  $\widetilde{\varepsilon}_{m}$ 
as a function of  magnetic field $b$.  
The two-fold degenerate states at the center,   
 $\widetilde{\varepsilon}_{2} = \widetilde{\varepsilon}_{3}=0$,  
remain just on the Fermi level at arbitrary magnetic fields.   
The other two levels, 
$\widetilde{\varepsilon}_{1}$ and $\widetilde{\varepsilon}_{4}$ 
move away from the Fermi level as $b$ increases. 
Slopes of them
are steeper than those for the noninteracting electrons $2b$ (dashed line).
In the large field limit $b\to\infty$,
the renormalized level positions approach the one described in the mean-field theory, 
i.e.,  
 $\varepsilon_{1}^\mathrm{HF}=-(2b+U/2)$ and 
$\varepsilon_{4}^\mathrm{HF}=(2b+U/2)$.  
These asymptotic form can be obtained as follows, substituting the mean values 
 $\langle n_{d1}^{}\rangle=1$, $\langle n_{d4}^{}\rangle =0$ and 
$\langle n_{d2}^{}\rangle = \langle n_{d3}^{}\rangle=1/2$
into the dot-part of the Hamiltonian with $\varepsilon_d=-(3/2) U$;
\begin{align}
\!\! 
\mathcal{H}_{d}^{0} + \mathcal{H}_U
=& \ 
2b 
\left(n_{d4}^{}
- n_{d1}^{}\right)
-\frac{3U}{2} 
\left(  n_{d2}^{}+ n_{d3}^{}+n_{d1}^{}+n_{d4}^{}\right)
\nonumber \\
& 
\!\!\!\!\!\!\!\!\!
+ U \Bigl[
 n_{d2}^{}n_{d3}^{}
+  n_{d1}^{}n_{d4}^{}
+ ( 
 n_{d2}^{}+n_{d3}^{}
)
 ( 
n_{d1}^{}+n_{d4}^{}
)
\Bigr]
\nonumber 
\\
\xrightarrow{\,b\to \infty\,}&   
\ 
U \left[ n_{d2}^{}n_{d3}^{} -\frac{1}{2} 
\left(n_{d2}^{}+n_{d3}^{}\right) \right]
\nonumber \\ 
&  
\!\! 
+ \left(2b + \frac{U}{2} \right) 
\Bigl(  n_{d4}^{} - n_{d1}^{} \Bigr)
+ \mathrm{const.}
 \label{eq:Hd_B_inf}
\end{align}
Here, the Coulomb interaction between the orbitals $m=2$ and $3$ is kept undecoupled.   
This asymptotic Hamiltonian also shows that the symmetric SU($2$) Anderson model describes
the Fermi-liquid properties of these two orbitals.

The magnetic field dependence of the wavefunction renormalization factors $Z_{m}$
plotted in Fig.\ \ref{b-dependent-FL-U2}(c) more clearly shows the crossover. 
At finite  magnetic fields,  only two of the four $Z_m$'s become 
independent:   $Z_2=Z_3$ and $Z_1=Z_4$  because of the particle-hole symmetry given in Eq.\ \eqref{extended_particle_hole_symmetry}.
The first one is for the degenerate levels remaining on the Fermi level, and the second one is for 
the levels moving away from the Fermi level.
At zero field, where the system has the SU($4$) symmetry, these two factors for the different orbitals become identical each other: $Z_{2}=Z_{1}=Z_\mathrm{SU(4)}^{}= 0.52$ for 
$U/(\pi\Delta)=2.0$.
Substituting this SU($4$) value into Eq.\ \eqref{T_star} gives the SU($4$) Kondo energy scale $T_{K}^{\mathrm{SU(4)}}/\Delta=0.41$.
Many-body effects significantly renormalize $Z_{2}$ from the SU($4$) value as magnetic field increases.
In the limit of $b\to\infty$, it approaches the SU($2$) symmetric value $Z_\mathrm{SU(2)}^{} =0.23$, which determines the SU($2$) Kondo energy scale $T_{K}^{\mathrm{SU(2)}}/\Delta=0.19$.
The many-body effects become less important for $Z_{1}$ with increasing field, and $Z_{1}$ approaches the non-interacting value $Z_{1}\to 1$ for the large magnetic field.

In order to clarify the many-body effects between electrons occupying the different orbitals,
we also examine the orbital dependent Wilson ratio $R_{m,m'}$ and corresponding residual interaction $\widetilde{U}_{m,m'}$ .
Figures \ref{b-dependent-FL-U2}(d) and \ref{b-dependent-FL-U2}(e) respectively show $R_{m,m'}-1$ and $\widetilde{U}_{m,m'}$ as functions of $b/T_{K}^{\mathrm{SU(4)}}$.
Magnetic field dependences of $\widetilde{U}_{m,m'}$ are plotted also in Fig.\ \ref{b-dependent-FL-U2}(f), where the magnetic field is scaled by $U$ 
to examine behaviors of $\widetilde{U}_{m,m'}$ at larger fields $b \gg T_{K}^{\mathrm{SU(4)}}$.
Owing to the particle-hole symmetry, only three of the six $\widetilde{U}_{m,m'}$ are independent: 
$\widetilde{U}_{2,3}$, $\widetilde{U}_{1,4}$, and $\widetilde{U}_{1,2}=\widetilde{U}_{1,3} = \widetilde{U}_{2,4}=\widetilde{U}_{3,4}$. 
Correspondingly, three independent parameters of the Wilson ratios, $R_{2,3}$, $R_{1,4}$, and $R_{1,2}$, can be deduced from Eq.\ \eqref{def_Wilson_ratio}:
\begin{align}
R_{2,3}\,-\,1\,&=\,\frac{1}{Z_2}\,\frac{\widetilde{U}_{2,3}}{\pi\Delta}, \label{R_23_U_tilde_23} \\
R_{1,4}\,-\,1\,&=\,\frac{\sin^{2}\delta_{1}}{Z_{1}}\,\frac{\widetilde{U}_{1,4}}{\pi\Delta}, \label{R_14_U_tilde_14} \\
R_{1,2}\,-\,1\,&=\,\sqrt{\frac{\sin^{2}\delta_{1}}{Z_{1}}\,\frac{1}{Z_{2}}}\,\frac{\widetilde{U}_{1,2}}{\pi\Delta}. \label{R_12_U_tilde_12}
\end{align}
Among the three independent parameters of $R_{m,m'}$ and $\widetilde{U}_{m,m'}$,
$R_{2,3}-1$ and $\widetilde{U}_{2,3}$ are for the doubly degenerate orbitals on the Fermi level. 
At zero field, $R_{2,3}$ and $\widetilde{U}_{2,3}$ take the SU($4$) values $R_{2,3}-1=0.31$ and $\widetilde{U}_{2,3}/(\pi\Delta)=0.16$ for $U/(\pi\Delta)=2.0$,
and $R_{2,3}-1$ already approaches very closely to the value for the infinite Coulomb interaction: $R_\mathrm{SU(4)}^\mathrm{max}- 1  \equiv 1/3$.
These parameters continuously evolve from the SU($4$) values to the SU($2$) symmetric values: $R_\mathrm{SU(2)}- 1 = 0.96$ and $\widetilde{U}_{2,3}/(\pi\Delta)=0.23$.

We also discuss the field dependence of the other parameters, $R_{1,2}$, $R_{1,4}$, $\widetilde{U}_{1,2}$, and $\widetilde{U}_{1,4}$.
$\widetilde{U}_{1,2}$ decreases from the SU($4$) value with increasing magnetic field,
and the corresponding Wilson ratio $R_{1,2}-1$ decreases to the non-interacting value $0$.
In contrast to $\widetilde{U}_{1,2}$, $\widetilde{U}_{1,4}$ increases from the zero field value and becomes larger than $\widetilde{U}_{2,3}$ and $\widetilde{U}_{1,2}$ for $b>T_{K}^{\mathrm{SU(4)}}$.  
It further increases at the larger magnetic field regions $b\gg T_{K}^{\mathrm{SU(4)}}$ as shown in Fig.\ \ref{b-dependent-FL-U2}(f).
This field dependence of $\widetilde{U}_{1,4}$ is similar to that of $\widetilde{U}$ for a single orbital Anderson model\cite{Hewson_1ch_mag_Anderson_part_I, Hewson_1ch_mag_Anderson_part_II}, 
although $\widetilde{U}_{1,4}$ does not approach to the bare value $U$.
We briefly discuss the field dependence of $\widetilde{U}$ and of the other Fermi liquid parameters for the single orbital Anderson model in Appendix \ref{Fermi-liquid parameters for single orbital Anderson impurity}.
This enhancement of $\widetilde{U}_{1,4}$ does not result in the enhancement of $R_{1,4}$.
In fact, $R_{1,4}-1$ as well as $R_{1,2}-1$ decreases to $0$ since the factor $\sin^{2}\delta_{1}$ goes to $0$.
We note that $R_{1,2}$ is slightly larger than $R_{1,4}$ at arbitrary $b$ in this case of $U/(\pi\Delta)=2.0$.  

\begin{widetext}

\begin{figure}[h]
\centering
\begin{minipage}{0.32\linewidth} 
\begin{center}
\includegraphics[width=\linewidth]{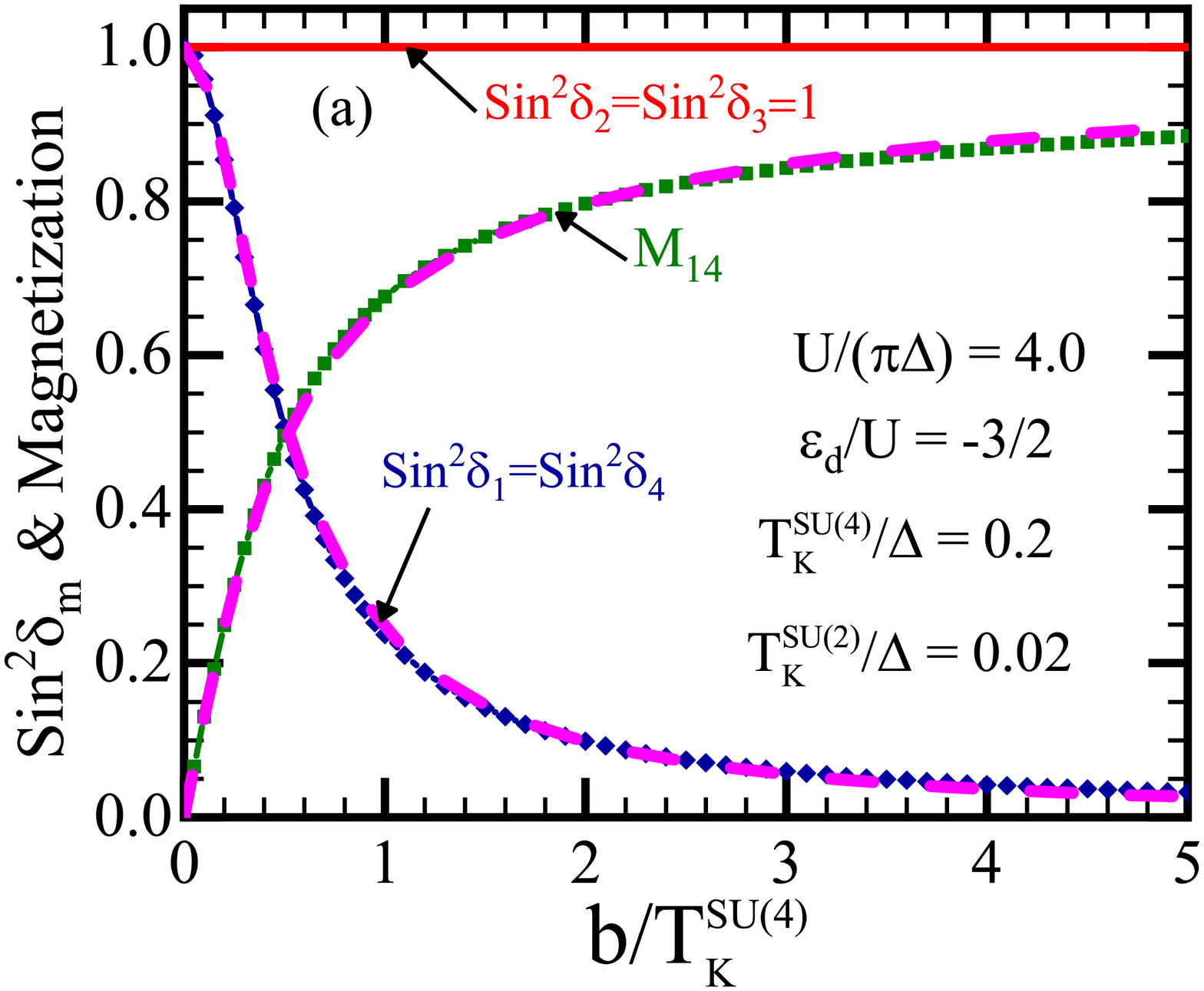}
\end{center}
\end{minipage}
\vspace{0.01\linewidth}
\begin{minipage}{0.32\linewidth} 
\begin{center}
\includegraphics[width=\linewidth]{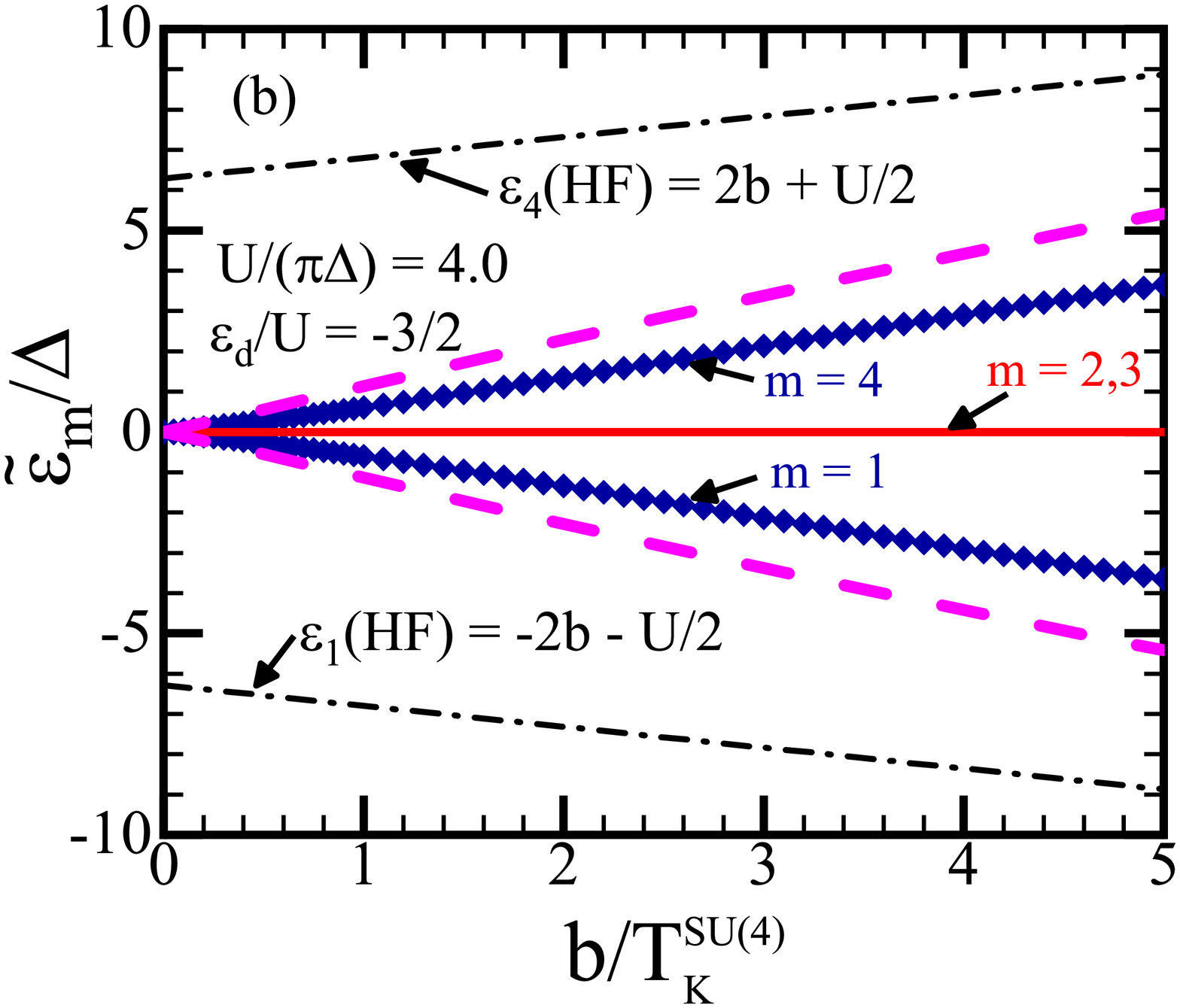}
\end{center}
\end{minipage}
\begin{minipage}{0.32\linewidth} 
\begin{center}
\includegraphics[width=\linewidth]{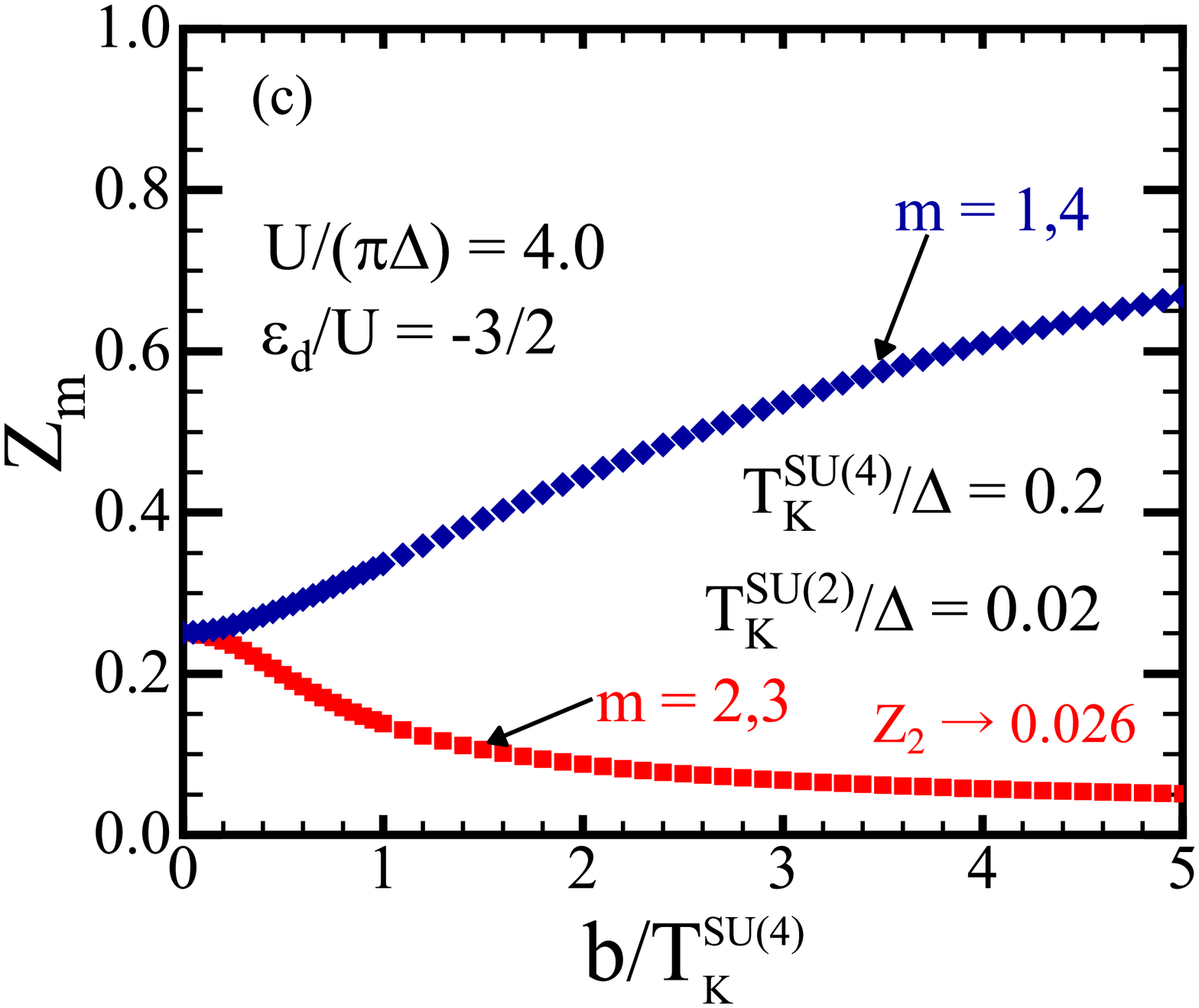}
\end{center}
\end{minipage}

\begin{minipage}{0.32\linewidth} 
\begin{center}
\includegraphics[width=\linewidth]{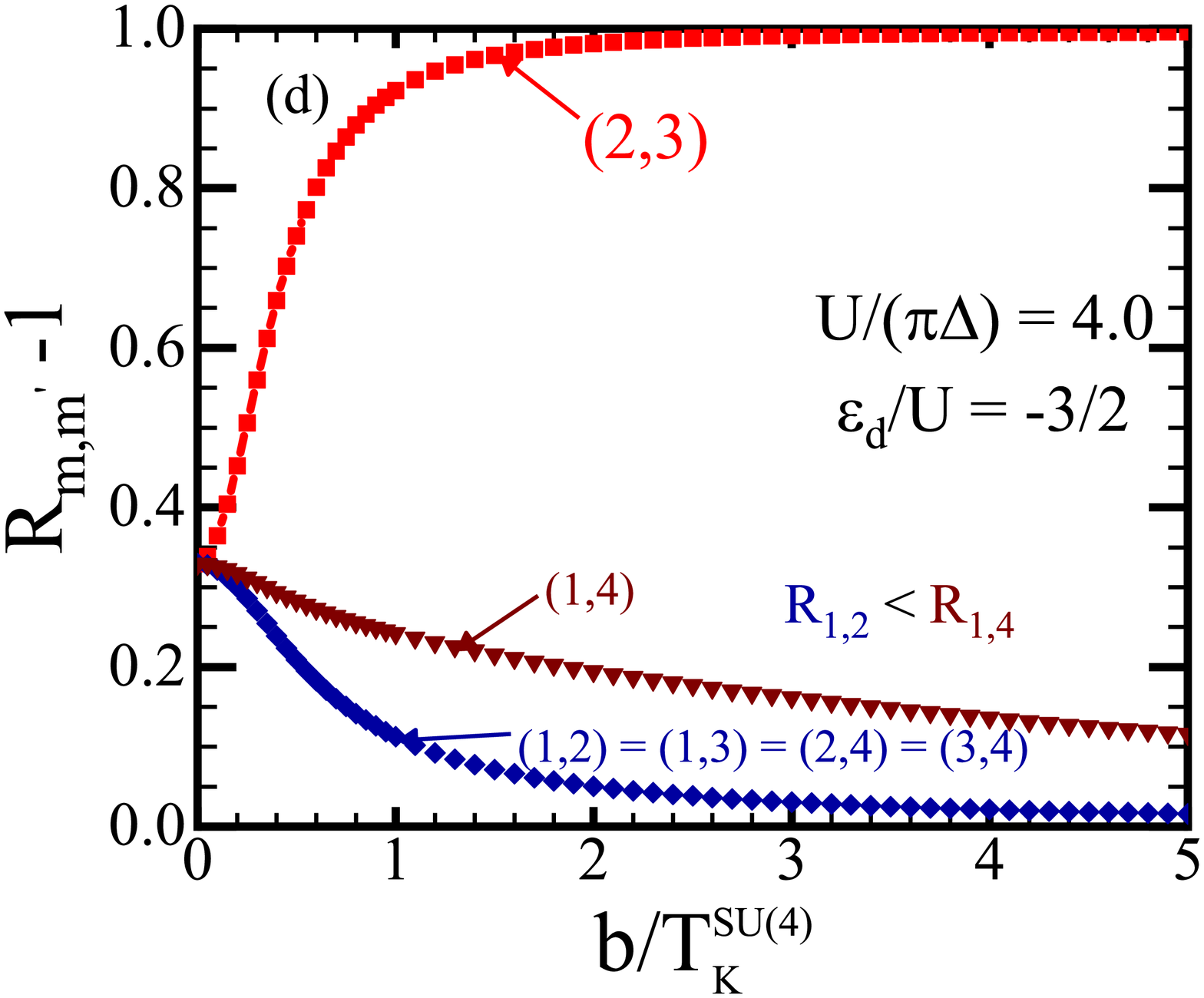}
\end{center}
\end{minipage}
\vspace{0.02\linewidth}
\begin{minipage}{0.32\linewidth} 
\begin{center}
\includegraphics[width=\linewidth]{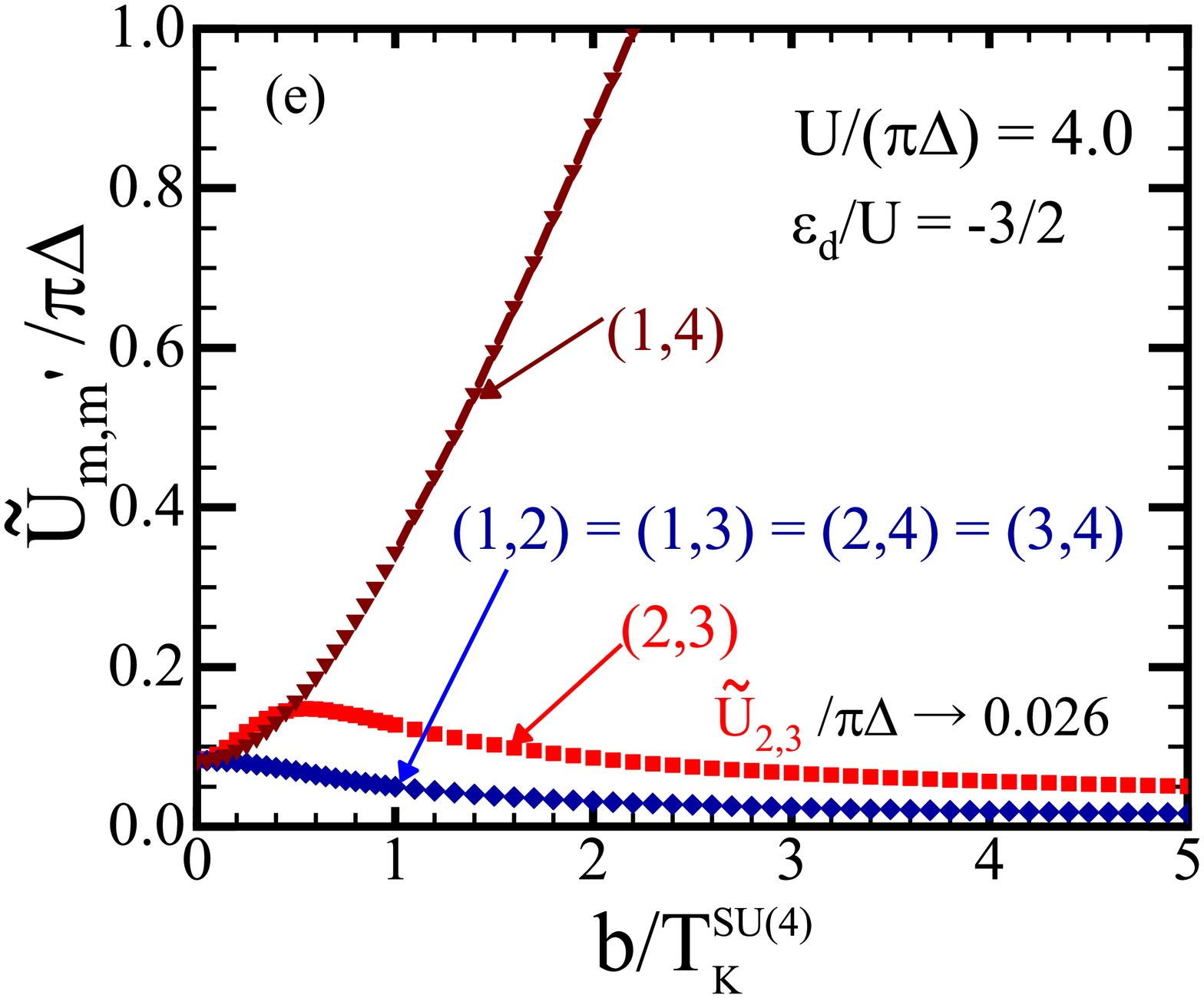}
\end{center}
\end{minipage}
\begin{minipage}{0.32\linewidth} 
\begin{center}
\includegraphics[width=\linewidth]{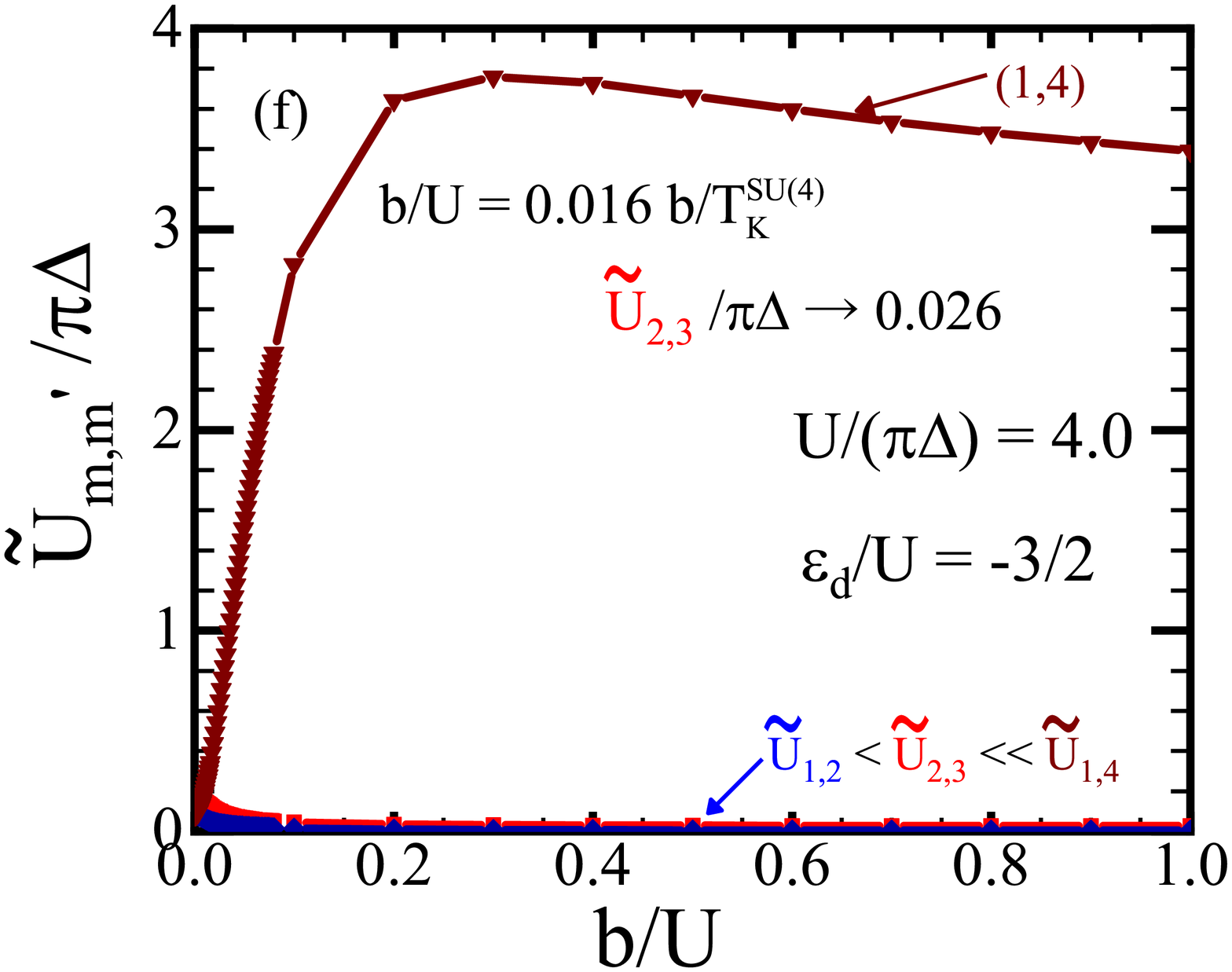}
\end{center}
\end{minipage}

\caption{(a) $\sin^{2}\delta_{m}$ and magnetization $\mathcal{M}_{14} 
=\langle n_{d1}\rangle  - \langle n_{d4} \rangle$, 
(b) renormalized level position $\widetilde{\varepsilon}_{m}$,
(c) renormalization factor $Z_{m}$,
(d) Wilson ratio $R_{m,m'}-1$, and (e) residual interaction $\widetilde{U}_{m,m'}$ 
are plotted as functions of magnetic field $b$ at half-filling $\varepsilon_{d}/U=-3/2$ for $U/(\pi\Delta)=4.0$. The x axis in (a)-(e) is scaled by the SU(4) Kondo temperature $T_{K}^{\mathrm{SU(4)}}=0.2\Delta=(0.016 U)$ determined at $b=0$.
The axis in (f) is scaled by $U$ for examining the behavior of $\widetilde{U}_{m,m'}$ at larger magnetic fields. 
The dot levels $\epsilon_{m}$ are chosen in a such way that is described in Eq.\ \eqref{eq:MatchingCondition}.  
In (a), $\sin^{2}\delta_{m}$ and $M_{d}$ for $U/(\pi\Delta)=2.0$ are also plotted by dashed lines to compare them with those for the present case $U/(\pi\Delta)=4.0$. 
Similarly, $\widetilde{\varepsilon}_{m}$ for $U/(\pi\Delta)=2.0$ are plotted in (b).
The dash-dotted lines indicate the mean-field splitting $\varepsilon_{1}^\mathrm{HF}=-(2b+U/2)$ and 
$\varepsilon_{4}^\mathrm{HF}=(2b+U/2)$. 
In the limit of $b\to\infty$, $Z_{2}$, $R_{2,3}-1$, and $\widetilde{U}_{2,3}$ approach the SU($2$) values for $U/(\pi\Delta)=4.0$: $Z_{2}\to 0.026$, 
$R_{2,3} \to 1.99$, and $\widetilde{U}_{2,3}/\pi\Delta \to 0.026$.
} 

\label{b-dependent-FL-U4}
\end{figure}

\end{widetext}

\subsection{Fermi-liquid parameters for larger $U$}
\label{quasi_for_strong_U}

We next consider a strong coupling case, 
taking the Coulomb  repulsion  to be $U/(\pi \Delta)=4.0$, 
which is twice as large as the one studied in the above.  
For this case, effects of the interactions 
on the field-induced SU($4$) to SU($2$) Kondo crossover emerges in a pronounced way. 
Such a situation is also realistic because the experimental values of $U$ and $\Delta$ 
depend on individual quantum dots and on the valleys to be measured.

Figure \ref{b-dependent-FL-U4}(a) plots 
ground-state values of 
 $\mathcal{T}_{m}(0)=\sin^{2}\delta_{m}$ 
 and of $\mathcal{M}_{14}$
as functions of magnetic fields $b/T_K^\mathrm{SU(4)}$
for both $U/(\pi \Delta)=4.0$ and $U/(\pi \Delta)=2.0$.
The results for $U/(\pi\Delta)=4.0$ and $U/(\pi\Delta)=2.0$ are plotted with solid lines and dashed lines, respectively.
The energy scale depends on the coupling constant as 
 $T_K^\mathrm{SU(4)}/\Delta=0.2$ for $U/(\pi\Delta)=4.0$   
and $T_K^\mathrm{SU(4)}/\Delta=0.41$ for $U/(\pi\Delta)=2.0$.  
We see that $\sin^{2}\delta_{m}$ and $M_{d}$ of $U/(\pi\Delta)=4.0$ show almost same $b$ dependences as those of $U/(\pi\Delta)=2.0$,
and thus they show the universality.
The universal behavior is determined by 
the $b$ dependence of a single parameter $\delta_{1}$ ($=\pi-\delta_{4}$).
Renormalized levels
$\widetilde{\varepsilon}_{m}$ for $U/(\pi\Delta)=4.0$ plotted in Fig.\ \ref{b-dependent-FL-U4}(b)  
show the different $b$ dependence from those for $U/(\pi\Delta)=2.0$.
Specifically, $\widetilde{\varepsilon}_{1}$ and $\widetilde{\varepsilon}_{4}$ stay closer to the Fermi level 
than those for  $U/(\pi\Delta)=2.0$.
However, this different $b$ dependence does not affect the universal behavior
of $\delta_{1}$ because a ratio of $\widetilde{\varepsilon}_{m}$ to $\widetilde{\Delta}_{m}$ determines the phase shift $\delta_{m}$, i.e.,
$\delta_{m}= \cot^{-1}(\widetilde{\varepsilon}_{m}/\widetilde{\Delta}_{m})$.

Figures \ \ref{b-dependent-FL-U4}(c) 
and \ref{b-dependent-FL-U4}(d) 
show  the renormalization factors 
 $Z_{m}=\widetilde{\Delta}_{m}/\Delta$ 
and the Wilson ratios  $R_{m,m^\prime}$, respectively.  
As in the $U/(\pi\Delta)=2.0$ case, 
 the quasi-particle parameters  $Z_{2}$ and $R_{23}$ 
for the doubly degenerate states at the Fermi level 
continuously evolve from the SU($4$) value to the SU($2$) value  
as $b$ varies from $0$ to $\infty$.
At zero field, these parameters take the SU($4$) values: 
 $Z_\mathrm{SU(4)}^{}=0.25$ and 
 $R_\mathrm{SU(4)}^{}-1=0.33$
for $U/(\pi \Delta)=4.0$.  
Note that the Wilson ratio is almost  saturated to the 
maximum possible value $R_\mathrm{SU(4)}^\mathrm{max}- 1  \equiv 1/3$
 at zero field. 
In the opposite limit $b \to \infty$, 
these parameters for the two-fold degenerate states ($m=2,3$)   
approach those for the symmetric SU($2$) Anderson model:  
$Z_\mathrm{SU(2)}\to 0.026$ and $R_{23}^\mathrm{SU(2)} - 1 \to 0.99$ 
for the same $U$.
These results show that the strong Coulomb interaction significantly affects 
the renormalization factor $Z_2$ or $\widetilde{\Delta}_{2}$.
$Z_{2}$ determines the energy scale for large field as 
 $T_K^\mathrm{SU(2)}=0.02\Delta$ with Eq.\ \eqref{T_star}.
The quasi-particle parameters  
 $Z_{1}$, $R_{1,2}$ and $R_{1,4}$, 
 for the states moving away from the Fermi level 
approach the noninteracting value  in the limit of $b\to \infty$;   
i.e,  $Z_{1} \to 1$,  $R_{12} \to 1$,  and $R_{14} \to 1$.  
Notably, $R_{1,4}$ becomes larger than $R_{1,2}$ 
for $U/(\pi \Delta)=4.0$ at finite $b$. 
This is quite different what we have found for 
the smaller interaction case $U/(\pi \Delta)=2.0$. 

In order to clarify this difference,
we plot the residual interactions $\widetilde{U}_{m,m'}$ as functions of magnetic fields in Figs.\ \ref{b-dependent-FL-U4}(e) and \ref{b-dependent-FL-U4}(f). 
The magnetic fields of Figs.\ \ref{b-dependent-FL-U4}(e) and \ref{b-dependent-FL-U4}(f) are respectively scaled by $T_{K}^{\mathrm{SU(4)}}$ and $U$. 
In these figures, especially in Fig.\ \ref{b-dependent-FL-U4}(f), we can see that $\widetilde{U}_{1,4}$ becomes much larger than the other two residual interactions $\widetilde{U}_{1,2}$ and $\widetilde{U}_{2,3}$ as $b$ increases. 
This field dependence of $\widetilde{U}_{1,4}$ clearly explain why the corresponding Wilson ratio $R_{1,4}$ becomes larger than $R_{1,2}$.
We also note that $\widetilde{U}_{2,3}$ for the doubly degenerate levels approach the SU($2$) symmetric value $\widetilde{U}_{2,3}/(\pi\Delta)\to0.026$.

All these results discussed in this section indicate that 
the quantum fluctuations and many-body effects are enhanced for large magnetic fields 
as the number of active channel decreases from 4 to 2.\cite{Teratani-JPSJ}.
We have also shown the enhancement of the fluctuations are more clearly seen for strong interactions 
by comparing the results for $U/(\pi\Delta)=4.0$ to those for $U/(\pi\Delta)=2.0$.

\section{Temperature dependence of magnetoconductance}
\label{Conductance_finT}

The above discussions about the Fermi-liquid parameters 
have mainly focused on the zero temperature properties of the crossover.
The results show that 
the quasi-particles are strongly renormalized as the ground state undergoes 
the crossover from the SU(4) Kondo state to the SU(2) Kondo state. 

We also study finite temperature properties of the crossover in this section by 
calculating each component of the magnetoconductance $g_{m}$ 
for $m=1,2,3,4$ and the total conductance $g_\mathrm{tot}$.
At half-filling, $\varepsilon_{d}=-\frac{3}{2} U$, only two components are independent: $g_{2}=g_{3}$ and $g_{1}=g_{4}$  due to the level structure described in Eq.\ \eqref{eq:MatchingCondition}.  
The  finite-temperature conductance, defined in Eq.\eqref{conductance_finT},  
 depends on the excited states whose contributions enter through 
the spectral function $A_{m}(\omega,T)$.  
We calculate the $T$-dependent $A_{m}(\omega,T)$, using the NRG with 
some extended methods for dynamic correlation functions 
described in Sec.\ \ref{sec:NRG_section} and Appendix \ref{sec:NRG_calculations},  to obtain $g_{m}$.
We examine two different interactions,  
 $U/(\pi\Delta)=2.0$ and  $4.0$,  
also for these components of the conductance 
assuming symmetric couplings $\Delta_L=\Delta_R=\Delta/2$. 

\begin{widetext}

\begin{figure}[h]
\centering
\begin{minipage}{0.42\linewidth} 
\begin{center}
\includegraphics[width=\linewidth]{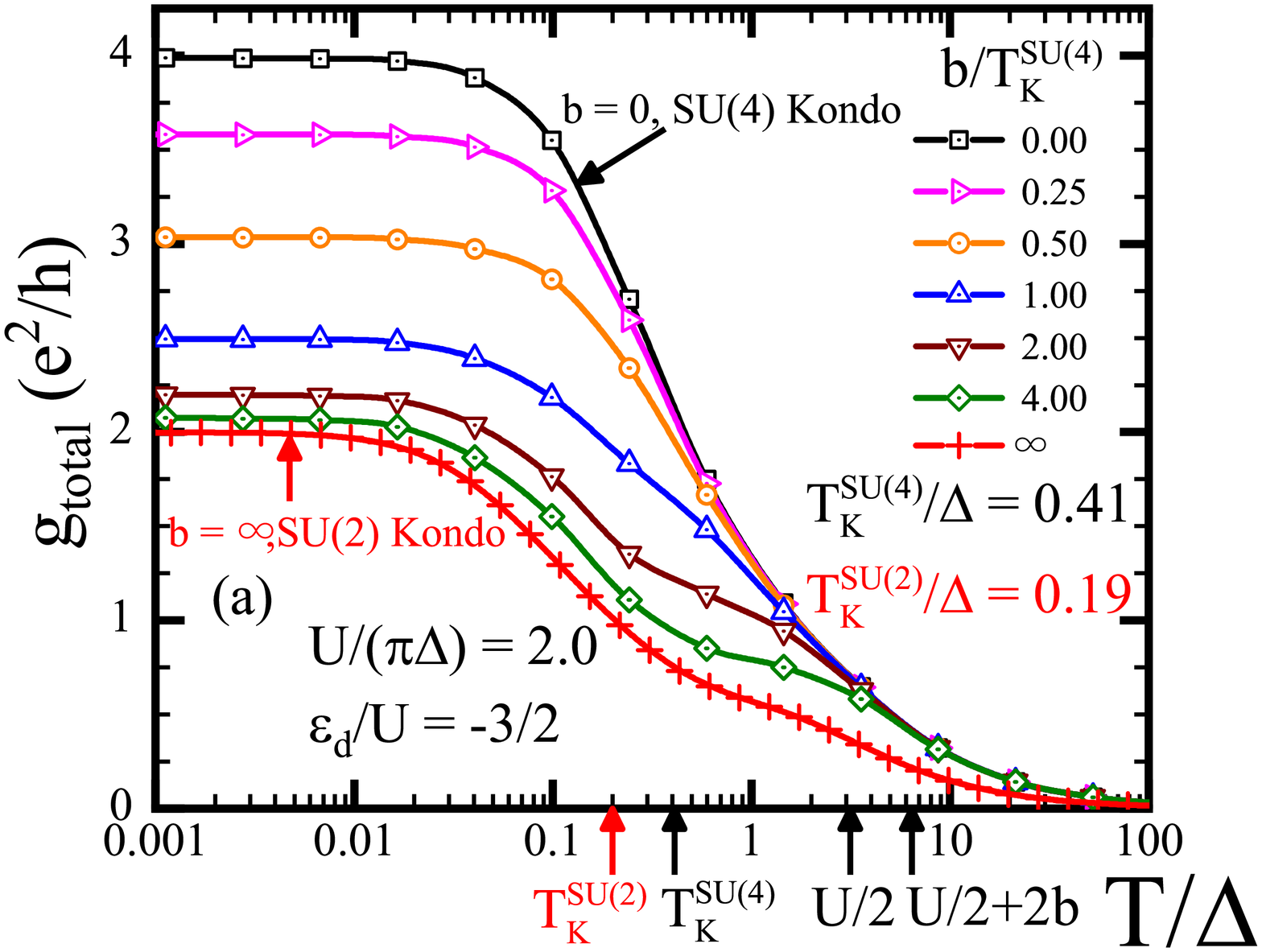}
\end{center}
\end{minipage}
\vspace{0.04\linewidth}
\hspace{0.02\linewidth}
\begin{minipage}{0.42\linewidth} 
\begin{center}
\includegraphics[width=\linewidth]{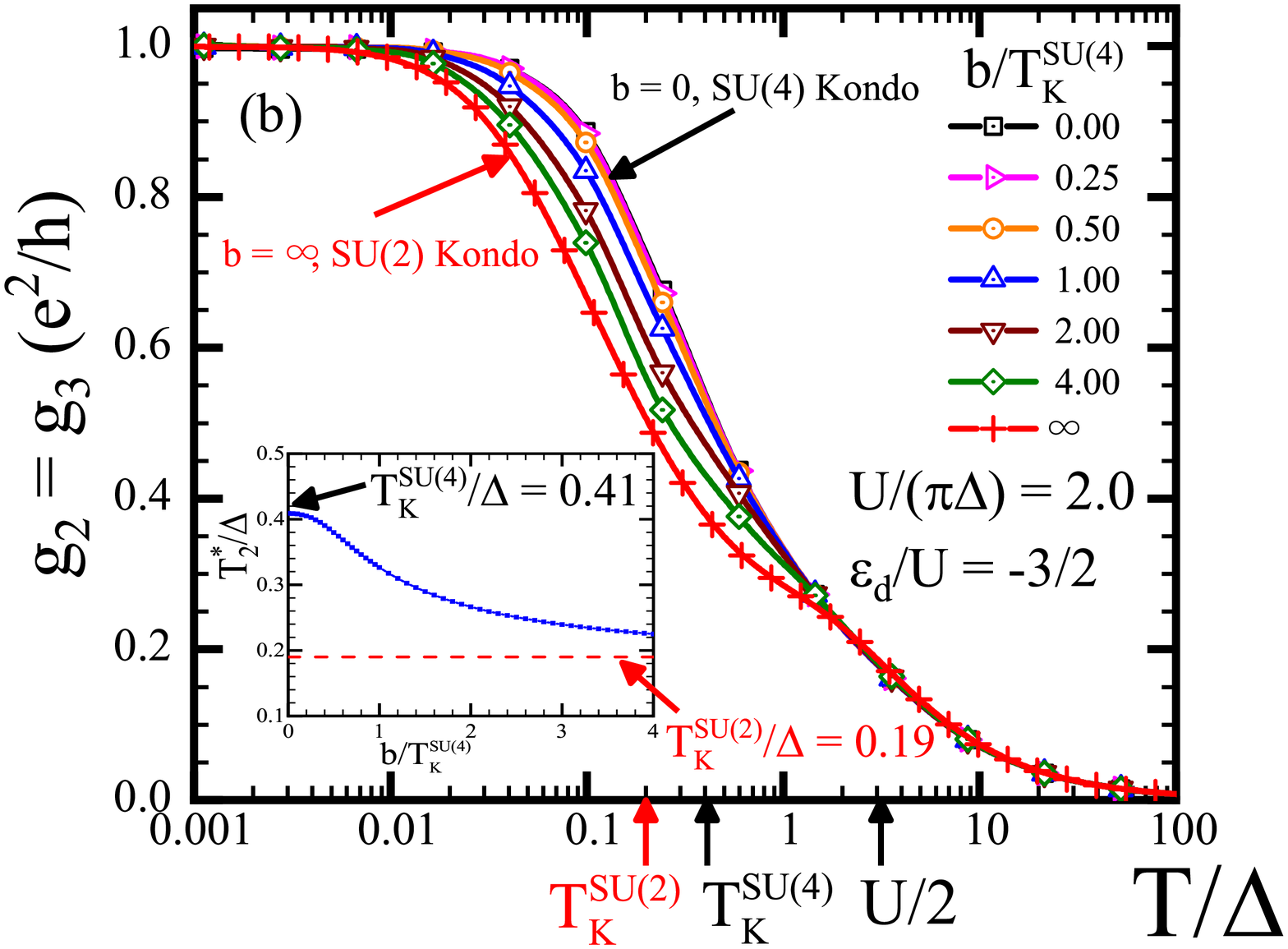}
\end{center}
\end{minipage}

\begin{minipage}{0.42\linewidth} 
\begin{center}
\includegraphics[width=\linewidth]{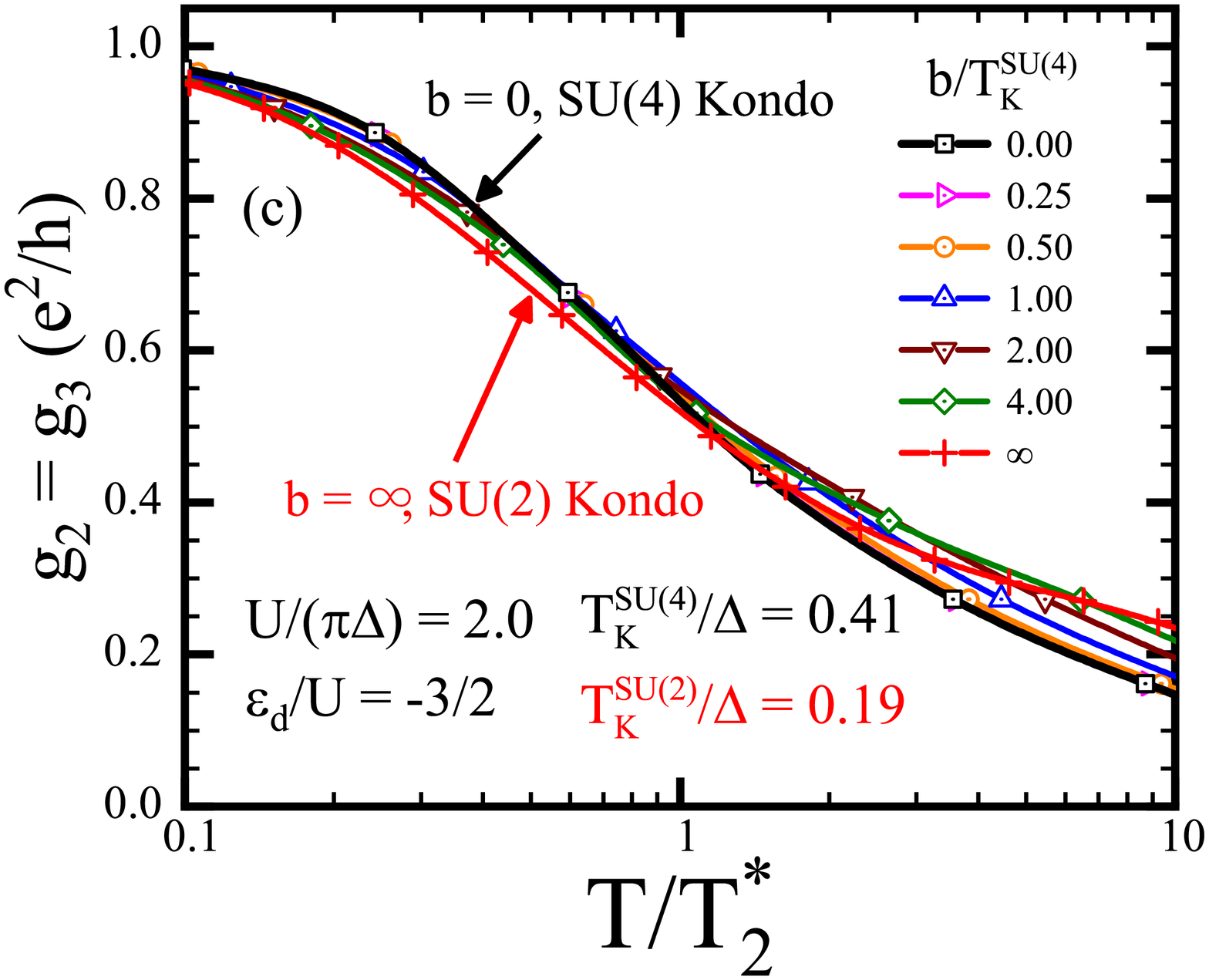}
\end{center}
\end{minipage}
\vspace{0.04\linewidth}
\hspace{0.02\linewidth}
\begin{minipage}{0.42\linewidth} 
\begin{center}
\includegraphics[width=\linewidth]{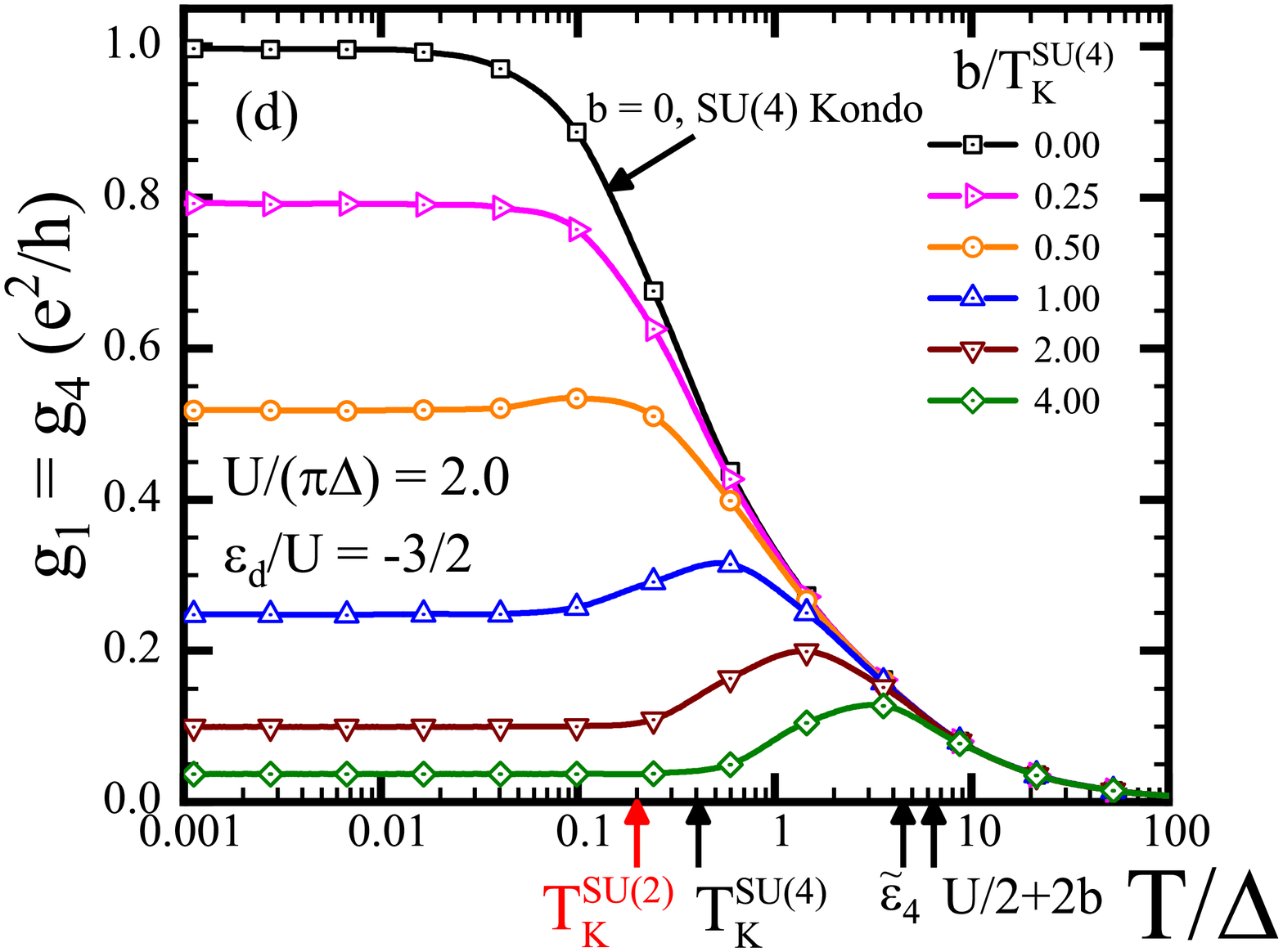}
\end{center}
\end{minipage}

\caption{Temperature dependence of  the linear conductance 
for  $U/(\pi\Delta)=2.0$  
are plotted  for six values of magnetic fields 
$b/T_{K}^\mathrm{SU(4)} = 0.0,\, 0.25,\, 0.5,\, 1.0,\, 2.0,$ and  $4.0$ 
at half-filling $\varepsilon_d^{}=-3U/2$.  
(a) shows the total conductance $g_\mathrm{tot}=\sum_{m=1}^{4}g_{m}$.
 The conductance consists of two components, i.e., $g_{2}=g_{3}$ and $g_{1}=g_{4}$.
(b) and (c) show the first one $g_{2}$, and (d) shows the second one $g_{1}$.
(a)-(c) also show the results for the SU(2) symmetric case by the symbols (+). 
In (a), (b), and (d), the x-axis is  normalized the bare resonance width $\Delta$, 
and the axis in (c) is normalized by a field dependent energy scale $T_{2}^{*}= (\pi/4) Z_{2}^{}\Delta$.
The inset of (b) shows $T_{2}^{*}$ as functions of $b/T_{K}^{\mathrm{SU(4)}}$.
At $b=0$, $T_{2}^{*}$ takes the SU($4$) symmetric value, $T_{K}^{\mathrm{SU(4)}}/\Delta = 0.41$.
In the opposite limit $b=\infty$, it takes SU($2$) value, $T_{K}^{\mathrm{SU(2)}}/\Delta = 0.19$ which is indicated by the dashed line. 
The vertical arrows at the bottom of the panels 
indicate  $T_{K}^\mathrm{SU(2)}$,
$T_{K}^\mathrm{SU(4)}$, $U/2$, $\widetilde{\varepsilon}_{4}^{}$ 
and $U/2+2b$; 
specifically the last two,  $\varepsilon_{4}$ and $U/2+2b$, 
are defined  with respect to  $b/T_{K}^\mathrm{SU(4)}=4.0$.  
} 

\label{GU2HalfFilling_All}
\end{figure}

\end{widetext}

\subsection{Conductance for $U/(\pi \Delta)=2.0$ at half-filling}    

Figures \ref{GU2HalfFilling_All}(a)-(d) plot
the total conductance $g_\mathrm{tot}$ 
and the components $g_{2}=g_{3}$ and $g_{1}=g_{4}$ as  functions of the temperature
for six values of magnetic fields, $b/T_{K}^\mathrm{SU(4)} = 0.0,\, 0.25,\, 0.5,\, 1.0,\, 2.0,$ and $4.0$.
The SU(4) Kondo energy scale, determined at $b=0$ for $U/(\pi \Delta)=2.0$, 
is estimated to be $T_{K}^\mathrm{SU(4)}=0.41\Delta$ as mentioned in Sec.\ \ref{quasi_for_weak_U}.

Figure \ref{GU2HalfFilling_All}(a) shows that
the total counductance at $b=0$ logarithmically increases around $T \sim T_{K}^\mathrm{SU(4)}$.
This logarithmic temperature dependence is a hallmark 
of the SU(4) Kondo effect.
$g_{\mathrm{tot}}$ increases to the unitary-limit value $4e^2/h$ 
as temperature goes down to $T\rightarrow 0$.
As the magnetic fields increase, 
the conductance   
at low-temperatures $T \ll T_{K}^\mathrm{SU(2)}$
decreases from the SU(4) unitary limit value 
$4e^{2}/h$ to  the SU(2) one  $2e^{2}/h$. 
We can also see 
that in a temperature range of 
  $ 0.1T_{K}^\mathrm{SU(2)} \lesssim T \lesssim T_{K}^\mathrm{SU(4)}$ 
the conductance curve deforms continuously into 
the curve for the SU(2) symmetric case.
The SU(2) Kondo state emerges for $b \to \infty$ 
where the characteristic  energy scale becomes $T_{K}^\mathrm{SU(2)}=0.19\Delta$. 
Therefore, the finite temperature conductance also shows 
the crossover behavior.
To examine how the magnetoconductance $g_\mathrm{tot}$ evolves with increasing $b$ in more detail,
we discuss the two components, $g_{2}$ and $g_{1}$. 

Figure \ref{GU2HalfFilling_All}(b) plots the component $g_{2}$ 
which represents contributions of the state $m=2$ and  $m=3$ 
remaining at the Fermi level. 
This component  $g_{2}$ decreases as $b$ increases  in the temperature range of 
  $0.1T_{K}^\mathrm{SU(2)} \lesssim T \lesssim T_{K}^\mathrm{SU(4)}$.
In order to investigate the behavior of $g_{2}$ in this range in more detail, 
we introduce an energy scale $T_{2}^{*}$:  
\begin{align}
T_{2}^{*}\,\equiv\,\frac{\pi}{4}\,\widetilde{\Delta}_{2},
\qquad \widetilde{\Delta}_{2}\,=\,Z_{2}\,\Delta.
\label{T_star_field_depend}
\end{align}
Aside from a numerical factor $\pi/4$, 
this energy  $T_2^*$ corresponds to 
the width  $\widetilde{\Delta}_{2}^{}\,(=\widetilde{\Delta}_{3}^{})$ 
of the Kondo resonance for  $m=2$ and  $m=3$, 
locked at the Fermi level even at finite magnetic fields. 
We use this energy scale to examine the scaling behavior.
 $T_{2}^{*}$ coincides with  $T_{K}^{\mathrm{SU(4)}}=0.41\Delta$ at $b=0$, 
and with $T_{K}^{\mathrm{SU(2)}}=0.19\Delta$  in the opposite limit $b\to\infty$.
The inset of Fig.\ \ref{GU2HalfFilling_All}(b) shows the energy scale $T^{*}$ 
as functions of $b$.
We can see that $T_{2}^{*}$ decreases from $T_{K}^{\mathrm{SU(4)}}$ 
to $T_{K}^{\mathrm{SU(2)}}$ with increasing $b$.
Correspondingly, a region where $g_{2}$ shows  
the $\log T$ dependence 
moves towards a low temperature side as $b$ increases.
Although $g_{2}$ decreases with increasing $b$ at the finite temperatures,
it approaches the unitary limit $e^2/h$ for  $T \to 0$ for arbitrary magnetic fields.
This is because the matching of the spin and orbital Zeeman splitting 
locks the phase shifts at $\delta_{2}=\delta_{3} =\pi/2$ even for magnetic fields.

Another important aspect of the crossover is the scaling behavior of the conductance.
In Ref.\ \onlinecite{MantelliMocaZarandGrifoni}, Mantelli and his coworkers 
examine effects of the spin-orbit interaction on the scaling behavior 
at quarter filling,  $\varepsilon_{d}=-\frac{1}{2}U$. 
We examine how the magnetic field affects the scaling at half-filling, $\varepsilon_{d}=-\frac{3}{2}U$. 
To explore the scaling behavior, 
we rescale temperatures by the field dependent energy scale $T_{2}^{*}$ defined in Eq.\ \eqref{T_star_field_depend}
and replot $g_{2}$ for the six values of $b$ 
as functions of the rescaled temperatures $T/T_{2}^{*}$ in Fig.\ \ref{GU2HalfFilling_All}(c).
The six curves are almost overlapped each other in this figure.
At low fields $b\lesssim T_{K}^{\mathrm{SU(4)}}$, the curves remain the same curve as the SU($4$) universal one over a wide temperature range.
At high fields $b\gg T_{K}^{\mathrm{SU(4)}}$, the curves collapse into the SU($2$) universal curve.
At half-filling points $\epsilon_{d}=-\frac{N-1}{2}U$, 
the Wilson ratio determines the $T^{2}$ coefficient $C_{T}$ for the SU($N$) conductance: 
$C_{T}\,=\,(\pi^{2}/48)\,\left[1\,+\,2\,(N-1)\,(R-1)^{2} \right]$.
Substituting the Wilson ratio for the SU($4$) case $R^{\mathrm{SU(4)}}-1=0.32$ 
and for the SU($2$) case $R^{\mathrm{SU(2)}}-1=0.96$ into the formula of $C_{T}$ 
yields the $T^{2}$ coefficients $C_{T}$ of each case for $U/(\pi\Delta)=2.0$: 
$C_{T}^{\mathrm{SU(4)}}\simeq0.33$ and $C_{T}^{\mathrm{SU(2)}}\simeq0.59$.
Since $C_{T}^{\mathrm{SU(4)}}<C_{T}^{\mathrm{SU(2)}}$, 
the conductance for $N=4$ is larger than that for $N=2$ at the low-temperature regions $T/T_{2}^{*}\lesssim0.1$. 
Figure \ref{GU2HalfFilling_All}(c) shows this magnitude relation of the conductance, 
and thus demonstrates that the scaling behavior depends on the number of orbitals $N$ and 
the Wilson ratio $R$.
  
Figure \ref{GU2HalfFilling_All}(d) shows  
 the other component $g_{1}(=g_{4})$ which correspond to the contributions 
of the other two state  moving away from the Fermi level.
At low temperatures $T \lesssim T_{K}^\mathrm{SU(4)}$,
these components decrease as $b$ increases  
and eventually vanish  at the high magnetic fields $b\gg T_{K}^\mathrm{SU(4)}$. 
We can also see  that  $g_{1}$  has  
a peak at large mange fields  $b\lesssim 0.5 T_{K}^\mathrm{SU(4)}$.
The emergent peak is caused by thermal excitations 
from (to) the renormalized level 
$\widetilde{\varepsilon}_{1}$  ($\widetilde{\varepsilon}_{4}$) 
which situates deep inside (far above) the Fermi level for large fields 
as shown in Fig.\ \ref{b-dependent-FL-U4}(b).
Furthermore, tor the large fields, 
the level structure of the CNT dot 
 approaches the mean-field levels described in Eq.\  \eqref{eq:Hd_B_inf},
and the atomic-limit peak also  emerges at $U/2+2b$ [see also Appendix 
\ref{Spectral functions in atomic limit case}]. 

 These results obtained for  $U/(\pi\Delta)=2.0$ show 
a rather  moderate evolution of  the crossover and the Kondo energy scale $T_{2}^{*}$ 
as $T_{K}^\mathrm{SU(2)}$ is only half of  $T_{K}^\mathrm{SU(4)}$.

\begin{widetext}

\begin{figure}[t]
\centering
\begin{minipage}{0.42\linewidth} 
\begin{center}
\includegraphics[width=\linewidth]{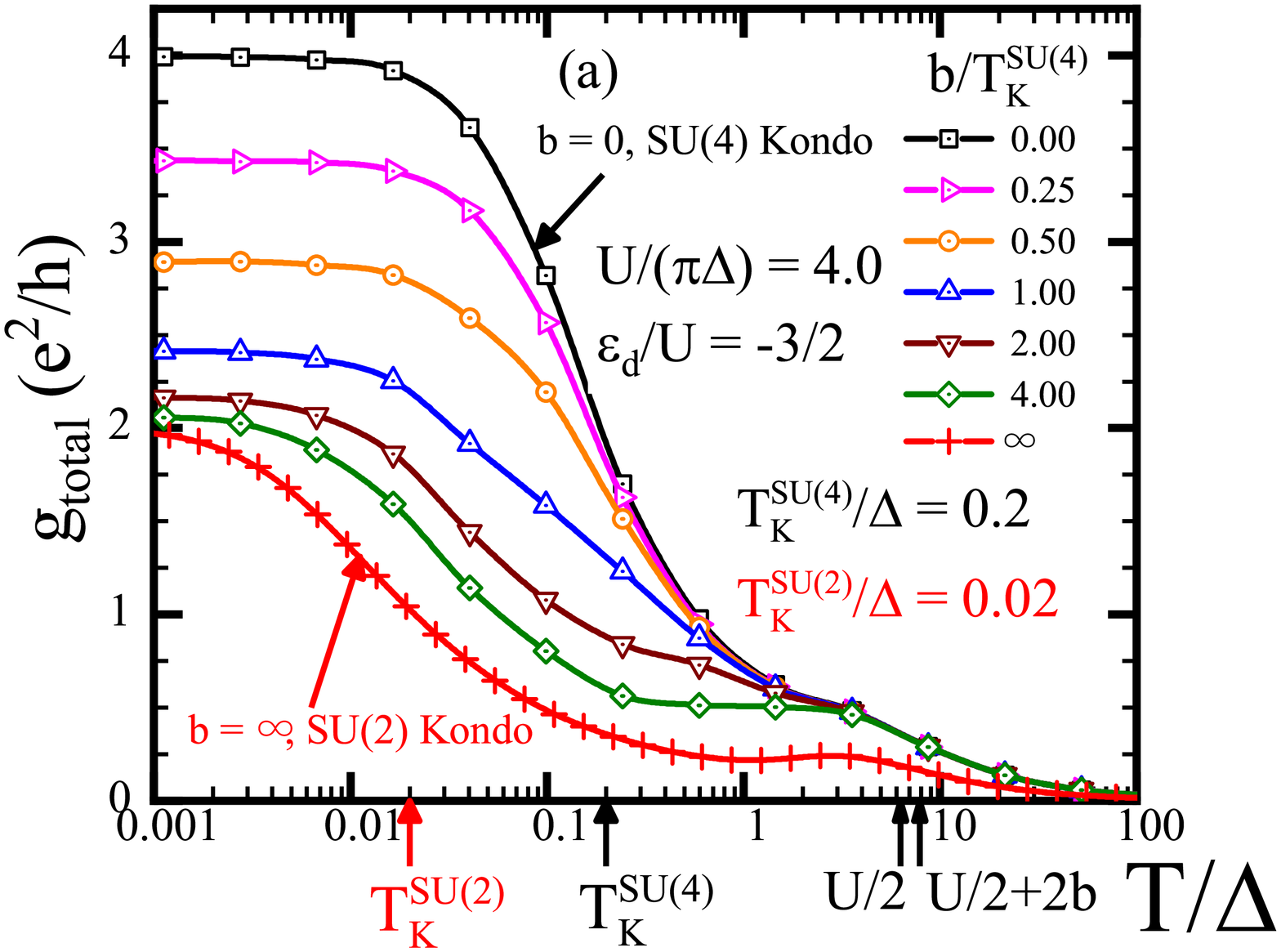}
\end{center}
\end{minipage}
\vspace{0.04\linewidth}
\hspace{0.02\linewidth}
\begin{minipage}{0.42\linewidth} 
\begin{center}
\includegraphics[width=\linewidth]{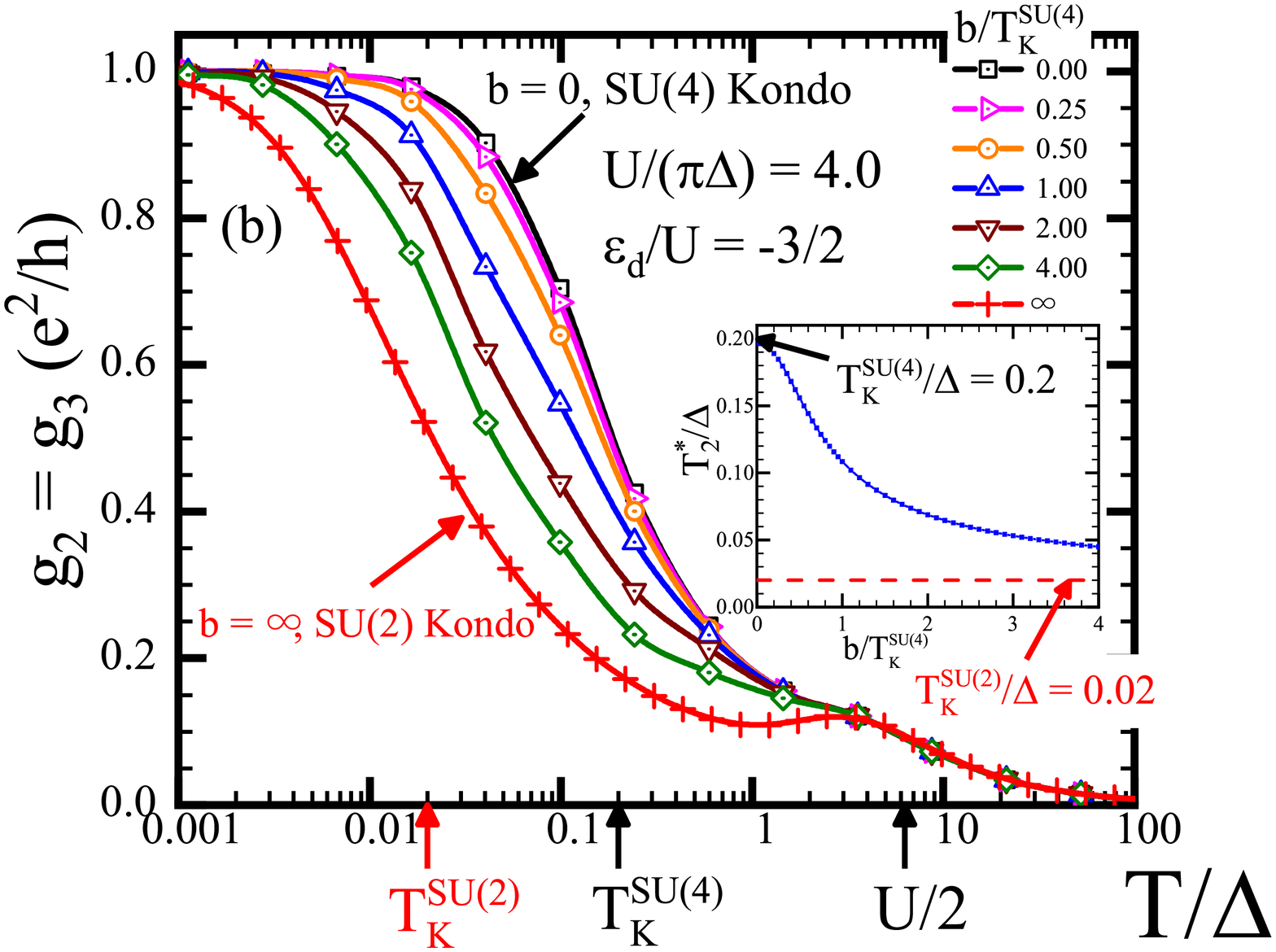}
\end{center}
\end{minipage}

\begin{minipage}{0.42\linewidth} 
\begin{center}
\includegraphics[width=\linewidth]{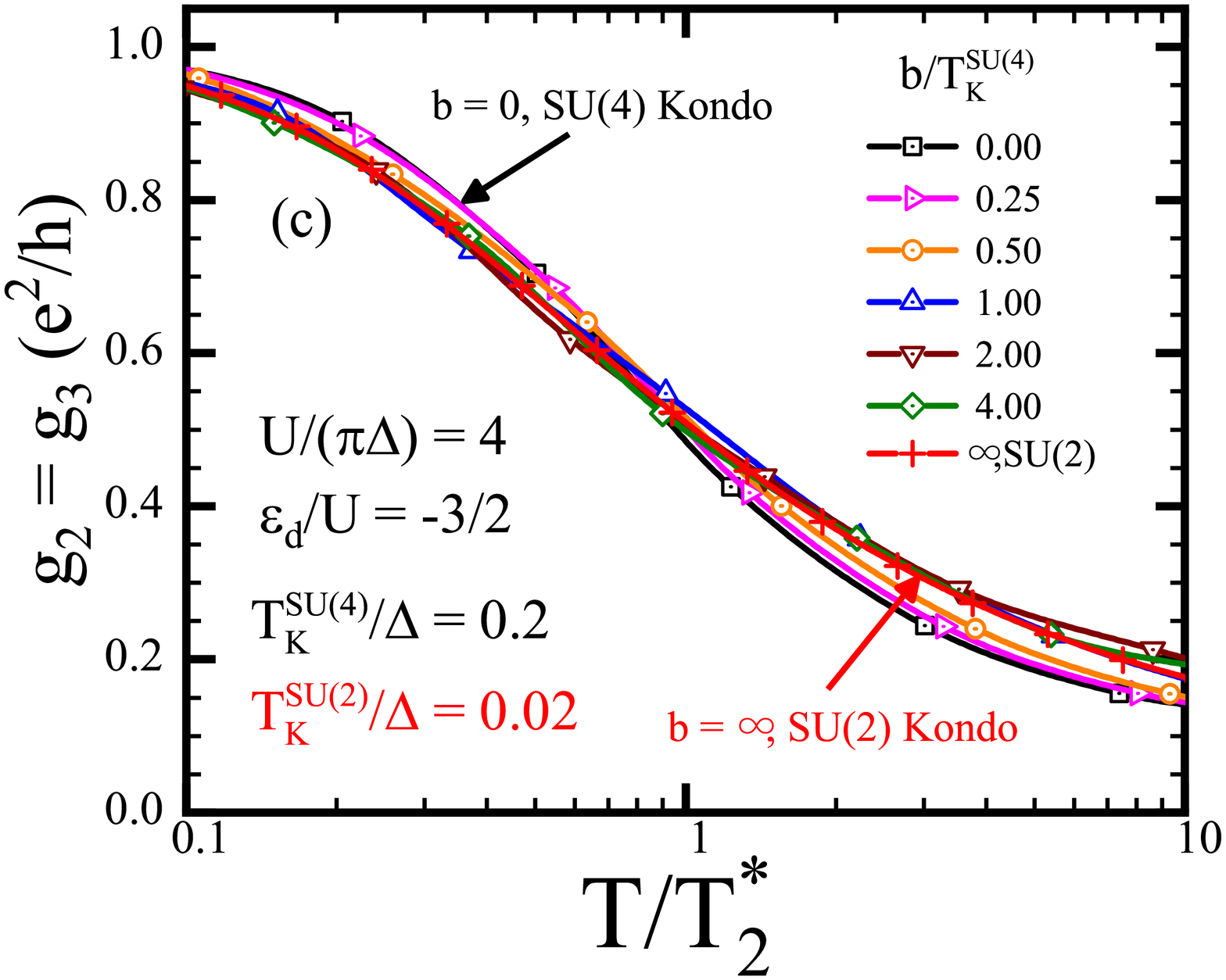}
\end{center}
\end{minipage}
\vspace{0.04\linewidth}
\hspace{0.02\linewidth}
\begin{minipage}{0.42\linewidth} 
\begin{center}
\includegraphics[width=\linewidth]{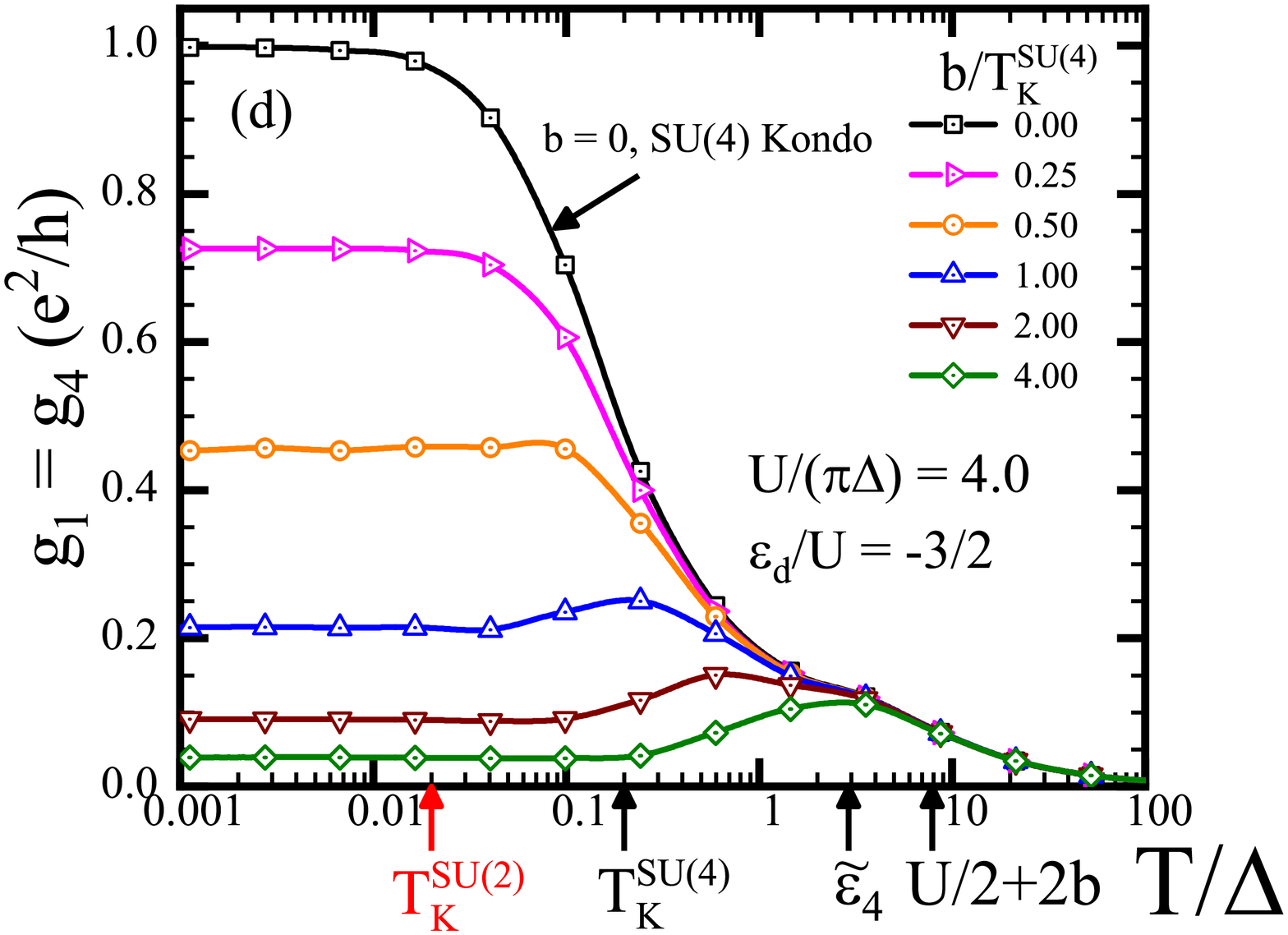}
\end{center}
\end{minipage}

\caption{Temperature dependence of  the linear conductance 
for  $U/(\pi\Delta)=4.0$  
are plotted  for six values of magnetic fields 
$b/T_{K}^\mathrm{SU(4)} = 0.0,\, 0.25,\, 0.5,\, 1.0,\, 2.0,\, 4.0$ 
at half-filling $\varepsilon_d^{}=-3U/2$.  
(a) shows the total conductance $g_\mathrm{tot}=\sum_{m=1}^{4}g_{m}$.
 The conductance consists of two components, i.e., $g_{2}=g_{3}$ and $g_{1}=g_{4}$.
(b) and (c) show the first one $g_{2}$, and (d) shows the second one $g_{1}$.
(a)-(c) also show the results for the SU(2) symmetric case by the symbols (+). 
In (a), (b), and (d), the x-axis is normalized the bare resonance width $\Delta$, 
and the axis in (c) is normalized by characteristic energy scales $T_{2}^{*}$.
The inset of (b) shows $T_{2}^{*}$ as functions of $b/T_{K}^{\mathrm{SU(4)}}$.
At $b=0$, $T_{2}^{*}$ takes the SU($4$) symmetric value, $T_{K}^{\mathrm{SU(4)}}/\Delta = 0.20$.
In the opposite limit $b=\infty$, it takes SU($2$) value, $T_{K}^{\mathrm{SU(2)}}/\Delta = 0.02$ which is indicated by the dashed line. 
The vertical arrows at the bottom of the panels 
indicate  $T_{K}^\mathrm{SU(2)}$,
$T_{K}^\mathrm{SU(4)}$, $U/2$, $\widetilde{\varepsilon}_{4}^{}$ 
and $U/2+2b$; 
specifically the last two,  $\varepsilon_{4}$ and $U/2+2b$, 
are defined  with respect to  $b/T_{K}^\mathrm{SU(4)}=4.0$.  
}
\label{GU4HalfFilling_All}

\end{figure}

\end{widetext}

\subsection{Conductance for  $U/(\pi \Delta)=4.0$ at half-filling}

We next examine the conductance for a case of strong interaction $U/(\pi \Delta)=4.0$ 
to see more clearly the field-induced crossover at finite temperatures. 
In this case, the characteristic energy scale for the SU(2) case  
is significantly suppressed  $T_{K}^\mathrm{SU(2)} =0.02\Delta$, 
which becomes much  smaller than the SU(4) energy scale  
$T_{K}^\mathrm{SU(4)}=0.2\Delta$, i.e., 
the difference is about one order of magnitude.

Figure \ref{GU4HalfFilling_All}(a) shows 
that  the total conductance $g_\mathrm{tot}$
 for a temperature region $0.1T_{K}^{\mathrm{SU(2)}}\lesssim T \lesssim T_{K}^{\mathrm{SU(4)}}$ decreases as magnetic field increases.  
Since the characteristic energy scale $T_{2}^{*}$ defined in Eq.\ \eqref{T_star_field_depend} in this case becomes much smaller 
than that for $U/(\pi\Delta)=2.0$, 
the region where the crossover occurs moves towards a low-temperature region, $T\lesssim T_{K}^{\mathrm{SU(4)}}=0.2\Delta$. 
Furthermore, the shoulder structures emerging in the high temperature region are more pronounced. 

Figure \ref{GU4HalfFilling_All}(b) clearly shows that the curves of $g_{2}$ 
evolve from the SU($4$) curve to the SU($2$) curve during the crossover. 
Specifically, the energy scale $T_{2}^{*}$ around which $g_{2}$ shows $\mathrm{log} T$ dependence 
decreases with increasing magnetic field.
The inset of Fig.\ \ref{GU2HalfFilling_All}(b) shows the suppression of $T_{2}^{*}$:
it decreases from $T_{K}^{\mathrm{SU(4)}}=0.2\Delta$ to $T_{K}^{\mathrm{SU(2)}}=0.02\Delta$.

The scaling behavior of $g_{2}$ in Fig.\ \ref{GU2HalfFilling_All}(c) also becomes clear 
because of this suppression.
In a wide range of temperatures, we can see that the scaled results collapse into two different universal curves, i.e., 
the SU($4$) curve for small fields $b/T_{K}^{\mathrm{SU(4)}}\lesssim 0.25$, 
and SU($2$) curve for large fields $b/T_{K}^{\mathrm{SU(4)}}\gtrsim 1$.
Since the Wilson ratios for $N=4$ and $N=2$ are respectively saturated to 
the maximum possible values $R_{\mathrm{SU(4)}}^{\mathrm{max}}-1=1/3$ and $R_{\mathrm{SU(2)}}^{\mathrm{max}}-1=1$,
the $T^{2}$ coefficients $C_{T}$ for each $N$ 
are also saturated: $C_{T}^{\mathrm{SU(4)}}\simeq0.34$ and $C_{T}^{\mathrm{SU(2)}}\simeq 0.62$.
Figure \ref{GU2HalfFilling_All}(c) clearly shows that $g_{2}$ for $b=0$ is larger than that for $b\to \infty$ 
at $T < T_{2}^{*}$ because of $C_{T}^{\mathrm{SU(4)}}<C_{T}^{\mathrm{SU(2)}}$.

Furthermore, we can recognize that a broad peak emerges 
for $b \gg T_{K}^\mathrm{SU(4)}$ at $T\approx U/2$. 
It corresponds to an thermal energy needed 
to add an electron or a hole to the degenerate states.
Thus, this atomic limit peak and  the quasi-particle excitation peak 
in Fig.\ \ref{GU4HalfFilling_All}(d) at $T\approx 2b+U/2$  yield 
the shoulder structure of $g_\mathrm{tot}$ at the high temperatures, 
as shown in Fig.\ \ref{GU4HalfFilling_All}(a).

\section{Spectral properties along the field-induced crossover}
\label{Spectral_properties_along_the_field-induced_crossover}

Spectral functions 
at finite magnetic fields also reflect the crossover from 
 the SU(4) to SU(2)  Kondo states.
 In addition to the Kondo resonance near the Fermi level,  
the Zeemann splitting causes a shift of the 
atomic-limit peak to  $\pm (2b + U/2)$.  
The spectral functions for the doubly degenerate states 
remain  the same  $A_{2}(\omega,T) =  A_{3}(\omega,T)$ for finite magnetic fields 
owing to the dot level structure given in Eq.\ \eqref{eq:MatchingCondition}.
Following relations additionally hold in the particle-hole symmetric 
case $\varepsilon_{d}=-(3/2)U$,
\begin{align}
A_{2}^{}(\omega,T)\,=\,A_{2}^{}(-\omega,T)
\,, \qquad 
A_{1}^{}(\omega,T)\,=\,A_{4}^{}(-\omega,T)\,.
\label{A-conditon} 
\end{align}
The second relation shows that $A_{4}^{}$ is a mirror image of $A_{1}^{}$, 
and we discuss $A_{1}^{}$ and $A_{2}^{}$ in the following.
We examine the spectral functions at $T=0$, and hence 
we drop the second argument of the functions, namely, $A_{m}^{}(\omega)\equiv A_{m}^{}(\omega,T=0), (m=1,2,3,4)$. 
As in the previous sections, we consider 
the two cases for the interaction: 
 (i) $U/(\pi\Delta)=2.0$ and (ii) $U/(\pi\Delta)=4.0$. 
The spectral function obtained by the NRG procedure is a set of discrete $\delta$ functions. 
To obtain a continuous spectrum, the logarithmic Gaussian function is used (see Appendix \ref{Logarithmic-Gaussian function to broaden discrete spectral function}). 
\begin{figure}[htbp]
\centering
\begin{minipage}{0.95\linewidth} 
\begin{center}
\includegraphics[width=\linewidth]{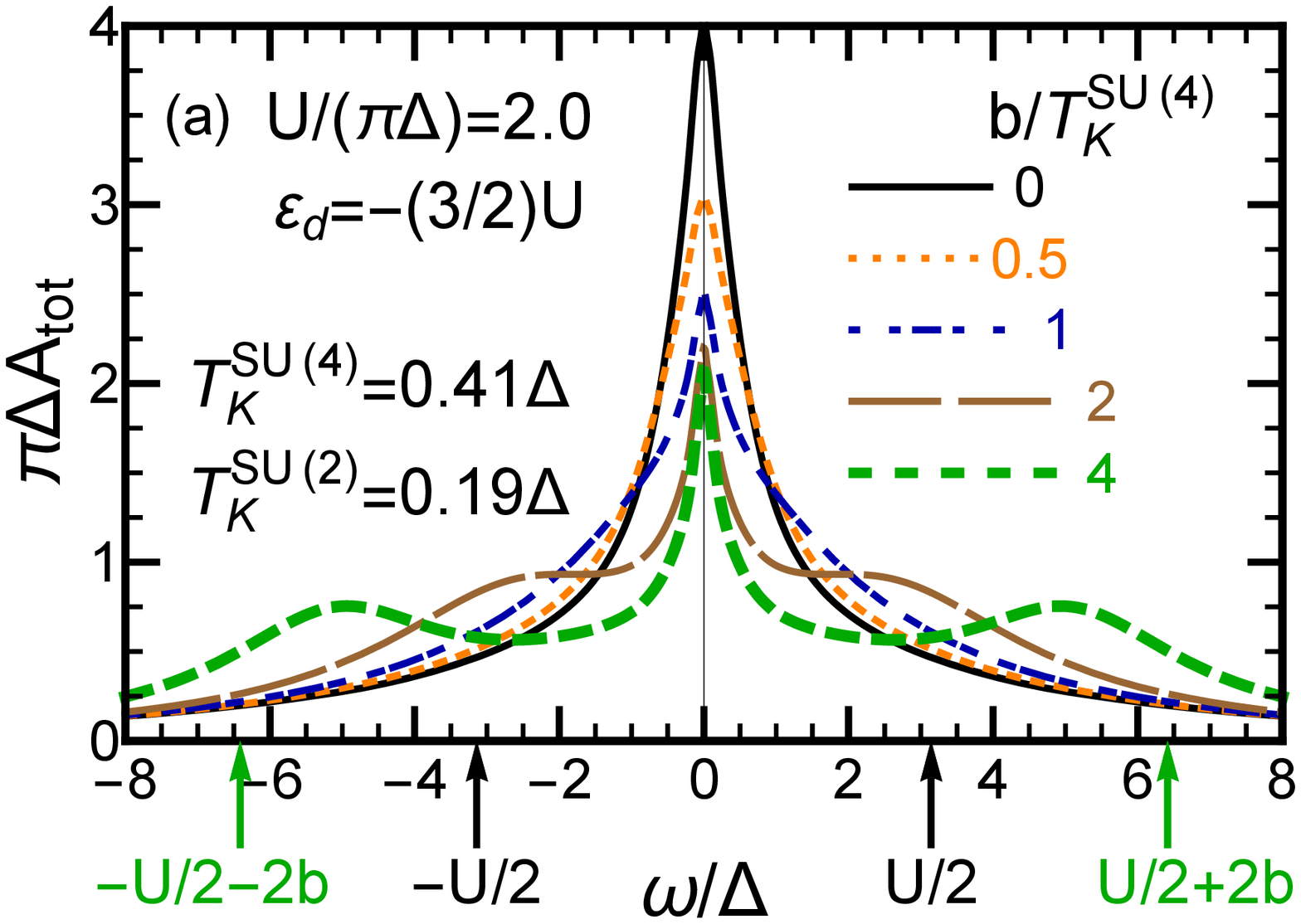}
\end{center}
\end{minipage}
\vspace{0.05\linewidth}

\begin{minipage}{0.95\linewidth} 
\begin{center}
\includegraphics[width=\linewidth]{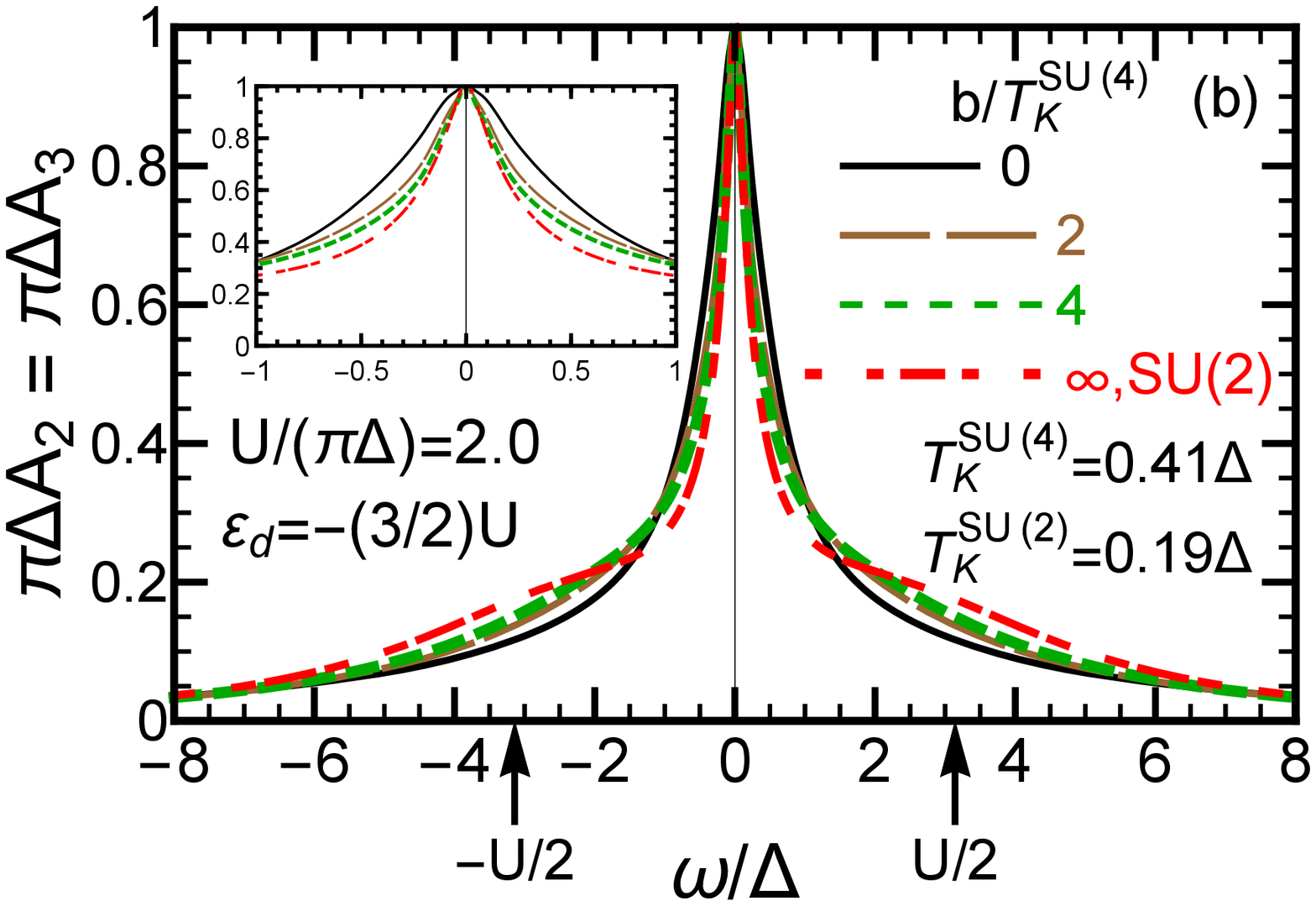}
\end{center}
\end{minipage}
\vspace{0.05\linewidth}

\begin{minipage}{0.95\linewidth} 
\begin{center}
\includegraphics[width=\linewidth]{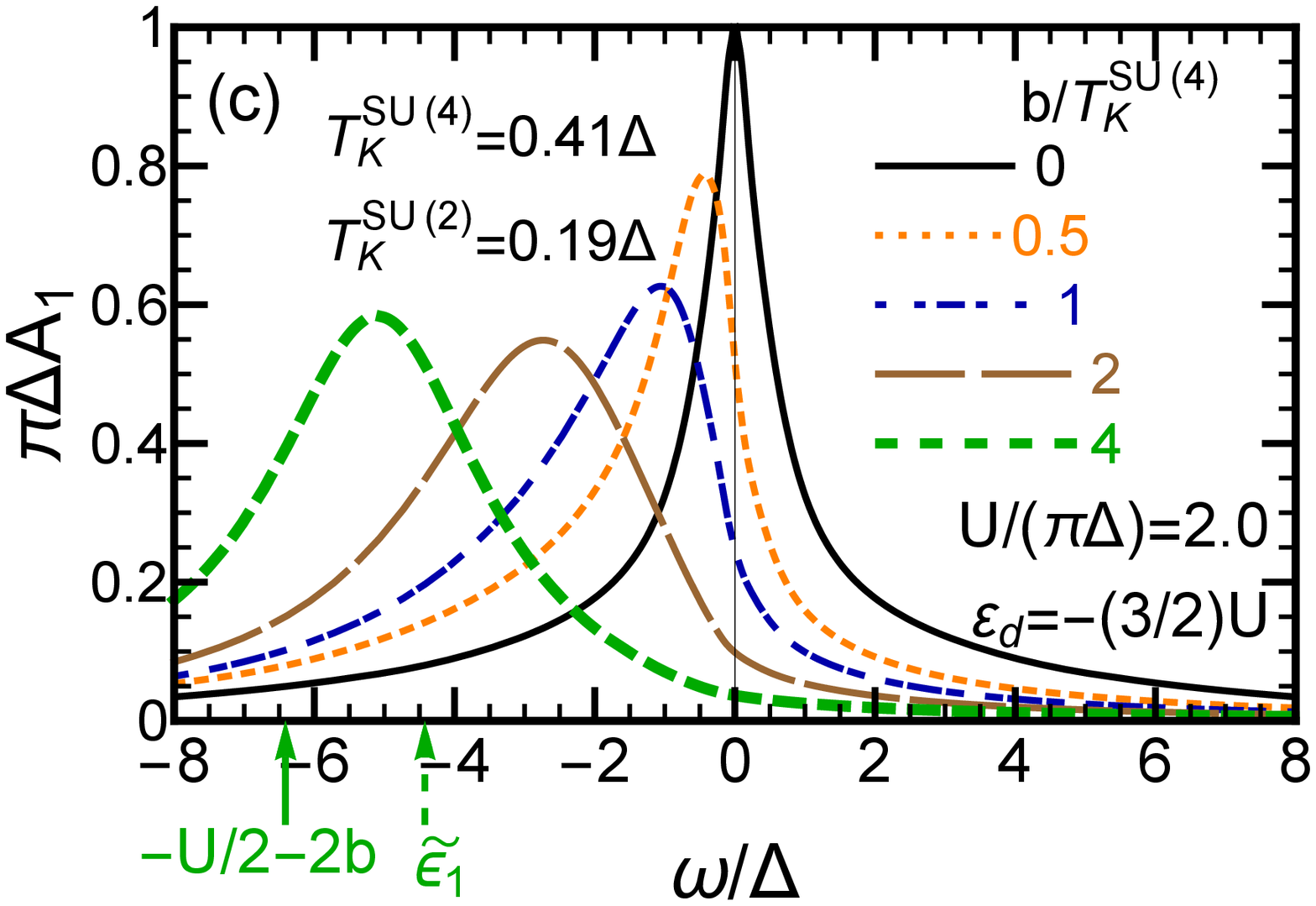}
\end{center}
\end{minipage}

\caption{
Zero temperature spectral functions for $U/(\pi\Delta)=2.0$ 
are plotted  for five values of magnetic fields, 
$b/T_{K}^\mathrm{SU(4)} = 0.0,\, 0.5,\, 1.0,\, 2.0,\, 4.0$   
at half-filling  $\varepsilon_{d}=-3U/2$:   
(a) $A_\mathrm{tot}^{}(\omega) =\sum_{m=1}^{4} A_{m}^{}(\omega)$,
(b) $A_{2}^{}(\omega)$, and
(c)   $A_{1}^{}(\omega)$.
Vertical arrows at the bottom of the panels indicate the points $\omega = \pm U/2$ and
 $ \pm(2b+U/2)$ where peaks emerge in the atomic limit. 
The peaks of $\omega=\pm(2b+U/2)$ are for the largest value of $b$ among the five, $b/T_{K}^{\mathrm{SU(4)}}=4.0$.
 The position of the renormalized resonance level $\widetilde{\varepsilon}_{1}$ 
for the same value of $b$ are also shown in the bottom. 
}
 \label{rhobz000-4Tk-U2_All}
\end{figure}

\subsection{Spectral function for $U/(\pi\Delta)=2.0$}

Figure \ref{rhobz000-4Tk-U2_All}(a) shows 
the total spectral function 
$A_\mathrm{tot}(\omega)=\sum_{m=1}^{4}A_{m}^{}(\omega)$ 
for five different values of magnetic fields and $U/(\pi\Delta)=2.0$.
At zero magnetic field, 
we can see that a single SU(4) Kondo resonance 
peak emerges on the Fermi level $\omega=0$.   
As the magnetic field increases in the range of $0<b<\infty$, 
the height of the Kondo peak decreases from 4 to 2 in units of $\pi\Delta$  
since the resonance peak positions for $m=1$ and $m=4$ move away 
from the Fermi level, leaving the other positions for $m=2$ and $m=3$ 
just on the Fermi level, 
as shown in Fig.\ \ref{b-dependent-FL-U2}(a).
This field dependence of the peak positions results 
in deforming the peak shape 
of the SU(4) Kondo resonance into that of the SU(2) Kondo resonance on the Fermi level.
Furthermore, two sub peaks emerge at higher energies, i.e., $\pm (2b + U/2)$.

Figure \ref{rhobz000-4Tk-U2_All}(b) plots $A_{2}$ ($=A_{3}$) for four values of $b$.
$A_{2}$ ($=A_{3}$) shows such deformation of the peak shape.
The deformation of $A_{2}$ is not so clear in the case of $U/(\pi\Delta)=2.0$ since 
$T_{K}^\mathrm{SU(2)}$ is only half as large as $T_{K}^\mathrm{SU(4)}$.
Nevertheless, the inset of Fig.\ \ref{rhobz000-4Tk-U2_All}(b) which is an enlarged 
view around the Fermi level shows that the resonance width on the Fermi level 
becomes sharper.
The sharpening of the width leads to the decrease of the spectral weight.
As the magnetic field increases, $A_{2}$ also develops two sub peaks 
at $\omega =\pm U/2$ to compensate the decrease of the weight. 
The two peak positions correspond to the excitation energies 
on adding an electron or hole to the dot.   
In Appendix \ref{Spectral functions in atomic limit case}, 
we provide analytic expressions of the spectral functions 
 in the atomic limit $v_{\nu} \to 0$, where 
the CNT dot is disconnected from the metallic leads.  
The remaining degenerate states turn into the SU(2) Kondo state in the limit of $b\rightarrow \infty$. 
indicating that the states undergo the crossover 
from the SU(4) to SU(2) Kondo state.

Figure \ref{rhobz000-4Tk-U2_All}(c) shows the other 
component  $A_{1}^{}(\omega)$,
which corresponds to the component of the level going down from the Fermi level.
Note that  $A_{4}^{}(\omega)=A_{1}^{}(-\omega)$, as mentioned.
We can see that the spectral weight 
transfers to the negative frequency region $\omega<0$, 
and an evolution of its resonance peak position shows good agreement with the 
field dependence of $\widetilde{\varepsilon}_{1}$ 
presented in Fig.\ \ref{b-dependent-FL-U2}(a).
This transfer leads to the development of the sub 
peaks and decrease of the Kondo peak of $A_\mathrm{tot}$.
With increasing magnetic fields, 
the resonance peak at $\omega = \widetilde{\varepsilon}_{1}$ merges 
with the atomic-limit peak at $\omega=-U/2-2b$ which shifts from the zero field 
position $-U/2$ in the presence of $b$.
This shift of the atomic-limit peak, 
which we discuss in the Appendix \ref{Spectral functions in atomic limit case}, 
results from the descent of the energy level $\varepsilon_{1}$ 
described in Eq.\ \eqref{eq:MatchingCondition}.
For much larger fields $b\gg T_{K}^\mathrm{SU(4)}$,
the curve of $A_{1}$ approaches the Lorentzian form. 
Therefore, the quasiparticle state of $m=1$ 
is unrenormalized from the 
correlated Kondo state to the bare state.

\subsection{Spectral function for $U/(\pi\Delta)=4.0$}
\label{Spectral functions in the strong interaction case}

\begin{figure}[htbp]
\centering
\begin{minipage}{\linewidth} 
\begin{center}
\includegraphics[width=0.95\linewidth]{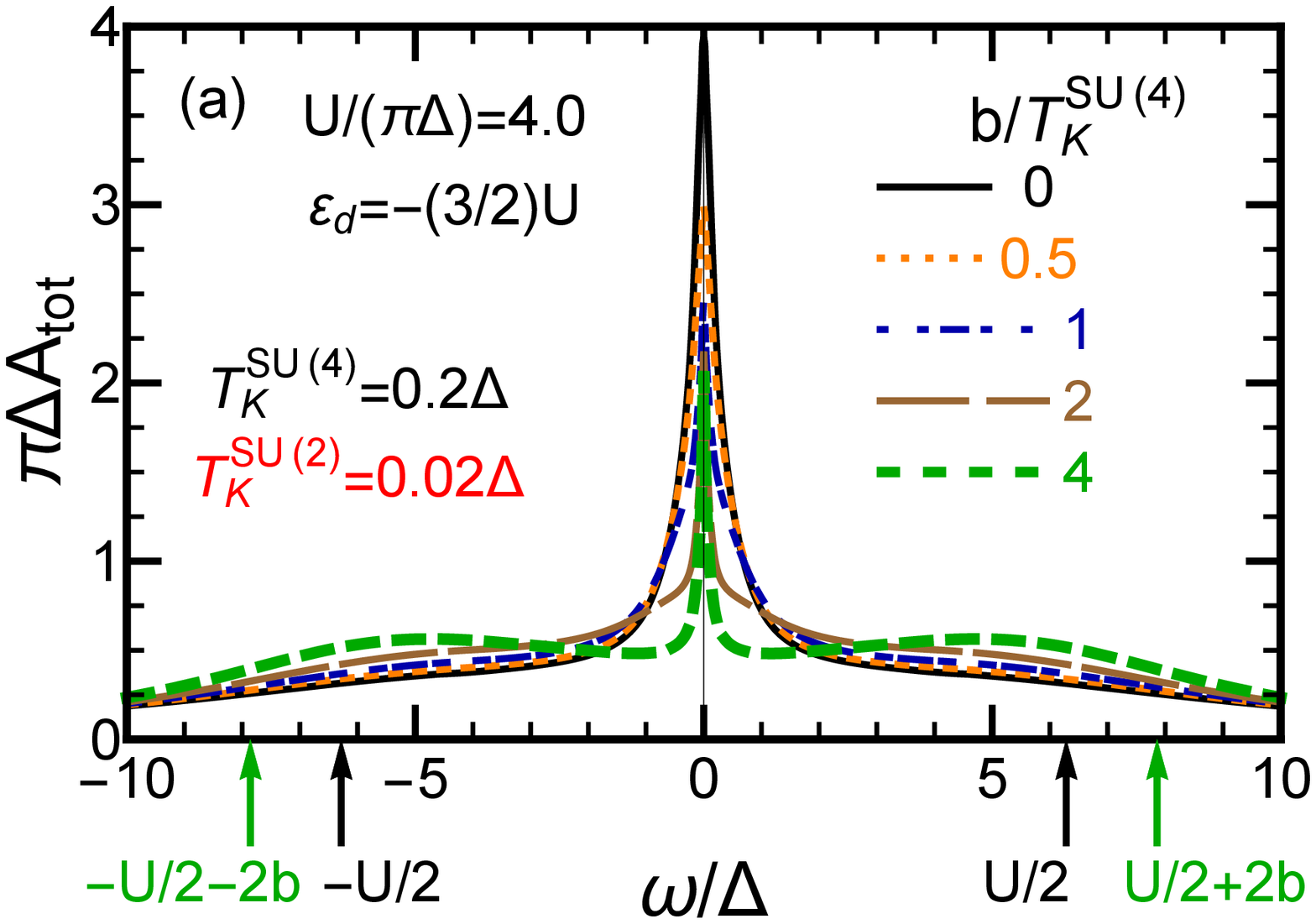}
\end{center}
\end{minipage}
\vspace{0.05\linewidth}

\begin{minipage}{\linewidth} 
\begin{center}
\includegraphics[width=0.95\linewidth]{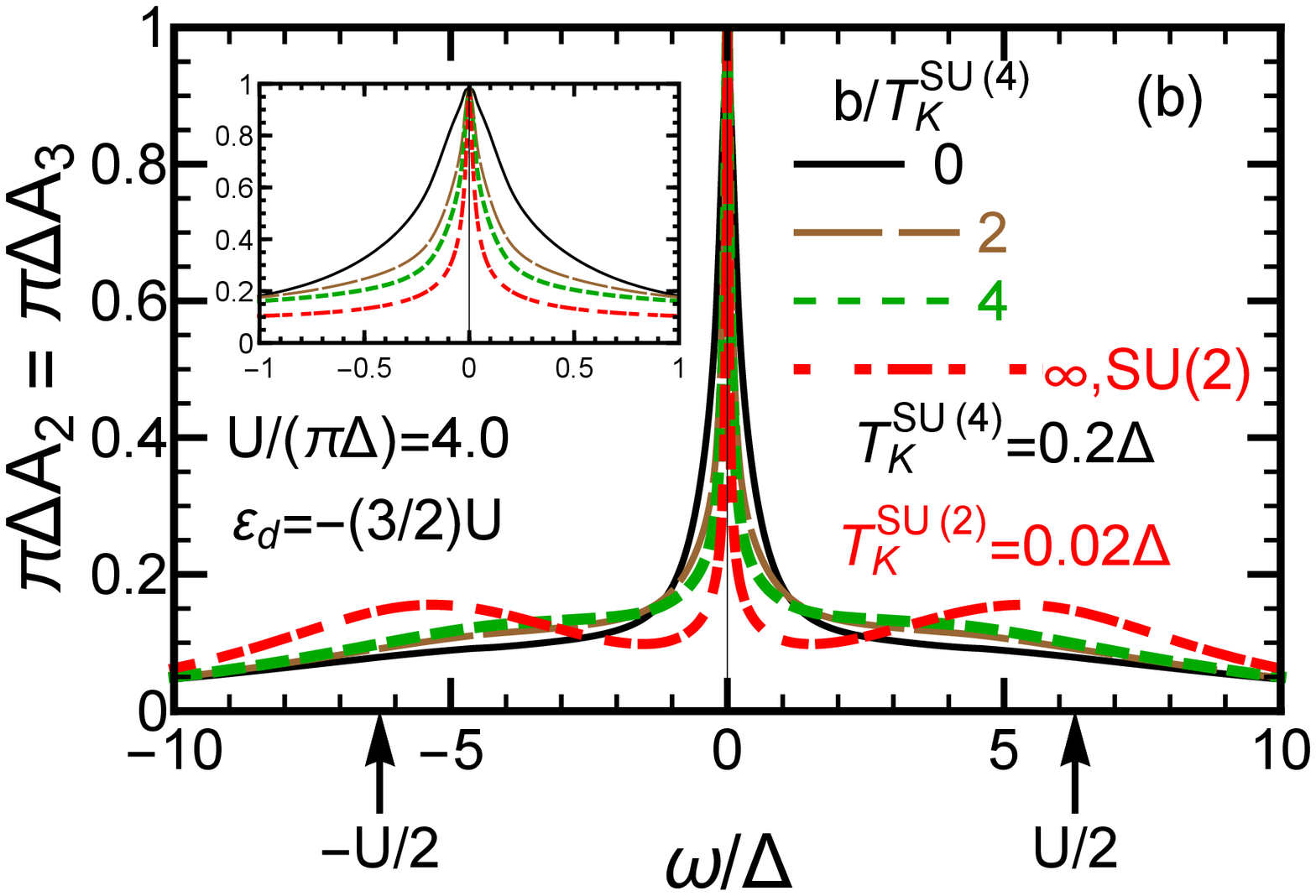}
\end{center}
\end{minipage}
\vspace{0.05\linewidth}

\begin{minipage}{\linewidth} 
\begin{center}
\includegraphics[width=0.95\linewidth]{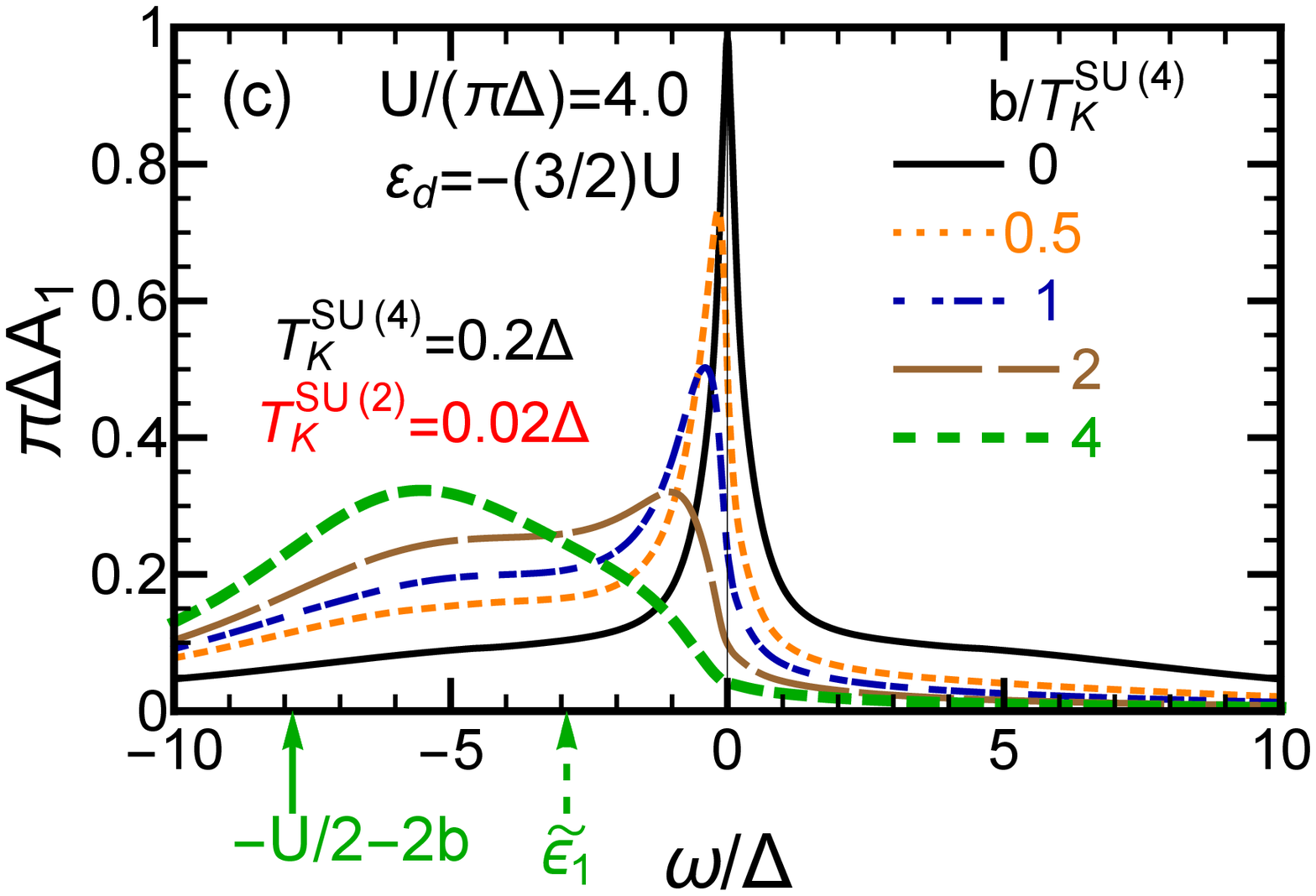}
\end{center}
\end{minipage}

\caption{
Spectral functions for $U/(\pi\Delta)=4.0$ 
are plotted  for five values of magnetic fields, 
$b/T_{K}^\mathrm{SU(4)} = 0.0,\, 0.5,\, 1.0,\, 2.0,\, 4.0$   
at half-filling  $\varepsilon_{d}=-3U/2$:   
(a) $A_\mathrm{tot}^{}(\omega) =\sum_{m=1}^{4} A_{m}^{}(\omega)$,
(b) $A_{2}^{}(\omega)$, and
(c)   $A_{1}^{}(\omega)$.
Vertical arrows at the bottom of the panels indicate the points $\omega = \pm U/2$ and
 $ -(2b+U/2)$ where peaks emerge in the atomic limit. 
The peaks of $\omega=\pm(2b+U/2)$ are for the largest value of $b$ among the five, $b/T_{K}^{\mathrm{SU(4)}}=4.0$.
 The position of the renormalized resonance level $\widetilde{\varepsilon}_{1}$ 
for the same value of $b$ are also shown in the bottom.
}
 \label{rhobz000-4Tk-U4_All}
\end{figure}

In order to investigate the effects of the strong interaction on the spectral functions, 
we next discuss the spectral functions
for $U/(\pi\Delta)=4.0$.
Figures \ref{rhobz000-4Tk-U4_All}(a)-(c) respectively show results of $A_\mathrm{tot}$, $A_{2}$ and $A_{1}$. 
We compare these results with the corresponding results for the weak interaction case 
in Fig.\ \ref{rhobz000-4Tk-U2_All}, 
$A_\mathrm{tot}$ for the strong interaction 
in Fig.\ \ref{rhobz000-4Tk-U4_All}(a) 
shows a similar trend
as that for $U/(\pi\Delta)=2.0$ in Fig.\ \ref{rhobz000-4Tk-U2_All}(a). 
However, the width of resonance for $U/(\pi\Delta)=4.0$ 
is smaller than that for $U/(\pi\Delta)=2.0$  
in arbitrary magnetic fields, because $U$ is larger. 
  
The component $A_{2}$ in Fig.\ \ref{rhobz000-4Tk-U4_All}(b) more clearly shows the 
narrowing of the resonance width than that for $U/(\pi\Delta)=2.0$,
because $T_{K}^\mathrm{SU(2)}$ is smaller than $T_{K}^\mathrm{SU(4)}$ by 
one order of magnitude  in this case i.e., $T_{K}^{\mathrm{SU(2)}}=0.02 \Delta$ and $T_{K}^{\mathrm{SU(4)}}=0.2 \Delta$.
The narrowing of the width leads to a loss of the spectral weight around the Fermi level, 
which is compensated by an enhancement of the atomic-limit peak at $\pm U/2$.

The atomic limit peak around $-U/2-2b$ of $A_{1}$ is broader in the strong interaction case shown in Fig\ \ref{rhobz000-4Tk-U4_All}(c) than in the weak interaction case, 
because the quasiparticle resonance position $\widetilde{\varepsilon}_{1}$ presented 
in Fig.\ \ref{b-dependent-FL-U4}(b) still remains around the Fermi level 
 at the higher fields, $b\gg T_{K}^\mathrm{SU(4)}$.
Owing to this remaining, the quasiparticle state is still renormalized even at the higher fields,
 and thus the shape of $A_{1}$ differs from the Lorentzian form.

\section{Summary}
\label{sec:summary}

We have studied the Kondo effect in a carbon nanotube quantum dot 
in a wide range of temperature and magnetic field
using the numerical renormalization group.

In the first half of the present paper, we have studied finite temperature properties of the SU($4$) Kondo state
by calculating the finite temperature conductance in a wide range of electron filling $N_{d}$.
The NRG results nicely agree with the experimental results in the wide range,
supporting an emergence of the SU(4) Kondo resonance at low-temperatures, $T\lesssim T_{K}^\mathrm{SU(4)}$.
Furthermore, we have precisely examined the temperature dependence of conductance especially at two fixed values of $N_{d}$: 
quarter-filling $N_{d} =1$ and half-filling $N_{d} =2$.
The obtained results show that the scaled conductance of $N_{d} =1$ is larger than that of $N_{d} =2$ at the low-temperatures. 
A microscopic Fermi-liquid theory, which is extended to arbitrary $N_{d}$, successfully explains such different behavior of the conductance depending on $N_{d}$.
The theory shows that a $T^{2}$ coefficient $C_{T}$ for the conductance vanishes at $N_{d} =1$.
In contrast to the quarter-filling case, $C_{T}$ does not become zero at half-filling, but saturates to a strong coupling limit value.
Thus, the universality depends on the electron filling.
We expect that this filling dependence of the universality can be observed. 

In the second half of the present paper, we have investigated how magnetic fields affect the ground state and also excited states in the course of the SU(4) to SU(2) Kondo crossover. 
Our previous papers show that quasiparticle states remaining the Fermi level 
are renormalized as the number of active levels decreases from four to two. 
The other two states become unrenormalized in the course of the crossover. 
The present paper has shown that the renormalization of the quasiparticle states 
more clearly appear in a strong interaction case because a characteristic energy scale $T_{2}^{*}$ clearly decreases from the SU(4) Kondo energy scale $T_{K}^\mathrm{SU(4)}$ to the SU(2) Kondo energy scale $T_{K}^\mathrm{SU(2)}$. 

The finite temperature conductance in the magnetic fields also shows such decrease of the energy scale. In addition, the scaling behavior at half-filling
shows that the excited states undergo the crossover.
Specifically, as soon as the magnetic fields $b$ become comparable to $T_{K}^\mathrm{SU(4)}$, 
the SU(4) universality is lost, and for the much larger fields, $b\gg T_{K}^\mathrm{SU(4)}$, the SU(2) universality emerges. 
Furthermore, the NRG results for both SU($2$) and SU($4$) symmetric cases indicate that the
Wilson ratio and the Kondo energy scale determine the low-temperature behavior of half-filled quantum dots.

We have also calculated total spectral function and their components in magnetic fields. 
The obtained spectral function shows that the resonance states remaining on the Fermi level become sharper as the magnetic field increases, showing a good agreement with the field dependence of the corresponding renormalized resonance width.
Furthermore, a spectral weight of the other two states transfers towards the higher frequency region,
because the Zeeman splitting shifts the two peak positions upward and downward from the Fermi level. Such transfer results in the emergence of two sub-peaks whose positions approach atomic-limit peak positions.

An interesting future work is to explore how the thermoelectric transport  of 
multilevel quantum dots depends on the electron filling\cite{Costi_Thermo_Power_Magnetic_Field,Svilans_Thermoelectric_Power_Experiments,Bas_Aligia_SU4_thermo}. 
We have already had the low-temperature expansions for the thermal conductance,\cite{teratani2020fermi} and the progress along this line will be discussed  elsewhere.

\begin{acknowledgments}
This work was supported by Grant-in-Aid for JSPS Fel- 1035
lows Grant No. JP18J10205 and JSPS KAKENHI Grant 1036
Nos. JP18K03495, JP19H00656, JP19H05826, JP16K17723, 1037
and JP19K14630, JST CREST Grant No. JPMJCR1876, 1038
the French program Agence nationale de la Recherche 1039
DYMESYS (ANR2011-IS04-001-01), and Agence nationale 1040
de la Recherche MASH (ANR-12-BS04-0016).
\end{acknowledgments}

\appendix

\section{Symmetries of Hamiltonian}
\label{sec:Symmetries_of_Hamiltonian}

In the case where  the impurity level is degenerate 
$\epsilon_{m}\equiv\epsilon_{d}$, 
the Hamiltonian $\mathcal{H}$ defined in Eq.\ \eqref{eq:Hamiltonian} 
has the $\mathrm{SU(\it{N})}$ symmetry.
This is also owing to the fact that the tunneling matrix element and the Coulomb interaction 
do not depend on the flavour $m$. 
It can also be confirmed  that  $\mathcal{H}$ 
commutes with each component of  the $\mathrm{SU(\it{N})}$ generators: 
\begin{align}
{\bm J}_{}^{\mu}\!\equiv\!\frac{1}{2}\!\!\sum_{m,m'=1}^{N}\!\!\left[\!d^{\dagger}_{m}{\bm \lambda}^{\mu}_{m,m'}d^{}_{m'}\!+\!\sum_{\nu=L,R}\!\int_{-D}^{D}\!\!\!d\varepsilon 
c_{\nu,\varepsilon,m}^{\dagger}{\bm \lambda}^{\mu}_{m,m'}c_{\nu,\varepsilon,m}^{}\!
\right].
\label{SU(4)_Operator}
\end{align}
Here, ${\bm \lambda}^{\mu}$ for $\mu=1,2,\cdots,N^2-1$
are the Gell-mann matrices that satisfy the commutation relations,
\begin{align}
\left[{\bm\lambda}_{}^{i}, {\bm\lambda}_{}^{j}  \right]=2\,i\,\sum_{k=1}^{N^2-1}\,f_{}^{ijk}\,{\bm\lambda}_{}^{k},
\label{commutation_relation_lambda}
\end{align}
with $f_{}^{ijk}$  the structure factor. 
The Hamiltonian  $\mathcal{H}$ has also 
an $\mathrm{U(1)_{tot}}$ symmetry and  commutes with the 
total number operator $Q_{\mathrm{tot}}=\sum_{m=1}^{N}Q_{m}$.
Here, $Q_{m}$ is the number operator for  flavour $m$: 
\begin{align}
Q_{m}&\,=\,\left[n_{dm} +\sum_{\nu=L,R}\int_{-D}^{D}\!d\varepsilon\,
c_{\nu,\varepsilon,m}^{\dagger}c_{\nu,\varepsilon,m} \right],
\label{Q_m_Operator_each}
\end{align}
for $m=1,2,\cdots,N$.
Perturbations that lift the degeneracy of the impurity level  
lower the symmetry from 
 $\mathrm{SU(\it{N})}$$\otimes$$\mathrm{U(1)_{tot}}$  to $\mathrm{U(1)}_{m=1}\otimes\mathrm{U(1)}_{m=2}\otimes\mathrm{U(1)}_{m=3}\otimes\mathrm{U(1)}_{m=4}$ as $\mathcal{H}$ still commutes with each number operator $Q_{m}$. 

For carbon nanotube dots, the two states of $m=2$ and $m=3$ remain degenerated even in magnetic fields when the matching condition Eq.\ \eqref{eq:MatchingCondition_ORG} 
holds.
In the condition, $\mathcal{H}$ commutes with the SU(2)-pseudospin operator,
\begin{align}
{\bm S}_{}^{\mu}\!\equiv\!\frac{1}{2}\!\!\sum_{m ,m' = \Uparrow, \Downarrow}^{}\!\!\left[\!d^{\dagger}_{m }{\bm \sigma}^{\mu}_{m ,m' }d^{}_{m' }\!+\!\sum_{\nu=L,R}\!\int_{-D}^{D}\!\!\!d\varepsilon 
c_{\nu,\varepsilon,m }^{\dagger}{\bm \sigma}^{\mu}_{m ,m' }c_{\nu,\varepsilon,m }^{}\!
\right]\!\!.
\label{SU(2)_Operator}
\end{align} 
acting on the two states. 
Here, ${\bm \sigma^{\mu}}$ for $\mu=x,y,$ and $z$ are the Pauli matrices.
The pseudospins $\Uparrow$ and $\Downarrow$ respectively denote the state of $m=2$ and $m=3$. 
Thus, the Hamiltonian has the $\mathrm{U(1)}_{m=1}\otimes[\mathrm{SU(2)}\otimes\mathrm{U(1)}]_{m=2,3}\otimes\mathrm{U(1)}_{m=4}$ symmetries. 
The SU(2) symmetry plays a central role in 
the field-induced SU(4) to SU(2) Kondo crossover.

At half-filling point $\epsilon_{d}/U=-3/2$, 
the Hamiltonian $\mathcal{H}$ is invariant under an extended electron-hole transformation:
\begin{align}
&d_{1}^{\dagger} \Rightarrow h_{4}^{}, \;\; d_{2}^{\dagger} \Rightarrow h_{3}^{}, \;\; d_{3}^{\dagger} \Rightarrow h_{2}^{},
\;\; d_{4}^{\dagger} \Rightarrow h_{1}^{},
\label{extended_particle_hole_symmetry}
\end{align}
and correspondingly, $c_{\nu, \varepsilon_{m}, m}^{\dagger} \Rightarrow -f_{\nu, -\varepsilon_{m'}, m'}^{}$ for $(m,m')=(1,4), (2,3), (3,2), (4,1)$. 
Here, $h_{m}$ and $f_{\nu, \varepsilon_{m'}, m'}^{}$ annihilate hole in the dot and the conduction bands, respectively.

\section{Overview of NRG iterations}
\label{Overview_of_NRG_iterations}

We provide a brief overview of an NRG iteration. 
The NRG Hamiltonian $\mathcal{H}$ for an Anderson impurity model is a semi-infinite tight-binding chain with the imuprity at a left end.
Specifically,
\begin{align}
&\mathcal{H}_{\mathrm{NRG}}^{}\,=\,\mathcal{H}_{d}^{0}\,+\,\mathcal{H}_{U}^{}\,+\,\mathcal{H}_{\mathrm{chain}}^{}\,+\,\mathcal{H}_{T}^{}, \label{SIAM_chain_infinite} \\
&\mathcal{H}_{d}^{0} \,= 
\sum_{m=1}^{N}
 \epsilon_m^{}  
d_{m}^{\dagger} d_{m}^{} , 
\qquad 
 \mathcal{H}_{U}^{} =
 \,U \! 
\sum_{m< m'} n_{dm} n_{dm'} , 
\label{eq:Hd_NRG} \\
&\mathcal{H}_{\mathrm{chain}}\,=\,\sum_{n\,=\,0}^{\infty}\,\sum_{m=1}^{N}\,t_{n}^{}\left(f_{n,m}^{\dagger} f_{n+1,m}^{}\,+\,\mathrm{H.c.} \right)\;, \label{eq:Hc_NRG} \\
&\mathcal{H}_{T}\,=\,V\,\sum_{m=1}^{N}\,\left(d_{m}^{\dagger}f_{m,0}^{}\,+\,\mathrm{H.c.}\right). 
\label{eq:Ht_NRG}
\end{align}
Here, $\Lambda$ is a logarithmic discretization parameter.
A hybridization matrix element $V$ couples the impurity site to a $0$th site of the chain.
Similarly, a hopping element $t_{n}$ couples the $n$th site to $(n+1)$th site.
The explicit forms of $V$ and $t_{n}^{}$ are respectively
\begin{align}
&\frac{V}{D}\,=\,\sqrt{2\Delta A_{\Lambda} \over \pi D}, \label{V_NRG} \\ 
&\frac{t_{n}^{}}{D}\,=\,\frac{\left(1+ \Lambda^{-1}\right)\,\left(1\,-\,\Lambda^{-n-1} \right)}{2\, \left(1\,-\,\Lambda^{-2n-1} \right)^{1/2}\,\left(1\,-\,\Lambda^{-2n-3} \right)^{1/2}}\,\Lambda^{-n/2}. \label{tn_NRG}
\end{align}
Note that the logarithmic discretization corrects the tunneling matrix element $V$.
$A_{\Lambda}$ represents such a correction and can be expressed as,
\begin{align}
A_{\Lambda}\,=\,\frac{1}{2}\left(\frac{1+1/\Lambda}{1-1/\Lambda} \right)\,\ln{\Lambda}.
\label{a_lam}
\end{align} 
This correction $A_{\Lambda}$ approaches $1$ in the continuum limit,
\begin{align}
\lim_{\Lambda\to1}A_{\Lambda}=1.
\label{A_lambda_infty}
\end{align}
The hopping element $t_{n}$ exponentially decays with increasing $n$ and 
its asymptotic form is 
\begin{align}
\,\frac{t_{n}}{D}\,\sim\,\frac{1+\Lambda^{-1}}{2}\Lambda^{-n/2}.
\label{t_n_infty}
\end{align}
To iteratively diagonalize the Hamiltonian $\mathcal{H}_{\mathrm{NRG}}$,
we introduce a Hamitonian $\mathcal{H}_{}^{L}$ which describe the chain ends at an $L$th site.
The explicit form of $\mathcal{H}_{}^{L}$ is
\begin{align}
&\mathcal{H}_{}^{L}\,=\Lambda^{(L-1)/2}\left[\mathcal{H}_{d}^{0}\,+\,\mathcal{H}_{U}^{}\,+\,\mathcal{H}_{\mathrm{chain}}^{L}\,+\,\mathcal{H}_{T}^{}  \right],\label{SIAM_U0_chain_L_truncate} \\
&\mathcal{H}_{\mathrm{chain}}^{L}\,=\,\sum_{n\,=\,0}^{L-1}\,\sum_{m=1}^{N}\,t_{n}^{}\left(f_{n,m}^{\dagger}f_{n+1,m}^{}\,+\,\mathrm{H.c.} \right).
\label{eq:Hc_NRG_L_truncate}
\end{align}
$\Lambda^{(L-1)/2}$ is multiplied to the right-hand side of Eq.\ \eqref{SIAM_U0_chain_L_truncate} to makes the hopping element $\Lambda^{(L-1)/2}\,t_{L-1}$ be order of 1. 
Eigenvalues of $\mathcal{H}_{}^{L}$ also become order of 1 due to the multiplication.
The full Hamiltonian $\mathcal{H}_{\mathrm{NRG}}$ is recovered as the limit of
\begin{align}
\mathcal{H}_{\mathrm{NRG}}^{}\,=\,\lim_{L\to\infty}\,\Lambda^{-(L-1)/2}\,\mathcal{H}_{}^{L}.
\label{relation_H_NRG_H_-1_L}
\end{align}
Here, the prefactor corresponds to the energy scale of 
low-lying excited states of  $\mathcal{H}_{}^{L}$: 
\begin{align}
\frac{T_{L}}{D}=\Lambda^{-(L-1)/2} \;.
\label{energy_scale_of_L}
\end{align}
To run the iteration, we deduce a recurrence relation for $\mathcal{H}^{L}$ from Eq.\ \eqref{SIAM_U0_chain_L_truncate}, 
\begin{align}
\mathcal{H}_{}^{L+1}\,=\,\Lambda^{1/2}\,\mathcal{H}_{}^{L}\,+\,\Lambda^{L/2}\,t_{L}\sum_
{m=1}^{N}\left(f_{L,m}^{\dagger} f_{L+1,m}^{}\,+\,\mathrm{H.c.} \right).
\label{recurrence_relation_H_L_H_L+1}
\end{align} 		
Using this relation, we can construct a matrix for $\mathcal{H}_{}^{L+1}$ from
eigenvalues and corresponding eigenstates for the Hamiltonian $\mathcal{H}_{}^{L}$.
The iterative procedure yields the eigenenergies $E_{L}(\omega)$ at each $L$th step.
we keep only the $N_{K}$ eigenstates with the lowest energies and discard the higher energy states
because a dimension of $\mathcal{H}_{}^{L}$ increases with increasing $L$. 

\begin{figure}[h]
\centering
\begin{minipage}{0.75\linewidth}
\begin{center}
\includegraphics[width=\linewidth]{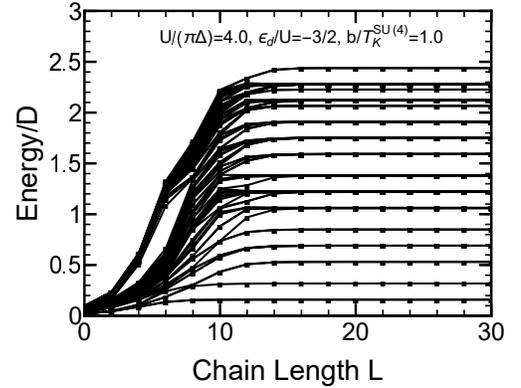}
\end{center}
\end{minipage}
\caption{
NRG energy flow of a CNT dot as functions of the even chain length $L$.
A CNT dot has $N=4$ internal degrees of freedom. 
We keep 4100 eigenvalues at each NRG step and plot the lowest 100 values in this figure.  
The logarithmic discretization parameter is $\Lambda=6$.   
Values of $U$, $\epsilon_{d}$, and magnetic field $b$ for the calculation are $U/(\pi\Delta)=4.0$, $\epsilon_{d}/U=-3/2$, and $b/T_{K}^{\mathrm{SU(4)}}=1.0$, respectively.
The impurity resonance width is $\Delta/D=1/(100\pi)$.
The SU(4) temperature determined at $b=0$ is $T_{K}^{\mathrm{SU(4)}}/\Delta = 0.2$. 
}
\label{NRG_Flow_of_energies}
\end{figure}

Figure \ref{NRG_Flow_of_energies} plots low-lying eigenenergies of a CNT dot as fucntions of even $L$ with typical parameters: $\Lambda=6.0$, $N_{K}=4100$, $U/(\pi\Delta)=4.0$, $\epsilon_{d}/U=-3/2$, and $b/T_{K}^{\mathrm{SU(4)}}=1.0$.  
Here, the bare resonance width is $\Delta /D=1/(100\pi)$.
The SU(4) temperature $T_{K}^{\mathrm{SU(4)}}/\Delta = 0.2$ determined at $b=0$ scales the magnetic field $b$. 
In the value of $b/T_{K}^{\mathrm{SU(4)}}=1.0$, the Fermi-liquid state of the dot is in the middle of the SU(4) to SU(2) Kondo crossover as shown in Fig.\ \ref{b-dependent-FL-U4}(a).
Specifically, Fig.\ \ref{b-dependent-FL-U4}(a) shows the magnetization $M_{d}$ is still not saturated to 1 at the value of $b$. 
 Figure \ref{NRG_Flow_of_energies} shows that the energies do not depend on $L$ in the region of $L\gtrsim 20$. 
At the last iteration $L=30$, the energy scale $T_{L}$ given in Eq.\ \eqref{energy_scale_of_L} 
is much smaller not only than $T_{K}^{\mathrm{SU(4)}}$ but also than the SU(2) Kondo temperature $T_{K}^{SU(2)}/\Delta=0.02$: $T_{L}/\Delta \approx 1.6\times 10^{-9}$, $T_{L}/T_{K}^{\mathrm{SU(4)}}\approx 8.2 \times 10^{-9}$, and $T_{L}/T_{K}^{SU(2)}\approx 1.0 \times 10^{-7}$. 
The value of $T_{K}^{SU(2)}$ is determined at $b\to\infty$.
Therefore, the low-lying energy states approach the Fermi-liquid states for $L$ greater than 20.  

\section{Calculation method of Fermi-liquid parameters}
\label{Calculation_method_of_Fermi-liquid_parameters}

\begin{figure}[h]
\centering
\begin{minipage}{0.7\linewidth}
\begin{center}
\includegraphics[width=\linewidth]{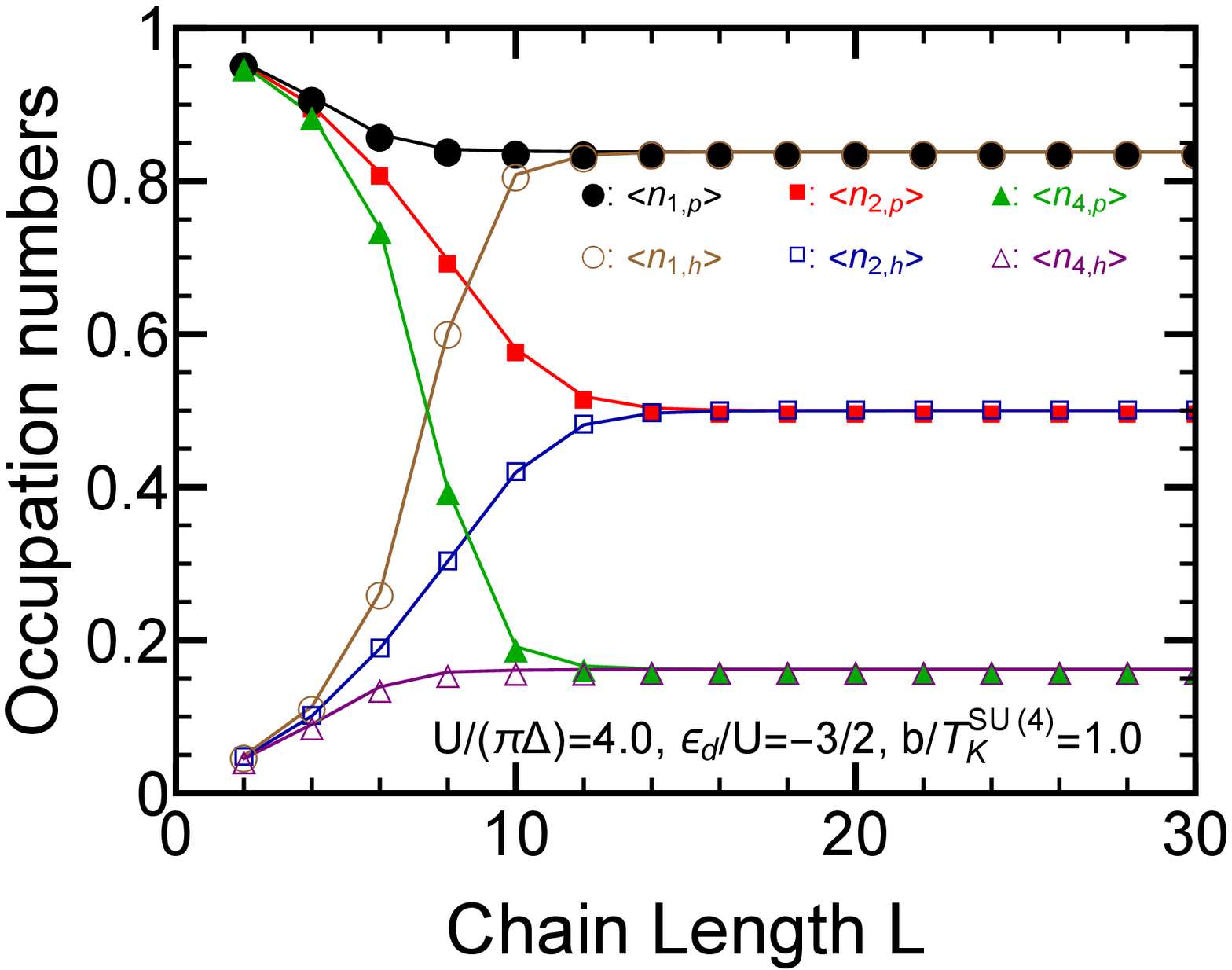}
\end{center}
\end{minipage}
\vspace{0.01\linewidth}

\begin{minipage}{0.7\linewidth}
\begin{center}
\includegraphics[width=\linewidth]{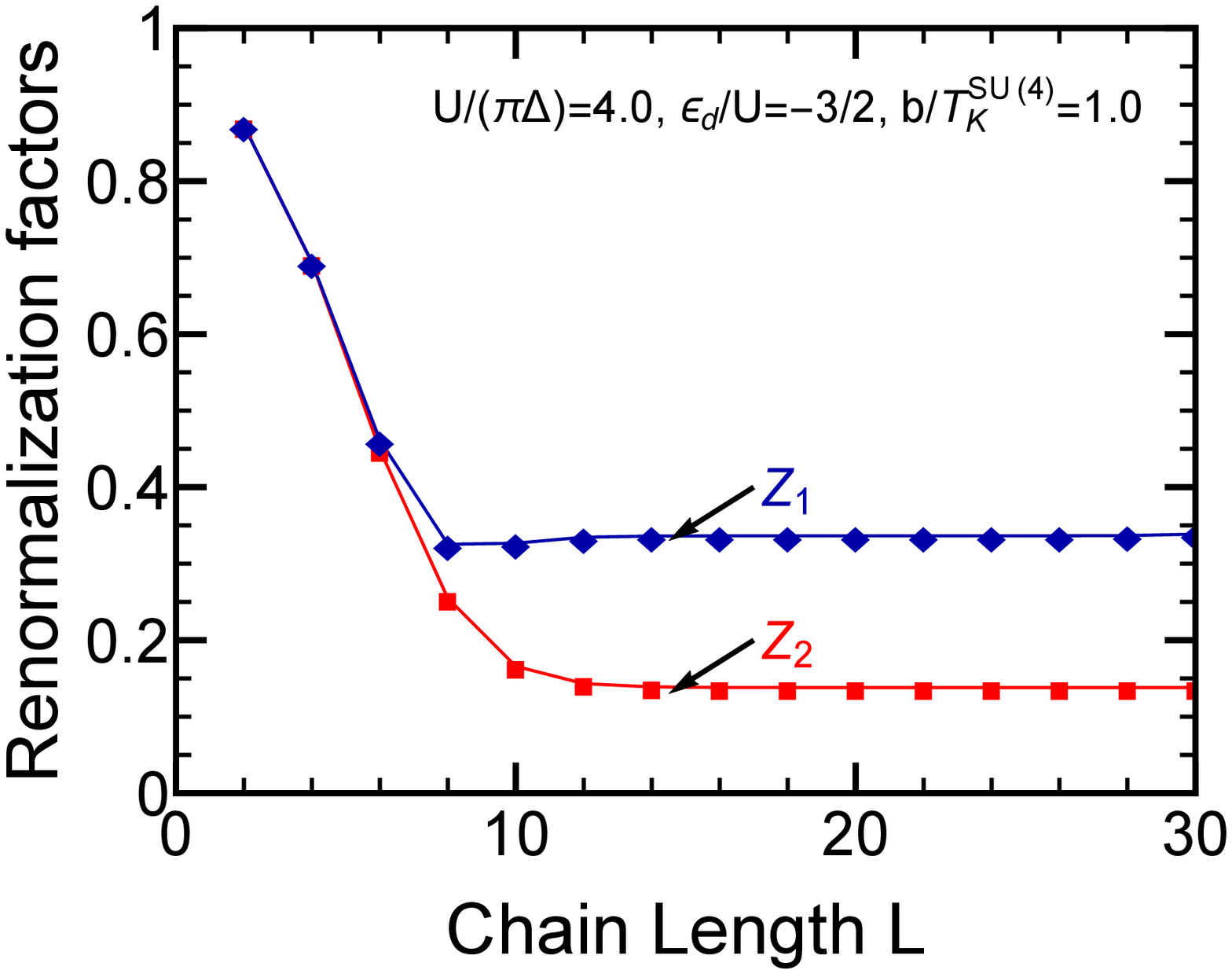}
\end{center}
\end{minipage}

\caption{
(a) shows flows of occupation numbers: $\langle n_{1,p} \rangle$, $\langle n_{1,h} \rangle$, $\langle n_{2,p} \rangle$, $\langle n_{2,h} \rangle$, 
$\langle n_{4,p} \rangle$, and $\langle n_{4,h} \rangle$. 
(b) shows those of renormalization factors: $Z_{1}$ and $Z_{2}$.
The x-axis is the NRG chain length $L$ in each figure.
In (a), Following six symbols, $\bullet$, $\circ$, $\blacksquare$, $\square$, $\blacktriangle$, and $\triangle$ respectively denote the values of $\langle n_{1,p} \rangle$, $\langle n_{1,h} \rangle$, 
$\langle n_{2,p} \rangle$, $\langle n_{2,h} \rangle$, $\langle n_{4,p} \rangle$, and $\langle n_{4,h} \rangle$. 
In (b), $\bullet$ and $\blacksquare$ respectively denote the values of $Z_{1}$ and $Z_{2}$. 
The parameters are $U/(\pi\Delta)=4.0$, $\epsilon_{d}/U=-3/2$, and $b/T_{K}^{\mathrm{SU(4)}}=1.0$ for each figure.  
These parameters are the same as those for Fig.\ \ref{NRG_Flow_of_energies}.
The SU(4) Kondo temperature is $T_{K}^{\mathrm{SU(4)}}/\Delta=0.2$ for the values of $U$ and $\epsilon_{d}$.
Eq.\ \eqref{eq:MatchingCondition} describes the magnetic filed $b$ dependence of the four dot levels $\epsilon_{1}$, $\epsilon_{2}$, $\epsilon_{3}$, and $\epsilon_{4}$.
}
\label{NRG_Flow_of_FL_parameters}
\end{figure}

\begin{figure}[h]
\centering
\begin{minipage}{0.75\linewidth}
\begin{center}
\includegraphics[width=\linewidth]{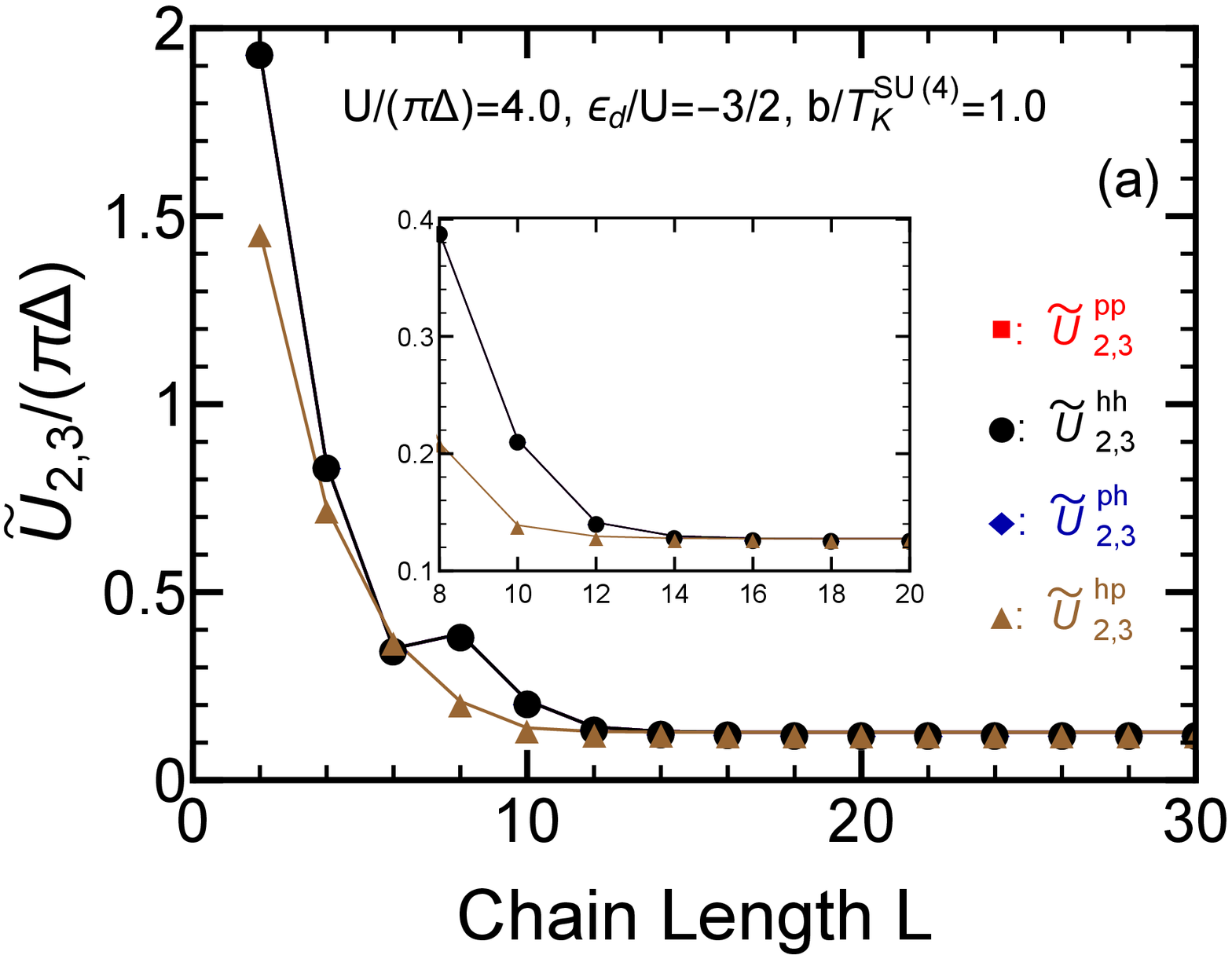}
\end{center}
\end{minipage}
\vspace{0.01\linewidth}

\begin{minipage}{0.75\linewidth}
\begin{center}
\includegraphics[width=\linewidth]{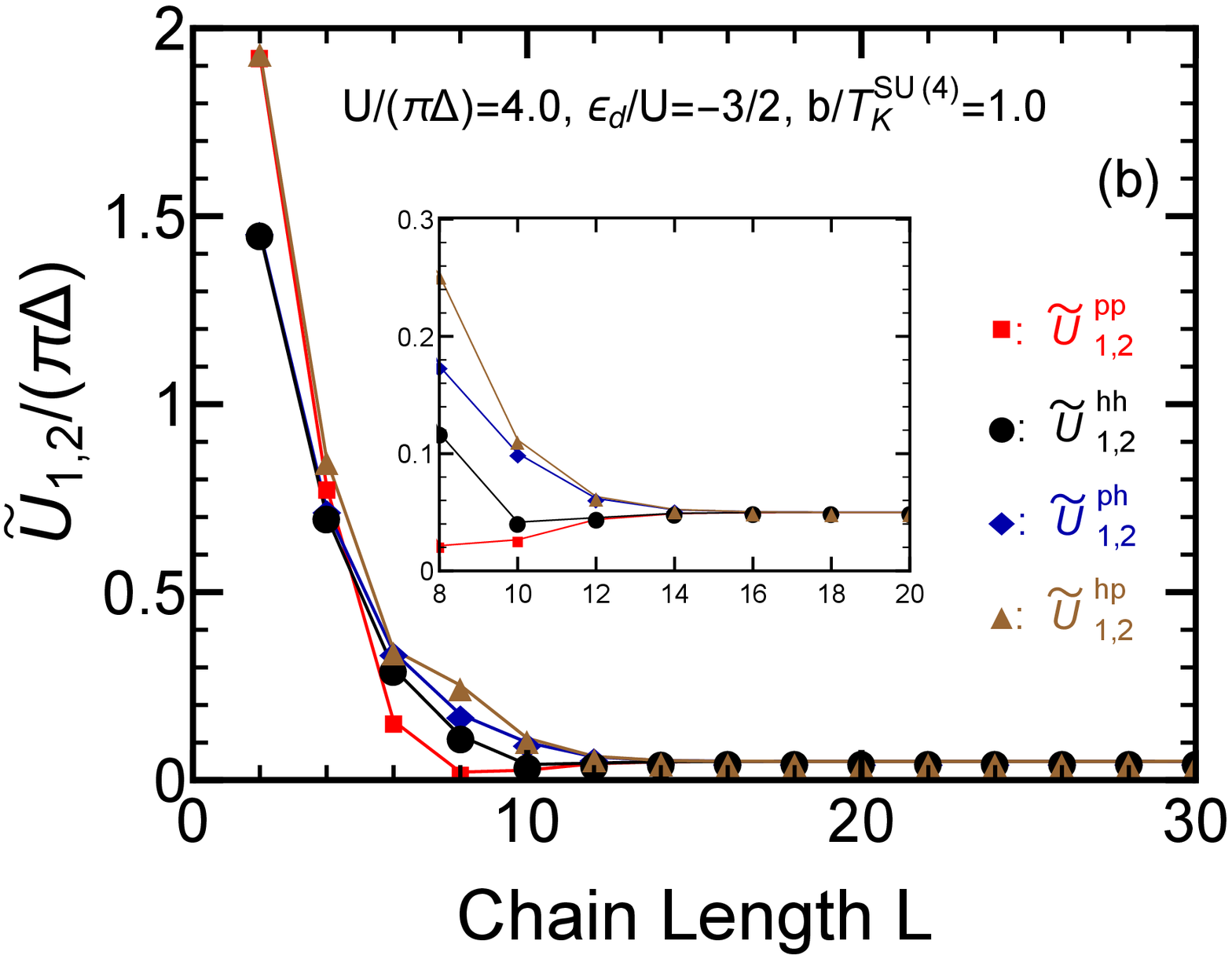}
\end{center}
\end{minipage}
\vspace{0.01\linewidth}

\begin{minipage}{0.75\linewidth}
\begin{center}
\includegraphics[width=\linewidth]{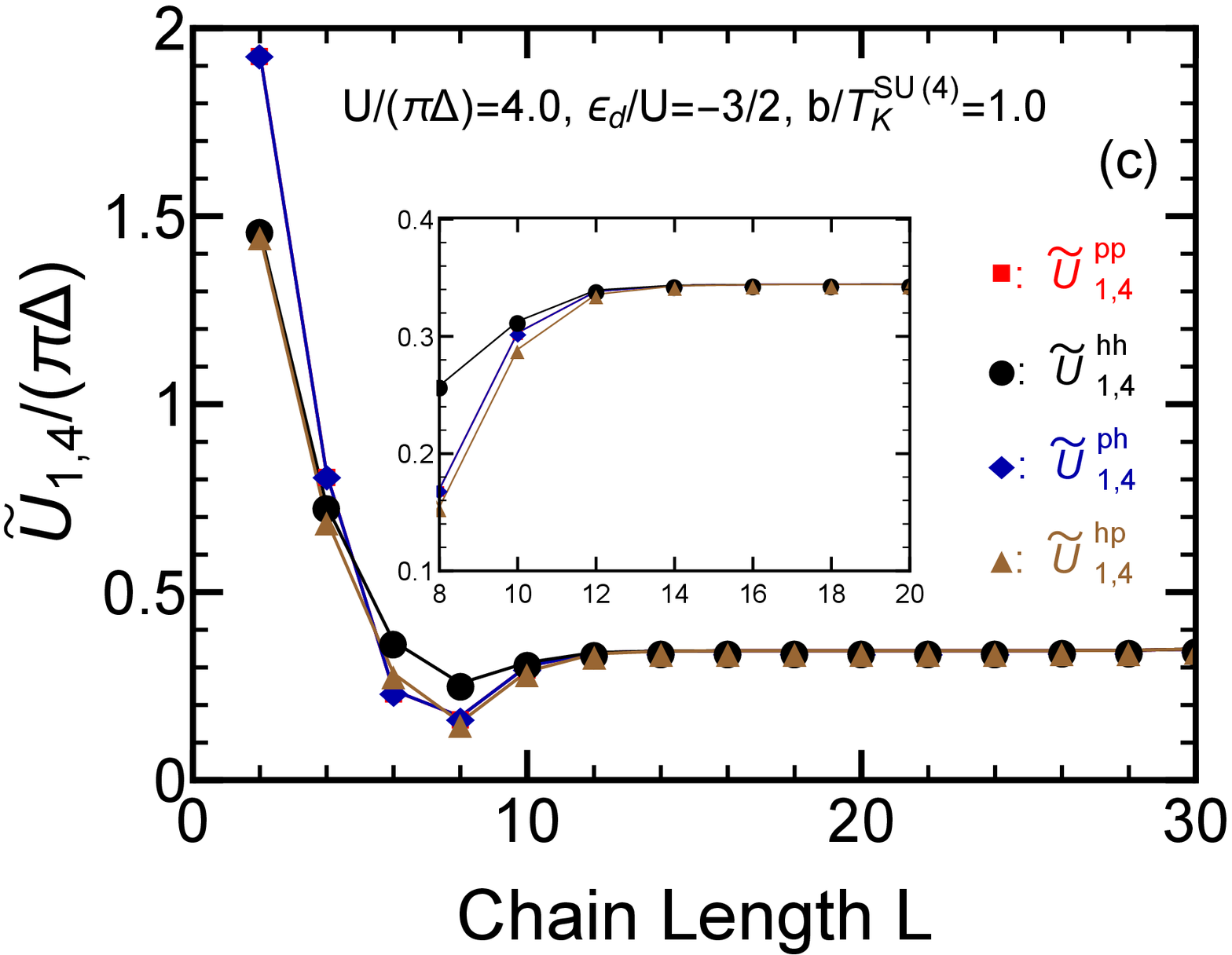}
\end{center}
\end{minipage}
\caption{Residual interactions $\widetilde{U}_{m,m^\prime}$ are plotted as functions of the chain length $L$.
(a), (b), and (c) respectively show $(m,m^\prime)=(2,3)$, $(1,2)$, and $(1,4)$ element of $\widetilde{U}_{m,m^\prime}$.
The particle-hole symmetry given by Eq.\ \eqref{extended_particle_hole_symmetry} imposes following relations to the residual interactions: $\widetilde{U}_{1,2}=\widetilde{U}_{1,3} = \widetilde{U}_{2,4}=\widetilde{U}_{3,4}$. 
The parameters are $U/(\pi\Delta)=4.0$, $\epsilon_{d}/U=-3/2$, and $b/T_{K}^{\mathrm{SU(4)}}=1.0$ for every three figure.
The SU(4) Kondo temperature is $T_{K}^{\mathrm{SU(4)}}/\Delta=0.2$ for the values of $U$ and $\epsilon_{d}$.
Following four symbols: $\blacksquare$, $\bullet$, $\blacklozenge$, and $\blacktriangle$
respectively represent $\widetilde{U}_{m,m^\prime}^{pp}$, $\widetilde{U}_{m,m^\prime}^{hh}$, $\widetilde{U}_{m,m^\prime}^{ph}$, and 
$\widetilde{U}_{m,m^\prime}^{hp}$ in each figure.  
The insets show an enlarged view in the range of $8\le L \le 20$. 
}
\label{NRG_Flow_of_residual_interactions}
\end{figure}

\begin{figure}[h]
\centering
\begin{minipage}{0.75\linewidth}
\begin{center}
\includegraphics[width=\linewidth]{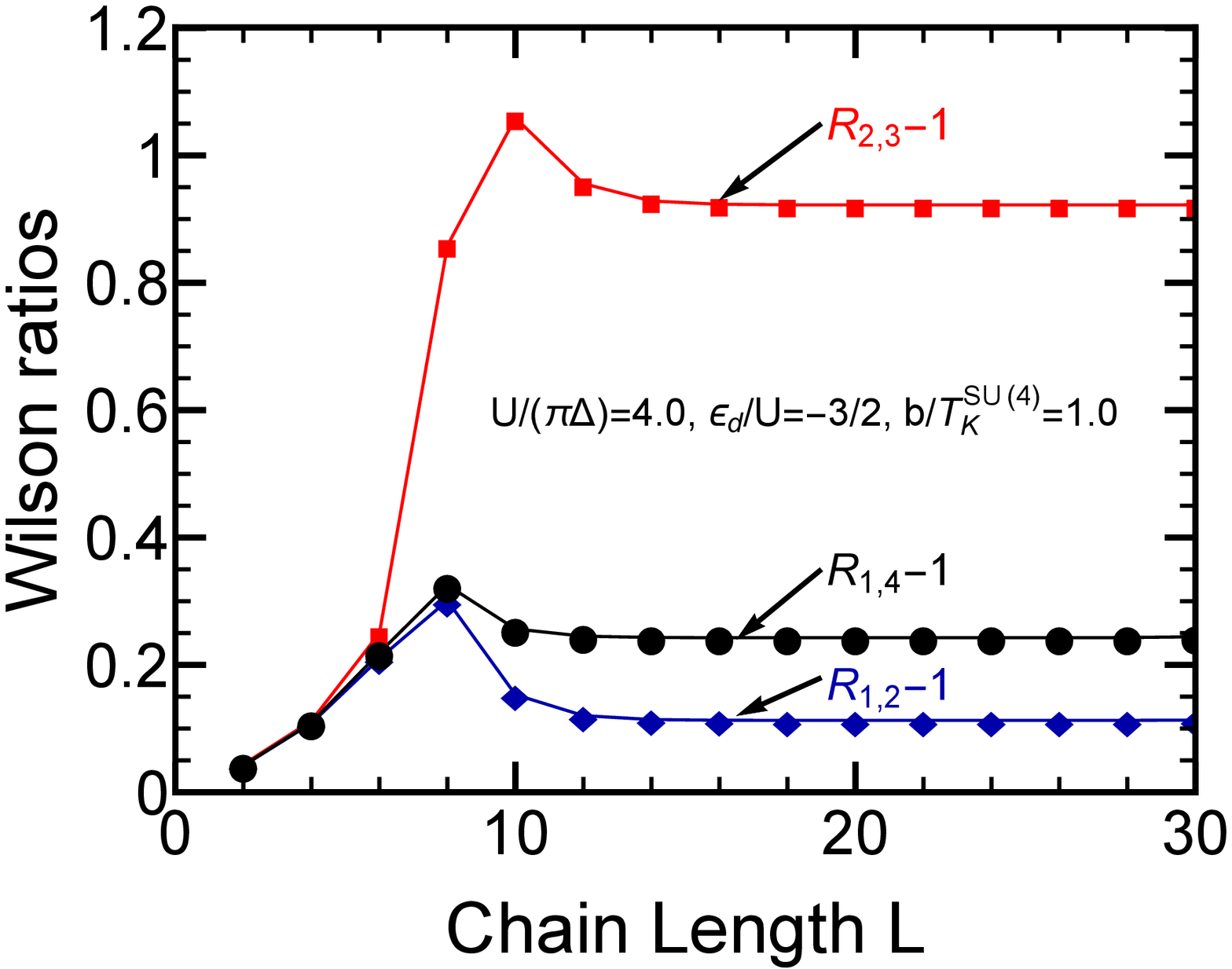}
\end{center}
\end{minipage}
\caption{Wilson ratios $R_{2,3}-1$, $R_{1,4}-1$, and $R_{1,2}-1$ are plotted
as functions of the chain length $L$ for the parameters: $U/(\pi\Delta)=4.0$, $\epsilon_{d}/U=-3/2$, and $b/T_{K}^{\mathrm{SU(4)}}=1.0$.
Each three symbol, i.e., $\blacksquare$, $\bullet$, and $\blacklozenge$ represents values of 
$R_{2,3}-1$, $R_{1,4}-1$, and $R_{1,2}-1$, respectively.
The SU(4) Kondo temperature is $T_{K}^{\mathrm{SU(4)}}/\Delta=0.2$ for the values of $U$ and $\epsilon_{d}$.
}
\label{NRG_Flow_of_Wilson_Ratios}
\end{figure}

We show how to calculate Fermi-liquid parameters such as phase shifts $\delta_{m}$,
renormalization factors $Z_{m}$, and residual interactions $\widetilde{U}_{m,m^{\prime}}$.
Let us consider a non-interacting case of the NRG Hamiltonian, i.e., $\mathcal{H}_{\mathrm{NRG}}$ at $U=0$
since a single particle and hole excitation from the ground state determine the phase shifts and the renormalization factors.
In such a case, a matrix form of the Hamiltonian $\mathcal{H}_{}^{L}$ is  
{\small
\begin{align}
\mathcal{H}_{U=0}^{L}=\Lambda^{(L-1)/2}
\begin{bmatrix} 
\epsilon_{m}\, & V\, 
& & & & \mbox{\Large \bf 0} \cr
V\, &0 & t_0\, & & &  \cr
0& t_0\, &0 
& t_1\,  &  &  \cr
& & \!\!\!\!\!\!\!\!\!\!\!\!\!\! \ddots 
  & \!\!\!\!\!\!\!\!\!\!\!\!\!\! \ddots 
  & \!\!\!\!\!\!\!\!\!\!\!\!\!\! \ddots& \cr
&  & & t_{L-2} &0& t_{L-1} 
\rule{1.0cm}{0cm} \cr
 \mbox{\Large \bf 0} &  & & &  t_{L-1} & 0  \cr
\end{bmatrix}.
\label{SIAM_U0_chain_matrix_form}
\end{align}
}
The Green's function $G_{d}(\omega)$ at the impurity site $(i=-1)$  is the $(1,1)$ 
element of the resolvent of this Hamiltonian matrix,  
\begin{align}
G_{m}(\omega) \,&\equiv\,
\left(\omega\,\bold{1}\,-\,\mathcal{H}_{U=0}^{L} \right)^{-1}_{1,1} \notag \\ 
&=\,\frac{1}{\omega - \epsilon_m\,\Lambda^{(L-1)/2} 
- V^2 \, \Lambda^{(L-1)} \, g_{00,m}(\omega)}. 
\label{relations_Gm1m1_g00}
\end{align}
An eigenvalue $E$ of $\mathcal{H}_{}^{L}$ is a pole of $G_{d}$.
Specifically, the value is a solution of the below equation,
\begin{align}
E - \epsilon_m\,\Lambda^{(L-1)/2} 
- V^2 \, \Lambda^{(L-1)} \, g_{00,m}(E)\,=\,0.
\label{equation_eigenvalue_H_-1_L}
\end{align} 
In Eqs.\ \eqref{relations_Gm1m1_g00} and \eqref{equation_eigenvalue_H_-1_L}, 
$g_{00,m}(\omega)$ is the Green's function at a site $i=0$ for a chain 
which starts from the zeroth site and ends at the $L$th site.        
A Hamiltonian $\mathcal{H}_{0}^{L}$ describes such a chain.
$g_{00,m}(\omega)$ is the $(1,1)$ element of the resolvent of $\mathcal{H}_{0}^{L}$,
\begin{align}
g_{00,m}(\omega)\,&\equiv\,\left(\omega\,\bold{1}\,-\,\mathcal{H}_{0}^{L} \right)^{-1}_{1,1}\;. 
\label{def_g_00}
\end{align}
A ratio of the impurity level position $\epsilon_{m}$ to its width $\Delta$ is deduced from Eq.\ \eqref{equation_eigenvalue_H_-1_L}, 
\begin{align}
\frac{\epsilon_{m}}{\Delta}\,=\,\frac{E}{\Delta}\,\Lambda^{\frac{-(L-1)}{2}}-\frac{2\,A_{\Lambda}}{\pi}\,\lim_{L\to\infty}\,\Lambda^{\frac{L\,-\,1}{2}}\,D g_{00,m}(E).
\label{ratio_epsilon_d_to_Delta}
\end{align}
We here use the relation Eq.\ \eqref{V_NRG} between the hybridization $V$ and $\Delta$.

We henceforth discuss interacting cases, i.e., cases of $U\neq 0$,  by extending the Eq.\ \eqref{ratio_epsilon_d_to_Delta} to such cases.
Let $E_{p,m}(L)$ denote an excitation energy on adding an electron to the ground state.
Similarly, we define $E_{h,m}(L)$ as an excitation energy on adding a hole to the ground state.
The NRG iteration procedure calculates $E_{p,m}(L)$ and $E_{h,m}(L)$ as a function of the chain length $L$.
We note that $E_{p,m}(L)>0$ and $E_{h,m}(L)<0$. Here we set the energy of the ground state as $0$. 
Replacing the non-interacting eigenvalue $E$ in Eq.\ \eqref{ratio_epsilon_d_to_Delta}
to the corresponding $E_{p,m}(L)$ or $E_{h,m}(L)$ yields ratios of renormalized level position $\widetilde{\epsilon}_{m}$ to
its width $\widetilde{\Delta}_{m}$,  
\begin{align}
&\frac{\widetilde{\epsilon}_{m,p}}{\widetilde{\Delta}_{m,p}}=\frac{E_{p,m}(L)}{\widetilde{\Delta}_{m,p}}\,\Lambda^{\frac{-(L-1)}{2}}-\frac{2\,A_{\Lambda}}{\pi}\,\Lambda^{\frac{L-1}{2}}\,D g_{00,m}(E_{p,m}(L)), \label{E_particle_NRG_ratio_epsilon_d_to_Delta} \\
&\frac{\widetilde{\epsilon}_{m,h}}{\widetilde{\Delta}_{m,h}}=\frac{E_{h,m}(L)}{\widetilde{\Delta}_{m,h}}\,\Lambda^{\frac{-(L-1)}{2}}-\frac{2\,A_{\Lambda}}{\pi}\,\Lambda^{\frac{L-1}{2}}\,D g_{00,m}(E_{h,m}(L)).
\label{E_hole_NRG_ratio_epsilon_d_to_Delta}
\end{align} 
Here, $\widetilde{\epsilon}_{m,p}$ and $\widetilde{\epsilon}_{m,h}$ are the renormalized resonance level positions calculated from 
$E_{p,m}$ and $E_{h,m}$, respectively.
Similarly, 
$\widetilde{\Delta}_{m,p}$ and $\widetilde{\Delta}_{m,h}$ are the renormalized line widths from 
$E_{p,m}$ and $E_{h,m}$.
The first terms in the right hand of the equations can be neglected in the limit of $L\to \infty$ 
because $E_{p,m}(L)$ and $E_{h,m}(L)$ are of order 1.
Thus, the ratios become 
\begin{align}
\frac{\widetilde{\epsilon}_{p(h),m}}{\widetilde{\Delta}_{p(h),m}}\,=\,-\frac{2\,A_{\Lambda}}{\pi}\,\lim_{L\to\infty}\,\Lambda^{\frac{L\,-\,1}{2}}\,D g_{00,m}\left(E_{p(h),m}\left(L\right)\right).
\label{ratio_epsilon_m_to_Delta_m}
\end{align} 
We obtain occupation numbers $\langle n_{d m,p}\rangle$ and $\langle n_{d m,h}\rangle$ by substituting the ratios into the Friedel sum rule given by,
\begin{align}
\langle n_{m,p(h)}\rangle \,=\,\frac{1}{2}\,-\,\tan^{-1}\left(\frac{\widetilde{\epsilon}_{p(h),m}}{\widetilde{\Delta}_{p(h),m}}\right).
\label{Friedel_sum_rule_in_appendix}
\end{align} 
Furthermore, subtracting Eq.\ \eqref{E_hole_NRG_ratio_epsilon_d_to_Delta} from Eq.\ \eqref{E_particle_NRG_ratio_epsilon_d_to_Delta} gives 
the renormalized level width $\widetilde{\Delta}_{m}$, 
\begin{align}
 \widetilde{\Delta}_{m}\!=\!\frac{\pi}{2A_{\Lambda}}\!
\frac{
E_{p,m}(L)\!-\!E_{h,m}(L)
}{
\Lambda^{(L-1)} D g_{00,m}(E_{p,m}(L)) \!
-\!\Lambda^{(L-1)} D g_{00,m}(E_{h,m}(L)) 
}.
\label{level_width_NRG_procedure}
\end{align}

Figure \ref{NRG_Flow_of_FL_parameters}(a) plots the occupation numbers of the CNT dot
as functions of the NRG chain length $L$.  
The parameters are $U/(\pi\Delta)=4.0$, $\epsilon_{d}/U=-3/2$, and $b/T_{K}^{\mathrm{SU(4)}} = 1.0$ for 
the plots.
The SU(4) Kondo temperature is $T_{K}^{\mathrm{SU(4)}}/\Delta=0.2$ for the values of $U$ and $\epsilon_{d}$.
Figure \ref{NRG_Flow_of_FL_parameters}(a) clearly shows that $\langle n_{1,p} \rangle$ from the particle excitation and 
$\langle n_{1,h} \rangle$ from the hole excitation converge to a same value $\langle n_{1} \rangle = 0.84$ with increasing $L$.
$\langle n_{2,p} \rangle$ and $\langle n_{2,h} \rangle$ also converge to the half-filled value $\langle n_{2} \rangle = 0.5$. 
Since the extended electron-hole symmetry described by Eq.\ \eqref{extended_particle_hole_symmetry} imposes
relations between the occupation numbers: $\langle n_{4,p} \rangle = 1-\langle n_{1,h} \rangle$ and $\langle n_{4,h} \rangle =1-\langle n_{1,p} \rangle$,
$\langle n_{4,p} \rangle$ and $\langle n_{4,h} \rangle$ converge to a value $1-\langle n_{1} \rangle = 0.16$.
The values of $\langle n_{d,1} \rangle$ and $\langle n_{d,4} \rangle$ give a value of the magnetization $M_{d}=\langle n_{d,1} \rangle - \langle n_{d,4} \rangle = 0.67$.
Figure \ref{b-dependent-FL-U4}(a) shows this value of $M_{d}$ at the point of $b/T_{K}^{\mathrm{SU(4)}} = 1.0$.

Figure \ref{NRG_Flow_of_FL_parameters}(b) plots the chain length $L$ dependence of renormalization factors $Z_{m}$, which are ratios of $\widetilde{\Delta}_{m}$ to $\Delta$, i.e., $Z_{m}=\widetilde{\Delta}_{m}/\Delta$.
We note that the electron-hole symmetry also makes two of the four factors independent: $Z_{1}=Z_{4}$ and $Z_{2}=Z_{3}$ 
This figure shows that values of $Z_{1}$ and $Z_{2}$ respectively converge to $Z_{1}=0.34$ and $Z_{2}=0.14$.
Since the value of $Z_{2}$ is about ten times as large as the SU(2) limit value $0.026$,
the Fermi-liquid state at $b/T_{K}^{\mathrm{SU(4)}}=1.0$ is in the middle of the crossover between the SU(4) and SU(2) Kondo state.

The NRG iteration also yields higher excitation energies from the ground state. 
$E_{k,p,m}$ and $E_{k,h,m}$ respectively represent such electron and hole excitation energies.
We note that $E_{k,p,m}>0$ and $E_{k,h,m}<0$.   
They describe a free quasiparticle Hamiltonian, 
\begin{align}
\mathcal{H}^{0}\!=\!\sum_{m=1}^{N}\sum_{k}\left( E_{k,p,m}^{}\,p_{k,m}^{\dagger}p_{k,m}^{}-E_{k,h,m}^{}\,h_{k,m}^{\dagger}h_{k,m}^{}\right).
\label{free_quasi_hamiltonian}
\end{align}
Here, operators $p_{k,m}^\dagger$ and $h_{k,m}^{\dagger}$ respectively create the quasiparticle with $E_{k,p,m}^{}$ and the quasihole with $E_{k,h,m}^{}$.
Correspondingly, $p_{k,m}^{}$ and $h_{k,m}^{}$ annihilate the quasiparticle and quasihole.
The ground state $|0\rangle$ is defined such that $p_{k,m}^{} |0\rangle = 0$ and $h_{k,m}^{} |0\rangle = 0$.
A linear combination of $p_{k,m}^\dagger$ and $h_{k,m}^{}$ expresses the operator $d_{m}^{\dagger}$ which creates the impurity electron, 
\begin{align}
d_{m}^{\dagger}\,=\,\sum_{k}\,\left( p_{k,m}^{\dagger}\phi_{k,p,m}^{}(-1)\,+\,h_{k,m}^{}\phi_{k,h,m}^{}(-1) \right)\;.
\label{linear_combination_p_and_h}
\end{align} 

To calculate Wilson ratios $R_{m,m^{\prime}}$ defined in Eq.\ \eqref{def_Wilson_ratio}, 
we consider a residual interaction term,
\begin{align}
\mathcal{H}_{L}^{U}\,=\,\Lambda^{(L-1)/2}\,\sum_{m < m^\prime}\widetilde{U}_{m,m^\prime} 
 \, \mathrm{N}\left[\, 
n_{dm}\,n_{dm^\prime} 
 \,\right]. 
\label{residual_interaction_terms}
\end{align} 
Here, $\widetilde{U}_{m,m^\prime}$ is the strength of the residual interaction between the
quasiparticles. $\mathrm{N}$ is the normal ordering operator.
We examine how the interaction affects the low-lying many-particle states 
to a first order perturbation in $\widetilde{U}_{m,m^\prime}$. 
This order calculation accurately describes the effects of the interaction on the low-energy states
because the term vanishes in the limit of $L\to\infty$.
For the explicit calculation, we introduce a two-particle excited state from the ground state:
\begin{align}
|A\rangle\,=\,p_{1,m}^\dagger p_{1,m^{\prime}}^\dagger\,|0\rangle,
\label{two_particle_state}
\end{align} 
where $m\neq m^\prime$. 
The NRG iteration calculates a corresponding eigenvalue $E_{pp}(L)$.
The first order calculation of $\widetilde{U}_{m,m^{\prime}}$ also gives $E_{pp}$,
\begin{align}
E_{pp}(L)\,&=\,\langle A | \mathcal{H}^{0}\,+\,\mathcal{H}_{L}^{U} |A \rangle \notag \\
&=\,2E_{p,m}^{}(L)\,+\,\widetilde{U}_{m,m^{\prime}}\Lambda^{-\frac{L-1}{2}}|\beta_{1,p,m}|^{2}|\beta_{1,p,m^{\prime}}|^{2}.
\label{Residual_interaction_PP}
\end{align} 
In this equation, $|\beta_{1,p,m}|^{2}$ is given by 
\begin{align}
\left|\beta_{1,p,m}\,\right|^2 
\,=\, 
\frac{\Lambda^{(L-1)/2}} 
{1\,- \, 
\widetilde{V}_{m}^2\,\Lambda^{(L-1)}\, 
\left. {
  \displaystyle {\mathstrut dg_{00,m}(E)} \over \displaystyle {\mathstrut dE} }
\right|_{E=E_{1,p,m}}},
\end{align}
where $\widetilde{V}_{m}^{2}=2D\widetilde{\Delta}_{m}A_{\Lambda}/\pi$.
The $L$ dependent $\widetilde{U}_{m,m^{\prime}}(L)$ is deducible from Eq.\ \eqref{Residual_interaction_PP}
since the NRG iteration procedure determines $E_{pp}(L)$ in the left hand of the equation.
We can alternatively consider two hole excitations to calculate $\widetilde{U}_{m,m^\prime}$.
Considering particle-hole excitations and hole-particle excitations also yields $\widetilde{U}_{m,m^\prime}$. 
Here, the word ``particle-hole excitations'' denote one particle excitation with $m$ flavour and one hole excitation with $m^{\prime}$ flavour.
The word ``hole-particle excitations'' similarly denote one hole and one particle excitation with $m$ and $m^{\prime}$ .
Thus, $\widetilde{U}_{m,m^\prime}$ is calculable by four different ways. 
$\widetilde{U}_{m,m^\prime}^{pp}(L)$, $\widetilde{U}_{m,m^\prime}^{hh}(L)$, $\widetilde{U}_{m,m^\prime}^{ph}(L)$, and 
$\widetilde{U}_{m,m^\prime}^{hp}(L)$ 
denote the residual interaction deduced from the two-particle, two-hole, particle-hole and hole-particle excitation energies, respectively.

Figures\ \ref{NRG_Flow_of_residual_interactions}(a)-(c) respectively plot $L$ dependence of $\widetilde{U}_{2,3}$, $\widetilde{U}_{1,2}$, and $\widetilde{U}_{1,4}$
which are calculated by the four ways. 
The three figures show that $\widetilde{U}_{m,m^\prime}^{pp}$, $\widetilde{U}_{m,m^\prime}^{hh}$, $\widetilde{U}_{m,m^\prime}^{ph}$, and 
$\widetilde{U}_{m,m^\prime}^{hp}$ converge a same value $\widetilde{U}_{m,m^\prime}^{}$ for each $(m,m^\prime)$.  
Each inset clearly shows such convergence of the four values. 
The converged values are $\widetilde{U}_{2,3}^{}/(\pi\Delta) = 0.13$, $\widetilde{U}_{1,2}^{}/(\pi\Delta) = 0.05$, and $\widetilde{U}_{1,4}^{}/(\pi\Delta) = 0.35$. 

The Wilson ratios $R_{m,m^{\prime}}$ are deducible from the obtained values of 
$\delta_{m}=\langle n_{dm} \rangle/\pi$, $Z_{m}$, and  $\widetilde{U}_{m,m^\prime}$ using Eqs.\ \eqref{R_23_U_tilde_23}-\eqref{R_12_U_tilde_12}.
Figure \ref{NRG_Flow_of_Wilson_Ratios} plots $R_{2,3}-1$, $R_{1,4}-1$, and $R_{1,2}-1$ as functions of $L$.
The four values of $\widetilde{U}_{m,m^\prime}^{pp}$, $\widetilde{U}_{m,m^\prime}^{hh}$, $\widetilde{U}_{m,m^\prime}^{ph}$, and 
$\widetilde{U}_{m,m^\prime}^{hp}$ are averaged to calculate the ratios $R_{m,m^\prime}-1$.
Each of the three ratios converges for $L\gtrsim 20$.
The converged values are $R_{2,3}-1=0.92$, $R_{1,2}-1=0.11$, and $R_{1,4}-1=0.24$. 
$R_{2,3}$ becomes larger than the other two ratios because
the corresponding impurity levels of $R_{2,3}$ remain the Fermi level.

\section{Fermi-liquid parameters for single orbital Anderson impurity}
\label{Fermi-liquid parameters for single orbital Anderson impurity}
\begin{figure}[h]
\centering
\begin{minipage}{0.8\linewidth}
\begin{center}
\includegraphics[width=\linewidth]{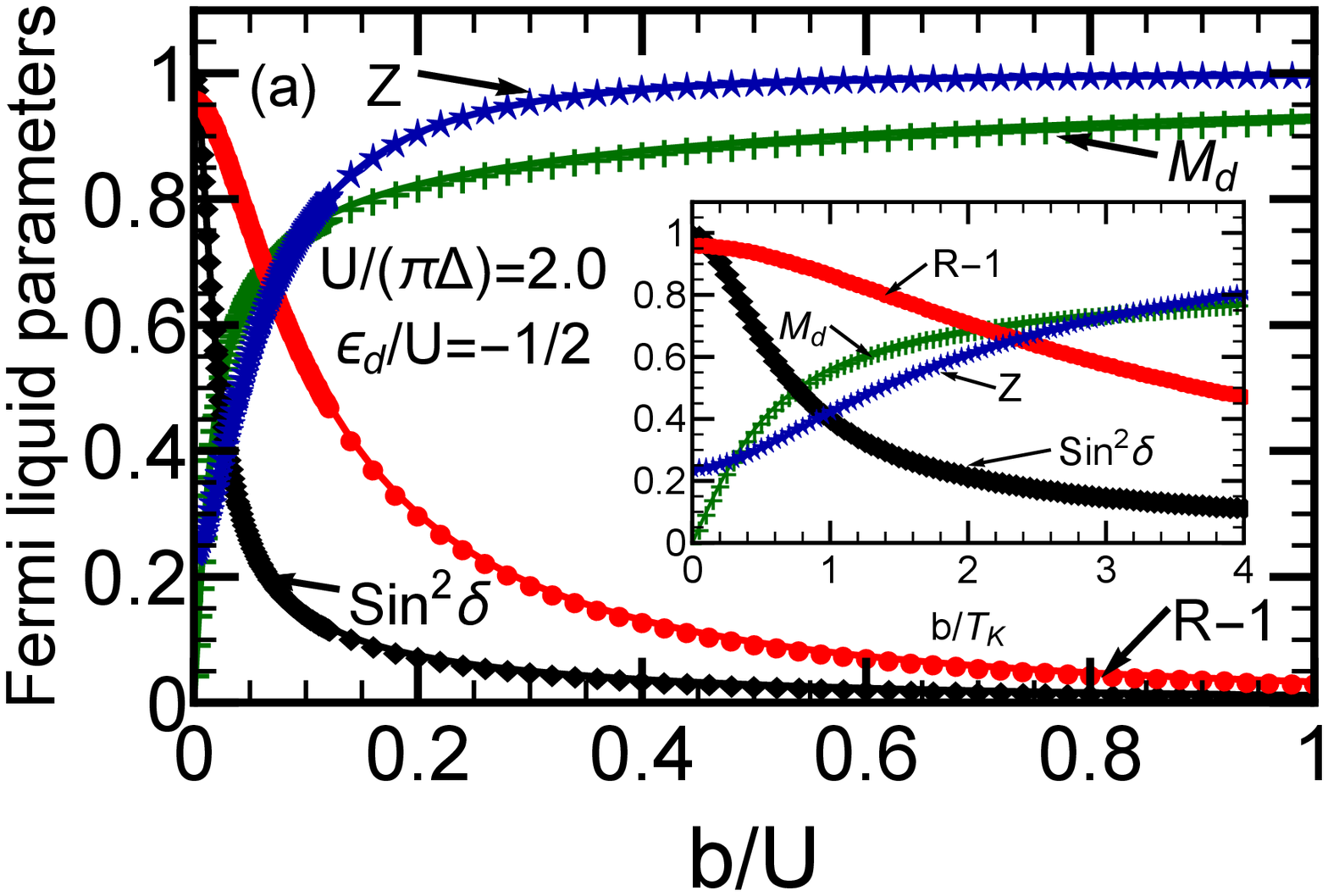}
\end{center}
\end{minipage}
\vspace{0.01\linewidth}

\begin{minipage}{0.8\linewidth}
\begin{center}
\includegraphics[width=\linewidth]{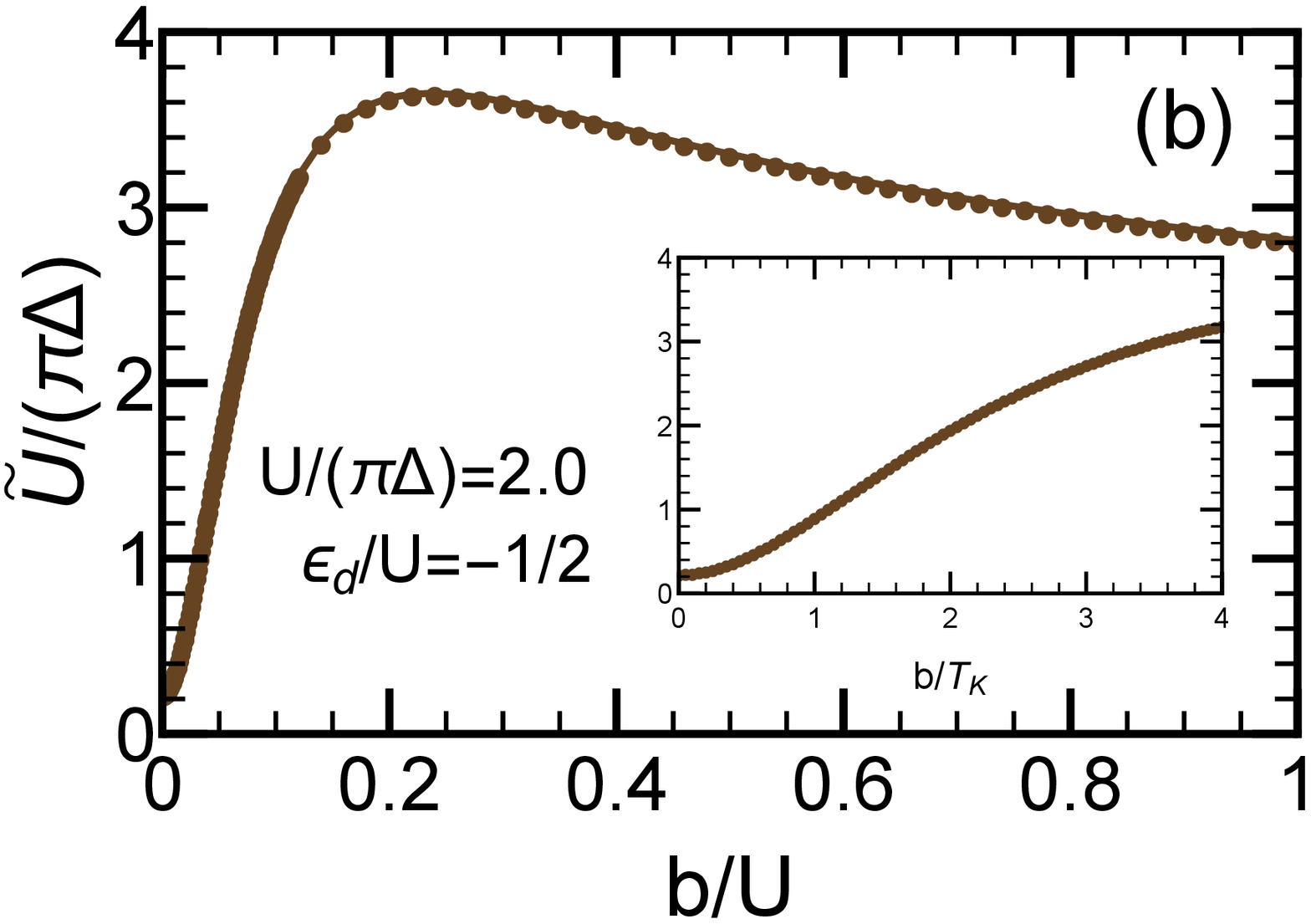}
\end{center}
\end{minipage}

\caption{(a) Magnetic field dependence of Fermi liquid parameters for single orbital Anderson impurity
at the electron-hole symmetric point $\varepsilon_{d}/U=-1/2$ for $U/(\pi\Delta)=2.0$.
At this point, a total occupation number is locked at one, i.e.,
$\langle n_{d\downarrow} \rangle +\langle n_{d\uparrow}\rangle=1$ at arbitrary magnetic fields, 
and thus $\langle n_{d\sigma} \rangle$ can be expressed in terms of the magnetization $M_{d}\equiv \langle n_{d\uparrow}\rangle - \langle n_{d\downarrow}\rangle$ 
as follows: $\langle n_{d\sigma}\rangle =(1+\mathrm{sgn}(\sigma) M_{d})/2$.
Furthermore, $\sin^{2}\delta_{\uparrow}=\sin^{2}\delta_{\downarrow}$ and $Z_{\uparrow}=Z_{\downarrow}$. 
(b) Residual interaction $\widetilde{U}$ as a function of $b$.
The x-axis is scaled by the Coulomb interaction $U$.
Each inset shows an enlarged view of the region around $b=0$.
In the insets, the axis is scaled by the Kondo temperature $T_{K}^{\mathrm{SU(2)}}=0.19\Delta=(0.03 U)$.}
\label{residual_interaction_1ch}
\end{figure}
We briefly discuss how the Fermi liquid state of the single Anderson impurity evolves with increasing magnetic field.
The field dependence of the Fermi liquid parameters have been discussed also in Refs.\ \onlinecite{Hewson_1ch_mag_Anderson_part_I, Hewson_1ch_mag_Anderson_part_II}.
Figures \ref{residual_interaction_1ch}(a) and \ref{residual_interaction_1ch}(b) plot NRG results of 
the Fermi liquid parameters and the residual interaction, respectively. In the NRG calculations,
the spin-dependent impurity level is chosen such that $\varepsilon_{d,\sigma}=\varepsilon_{d}+\mathrm{sgn}(\sigma)b$, 
and the center of the impurity level is locked at the half-filling, $\varepsilon_{d}=-U/2$.
Figure \ref{residual_interaction_1ch}(a) shows that
the transmission probability $T(0)=\sin^{2}\delta$ decreases as soon as magnetic fields become comparable with the 
Kondo temperature $T_{K}^{\mathrm{SU(2)}}$, and correspondingly, 
the induced magnetization is rapidly saturated to $1$, i.e., $M_{d}\to 1$.
In contrast, $Z$ and $R-1$ vary more slowly than $T(0)$ and $M_{d}$ with the scales of $U$. 
The inset of Fig.\ref{residual_interaction_1ch}(a) shows that 
$\widetilde{\Delta}$ and $R-1$ are still renormalized for small magnetic fields $b\lesssim T_{K}^{\mathrm{SU(2)}}$.
For large magnetic fields $b\gg T_{K}^{\mathrm{SU(2)}}$, these parameters approach the non-interacting values, $Z \to 1$ and $R-1 \to 1$.

The residual interaction plotted in Fig. \ref{residual_interaction_1ch}(b) also varies from a zero field value $4T_{K}^{\mathrm{SU(2)}}=0.23\pi\Delta$ with increasing $b$.
For small magnetic fields $b\lesssim 0.2 U$, $\widetilde{U}$ is enhanced and its value becomes larger than the bare Coulomb interaction $U$.
As magnetic fields further increase, it decreases from the enhanced value to the bare value $U$.   

\section{NRG  for dynamical correlations}

\label{sec:NRG_calculations}

\subsection{Spectral function and self-energy}

In this work, the spectral function $A_{m}^{}(\omega)$  has been calculated 
using  the ``{\it complete Fock-space basis set}",  
developed by Andes {\it et al.\/}\cite{AndersSchiller2005,Anders-Complete-Fock}
 and by Weichselbaum and von Delft.\cite{WeichselbaumVonDelft}
In this approach, 
contributions of the high energy states which are discarded 
at the NRG steps can be recovered to form a complete basis 
for the Wilson's NRG chain, by carrying out the backward iteration.
The merit of this approach is that the sum rule for the spectral weights can be fulfilled. 

We have also employed 
the method due to Bulla, Hewson, and Pruschke:\cite{Bulla-Self-Energy} 
we have calculated not only $G_{m}(\omega)$ but also 
the higher-order Green's function $F_{m}(\omega)$ 
\begin{align}
F_{m}(\omega)\,= & \ -i \int_0^{\infty}\! dt \, 
e^{i (\omega +i0^+ )t} 
\nonumber \\
& \quad 
\times \! 
 \sum_{m'(\neq m)}
\left \langle \left\{  n_{dm'}^{}(t) 
\,d_{m}^{}(t)\, , \,d_{m}^{\dagger}(0) \right\} \right\rangle \,.
\label{definition-Fm} 
\end{align}
Then, the self-energy can be determined directly through the relation 
 $\Sigma_m(\omega) = U F_{m}(\omega)/G_{m}(\omega)$. 
The final form of the Green's function has been obtained from
 $\Sigma_{m}(\omega)$ and  the noninteracting Green's function $G_{m}^0(\omega)$
using the Dyson equation given in  Eq.\ \eqref{Dyson_eq}.  
The merit to treat the self-energy as an input is that  the fully analytic expression 
which is not affected by the logarithmic discretization 
can be used for $G_{m}^0(\omega)$.

\subsection{The  {\it z averaging} }
\label{Z-averaging-Oliveira}

The size of the Hilbert space to be diagonalized at each NRG step increases  
as the number of conduction electron channels increases.
To ensure the accuracy of the NRG calculation, 
a large $\Lambda$ is used for quantum impurities with 
a number of internal degrees of freedom.

Oliveira and Oliveira 
 found that thermodynamic averages which are  
calculated for large $\Lambda$ show an artificial oscillation 
at low temperatures,\cite{MYoshidaWhitakerOliveira,Z-average-Oliveira} 
and they proposed  the {\it z averaging\/} for removing such an artificial oscillation. 
The parameter $z$ slides a set of discretization points from that of 
the standard Wilson chain,\cite{Wilson-NRG} 
\begin{align}
\pm \Lambda^{-n}\,\rightarrow\,\pm \Lambda^{-(n+1-z)}\,, 
\qquad    n\,=\,0,1,2,\cdots,  
\end{align}
with $0\,\leq \, z \, \leq \, 1.$
For $z = 1$, it coincides with the standard Wilson chain.
The discretized conduction band can be transformed 
into a $z$ dependent Wilson chain
 \begin{align}
\!\!
 \mathcal{H}_c \,\Rightarrow  \sum_{n=0}^{\infty} \sum_{m=1}^4
 t_n^{}(z) 
\left(
 f_{n,m}^{\dagger}  f_{n+1,m}^{} +  f_{n+1,m}^{\dagger} f_{n,m}^{}
 \right) . 
 \end{align}
The hopping matrix element  $ t_n^{}(z)$ 
that can be determined  using the Hauseholder algorithm 
summarized in Refs.\ \onlinecite{MYoshidaWhitakerOliveira,Z-average-Oliveira}.
We have carried out  NRG  calculations for some fixed  values of $z$, 
and calculate expectation values using the obtained eigenstates.
Then,  an average is taken  over ``$z$"  for  two different values $z=0.5$ and $1$, 
which is enough to eliminate the artificial oscillations in our case. 

\subsection{Logarithmic-Gaussian function to broaden discrete spectral function }
\label{Logarithmic-Gaussian function to broaden discrete spectral function}
We have calculated the spectral function $A_{m}$ using the Lehman representation given in Eq.\ \eqref{Lehmann-Am}.   
The resulting function is a set of discrete $\delta$ functions at frequencies $\omega_{n}$.
Replacing the $\delta$ function to the logarithmic-Gaussian function makes $A_{m}$ continuous since the NRG energy scale $T_{L}$ given in Eq.\ \eqref{energy_scale_of_L} exponentially falls of with increasing the chain length $L$.
The function  to broaden a particle excitation peak at $\omega_{n}>0$ is 
\begin{align}
X(\omega)\,=\,\frac{\mathrm{e}^{-b^{2}/4}}{b\,\omega_{n}\sqrt{\pi}}\,\mathrm{exp}\,\left\{-\left(\frac{\mathrm{ln}\left(\omega/\omega_{n}\right)}{b}\right)^{2} \right\},
 \label{log_gausian_particle}
\end{align}
which is defined for $\omega > 0$.
Correspondingly, a hole excitation peak at $\omega_{n}<0$ is broadened by 
\begin{align}
Y(\omega)\,=\,\frac{\mathrm{e}^{-b^{2}/4}}{b\,|\omega_{n}|\sqrt{\pi}}\,\mathrm{exp}\,\left\{-\left(\frac{\mathrm{ln}\left | \omega/\omega_{n}\right | }{b}\right)^{2} \right\},
 \label{log_gausian_hole}
\end{align}
for $\omega < 0$.
We set a broadening parameter $b=1.1$ for the spectral functions plotted in Figs.\ \ref{rhobz000-4Tk-U2_All} and \ref{rhobz000-4Tk-U4_All}.
Note that the spectral weight is conserved because the broadening functions are normalized to $1$.

\section{Spectral function in atomic limit}
\label{Spectral functions in atomic limit case}

We consider the atomic limit  
in order to show  how the spectral weight of the impurity states
evolves as magnetic increases at high-energies 
in the case that the Zeemann splittings of the impurity levels  
are given by Eq.\ \eqref{eq:MatchingCondition}.

At zero temperature, the flavour $m$-resolved single-particle spectral function 
can be written in the Lehmann representation as,
\begin{align}
A_{m}^{}(\omega)\, =  & \  
\frac{1}{M}\, \sum_{i=1}^M\, 
\sum_{n}\,  \Biggl[ 
\nonumber \\
& 
\ \ \ 
\left|\langle{n}|d_{m}^{\dagger}|\Psi_{\mathrm{GS},i}\rangle\right|^{2}\,
 \delta \Bigl( \omega\,- (\,E_{n}\,-\,E_\mathrm{GS} ) \Bigr) 
\nonumber 
\\ 
&   
+ \left|\langle{n}|d_{m}|\Psi_{\mathrm{GS},i}\rangle\right|^{2}\,
 \delta \Bigl ( \omega\,+ (\,E_{n}\,-\,E_\mathrm{GS} ) \Bigr)    
\,\Biggr]\,.
\label{Lehmann-Am}
\end{align}
Here,  $|{n}\rangle$ and  $E_{n}$ 
are the  eigenstate and eigenenergy 
of Hamiltonian $\mathcal{H}$,  respectively.  
The ground state $|\Psi_{\mathrm{GS},i}\rangle$  
with the energy $E_\mathrm{GS}$ can generally be degenerate, 
and the summation over $i$ represents an average 
over  $M$-fold degenerate states.

In the atomic limit, 
the CNT dot whose eigen energies are defined is Eq.\ \eqref{eq:MatchingCondition}   
is disconnected from the leads ($v_{\nu}=0$ for $\nu=L,R$),  
and there remains two-fold degeneracy for the ground state 
at half-filling $\varepsilon_d^{} =-3U/2$, 
\begin{align}
&|\Psi_{\mathrm{GS}^{},2}^{}\rangle
\,=\,d_{1}^{\dagger}d_{2}^{\dagger}|0\rangle \;, \\ 
&|\Psi_{\mathrm{GS}^{},3}^{}\rangle
\,=\,d_{1}^{\dagger}d_{3}^{\dagger}|0\rangle \;.
\label{gs-double}
\end{align}
Either of the two one-particle states,  $m=2$ or $3$ situated 
on the Fermi level,  is occupied, 
and  the lowest one-particle level with the energy $\varepsilon_{1}$ is occupied 
while the highest one with $\varepsilon_{4}$ is empty. 
 Therefore, the ground energy for these two-electron states becomes 
\begin{align}
E_\mathrm{GS}
 \,=\,\varepsilon_{1}+\varepsilon_{2}+U 
\,=\,2\varepsilon_{d}- 2 b + U  
\,.
 \label{gs-energy}
\end{align}
We next consider a single-particle excitation to add 
an electron into the level of $m=4$, 
and a single-hole excitation to remove the electron 
occupying the $m=1$ level, 
 \begin{align}
&|\Psi_{p4}^{}\rangle
\,=\,d_{4}^{\dagger} 
|\Psi_{\mathrm{GS}^{},i}^{}\rangle
\,, \qquad 
E_{p4}= 3\varepsilon_d+3U\,,
 \\ 
&|\Psi_{h1}^{}\rangle
\,=\,d_{1}^{} |\Psi_{\mathrm{GS}^{},i}^{}\rangle 
\,,
\qquad   E_{h1}=\varepsilon_d \,. 
\end{align}
The excitation energies  from the  ground state $i=2,3$  
to these two  states are given by 
\begin{align}
E_{p4} - E_\mathrm{GS} 
\,=& \, 
\varepsilon_{d}+ 2b + 2U 
\, = \,  
2 b + \frac{U}{2} 
\,,  \\
E_\mathrm{GS} 
- E_{h1} 
\,=& \, 
 \varepsilon_{d}- 2b + U 
\, = \,  
- 2b - \frac{U}{2} 
\;.
\end{align}
Therefore, the spectral weights of these processes are given by    
\begin{align}
&A_{4}^{}(\omega)\,=\,
\delta \left( \omega\,-\,\left(2b\,+\,\frac{U}{2}\right) \right)  \,,
\notag \\
&A_{1}^{}(\omega)\,=\,
\delta \left(  \omega\,+\,\left(2b\,+\,\frac{U}{2}\right) \right) \,.
\end{align} 
These weights shift towards high-energy region  
from the usual atomic limit position  $\pm U/2$.
We have observed the corresponding shifts of the spectral weight in the NRG results 
shown in Fig.\ \ref{rhobz000-4Tk-U2_All}(c) and Fig.\ \ref{rhobz000-4Tk-U4_All}(c)  
although these atomic peaks merge with the resonance peak 
which also moves away from the Fermi level as $b$ increases.

The other single electron (hole) excitation  from the ground state 
$|\Psi_{\mathrm{GS}^{},2}^{}\rangle$  corresponds to  
an addition of an electron 
to the level $m=3$ (annihilation of  an electron from $m=2$). 
The similar excitations also occur from $|\Psi_{\mathrm{GS}^{},3}^{}\rangle$.  
The peaks corresponding  to these excitations 
appear at $\omega=\pm U/2$ in the spectral functions for $m=2$ and $3$, 
\begin{align}
A_{2}^{}(\omega) =
A_{3}^{}(\omega) =  
\frac{1}{2} \delta \Bigl(\omega - \frac{U}{2}\Bigr)
 + \frac{1}{2}\delta \Bigl( \omega + \frac{U}{2}\Bigr)  .
\end{align} 
These two peaks are equivalent to the Hubbard peaks for the SU(2) symmetric case.

\end{document}